\documentclass[english]{article}
\usepackage[T1]{fontenc}
\usepackage{color}
\usepackage{graphicx}
\usepackage{amssymb}

\makeatletter

\providecommand{\tabularnewline}{\\}

\usepackage{a4}
\input{epsfig.sty}
\def\fnum@table{\tablename~{\bf\thetable}}
\def\fnum@figure{\figurename~{\bf\thefigure}}
\def\tablename{\footnotesize{\bf Table}}
\def\figurename{\footnotesize{\bf Figure}}

\def\be{\begin{equation}}
\def\ee{\end{equation}}

\makeatother

\usepackage{babel}

\begin{document}

\title{\textbf{Monte Carlo treatment of hadronic interactions in enhanced
Pomeron scheme: I.~QGSJET-II model}}

\author{\textbf{S. Ostapchenko}%
\thanks{e-mail: sergey.ostapchenko@ntnu.no%
} \\
\textit{\small NTNU, Institutt}{\small{} }\textit{\textcolor{black}{\small for
fysikk}}\textit{\small , 7491 Trondheim, Norway}\\
\textit{\small D.V. Skobeltsyn Institute of Nuclear Physics, Moscow
State University, 119992 Moscow, Russia}\textit{ }\\
}

\maketitle
\begin{center}
\textbf{\large Abstract}
\par\end{center}{\large \par}

The construction of a Monte Carlo generator for high energy hadronic
and nuclear collisions is discussed in detail. Interactions are treated
in the framework of the Reggeon Field Theory, taking into consideration
enhanced Pomeron diagrams which are resummed to all orders in the
triple-Pomeron coupling. Soft and ``semihard'' contributions to
the underlying parton dynamics are accounted for within the ``semihard
Pomeron'' approach. The structure of cut enhanced diagrams is analyzed;
they are regrouped into a number of subclasses characterized by
positively-defined contributions which define partial weights for
various ``macro-configurations'' of hadronic final states. An iterative
procedure for a Monte Carlo generation of the structure of final states
is described. The model results for hadronic cross sections and for particle
production are compared to experimental data.

\section{Introduction\label{intro.sec} }

Nowadays Monte Carlo (MC) generators of hadronic interactions are
standard tools for data analysis in high energy collider and cosmic
ray (CR) fields. The idea behind employing such MC models is twofold.
First of all, they provide a bridge between rigorous theoretical approaches
and corresponding experimental studies, thus allowing to confront
novel ideas against observations. On the other hand, MC simulations
are an inevitable part of contemporary experimental analysis procedures,
a measurement of new phenomena depending crucially on the understanding
of the corresponding detector response and of the contribution of
the ``standard'' hadronic physics which is mimicked with the help
of the MC tools. 

In particular, hadronic interaction models play an important role
in investigations of very high energy cosmic rays. Because of the extremely
low flux of such ultra-energetic particles, they can not be detected
directly. Instead one infers their properties from measured characteristics
of nuclear-electro-magnetic cascades, so-called extensive air showers
(EAS), induced by them in the atmosphere. The corresponding analysis
relies crucially on the MC treatment of the cascade development, most
importantly, of its backbone - the cascade of hadron-nucleus (nucleus-nucleus)
interactions in the atmosphere. The peculiarity of cosmic ray applications
of hadronic interaction generators is related to the fact that one
has to treat hadronic collisions at energies orders of magnitude higher
than ones of present day colliders and that EAS characteristics depend
strongly on model predictions for very forward spectra of secondary
particles. As a consequence, CR interaction models, like DPMJET \cite{aur92},
EPOS \cite{wer06},  QGSJET \cite{qgs97},
or SYBILL \cite{ahn09}, which are designed to treat general inelastic
hadronic collisions, are developed in the framework of the Reggeon
Field Theory (RFT) \cite{gri68}, which allows one to take into
consideration contributions from both ``soft'' and ``hard'' parton
dynamics to the interaction mechanism. 

Soft nonperturbative interactions are described as soft Pomeron
exchanges and dominate hadronic collisions at large impact parameters,
thus giving important contributions to total, inelastic, and diffractive
hadron-nucleus (nucleus-nucleus) cross sections. On the other hand,
at sufficiently high energies the role of so-called semihard hadronic
collisions which involve partons of moderately large virtualities
is significantly enhanced, the smallness of the corresponding strong
coupling being compensated by large collinear and infra-red logarithms
and by high density of small $x$ partons. A convenient way to include
such processes in the RFT treatment is provided by the ``semihard
Pomeron'' approach \cite{qgs94,dre99} where the perturbative
part of an ``elementary'' semihard rescattering is described within
the DGLAP formalism, which is preceded by nonperturbative parton cascades
(``soft preevolution'') described as soft Pomeron emissions.

Additionally, high parton densities reached in ``central'' collisions
of hadrons and, especially, nuclei result in significant nonlinear
corrections to the interaction dynamics, related to parton shadowing
and saturation \cite{glr}. In MC generators, such effects are
typically accounted for in a phenomenological way, via energy-dependent
parametrizations of some model parameters. The drawback of such constructions
is evident: with nonlinear effects dominating the interaction mechanism
in the very high energy limit, model predictions are governed by the
choice of the corresponding empirical parametrization, rather than
by the underlying theoretical approach.

In this work, we choose an alternative way, treating nonlinear interaction
effects in the RFT framework as Pomeron-Pomeron interactions
 \cite{kan73,car74,kai86},
based on the recent progress in the resummation of the corresponding,
so-called enhanced, RFT diagrams \cite{ost06,ost08,ost10}.
A MC implementation of such an approach has been hampered for a long
time by two factors. First, with the energy increasing, enhanced graphs
of more and more complicated topologies start to contribute significantly
to the scattering amplitude and to partial cross sections for particular
hadronic final states. Thus, dealing with enhanced diagrams, all-order
resummation of the corresponding contributions is a must, both for
elastic scattering diagrams and for the cut diagrams representing
particular inelastic processes. Secondly, it is quite nontrivial to
split the complete set of cut enhanced diagrams into separate classes
characterized by positively-defined contributions which could be interpreted
probabilistically and employed in a MC simulation procedure. While
the first problem has been addressed in \cite{ost06,ost08,ost10},
the MC implementation of the approach is discussed in the present
work. Here we mainly address the construction of the model while the
results for various particle production processes and applications
of the model for calculations of EAS development will be the subject
of the forthcoming publication \cite{ost10a}.

The outline of the paper is as follows. In Section \ref{sec:Hadron-hadron-scattering-amplitude},
the calculation of hadron-hadron scattering amplitude is discussed,
taking into account enhanced diagram contributions. 
 In Section \ref{sec:Configurations-of-final}, we consider
unitarity cuts of elastic scattering diagrams and define partial contributions
for various ``macro-configurations'' of the interaction, which are
employed in the corresponding MC procedure, as described in Section
\ref{sec:Monte-Carlo-sampling}. Finally, in Section \ref{sec:Discussion},
we discuss characteristic features of the developed model, calibration
of model parameters, and present the model results for various
hadronic cross sections.

\section{Hadron-hadron scattering amplitude \label{sec:Hadron-hadron-scattering-amplitude}}

\begin{figure}[htb]
\begin{centering}
\includegraphics[width=7cm,height=3cm]{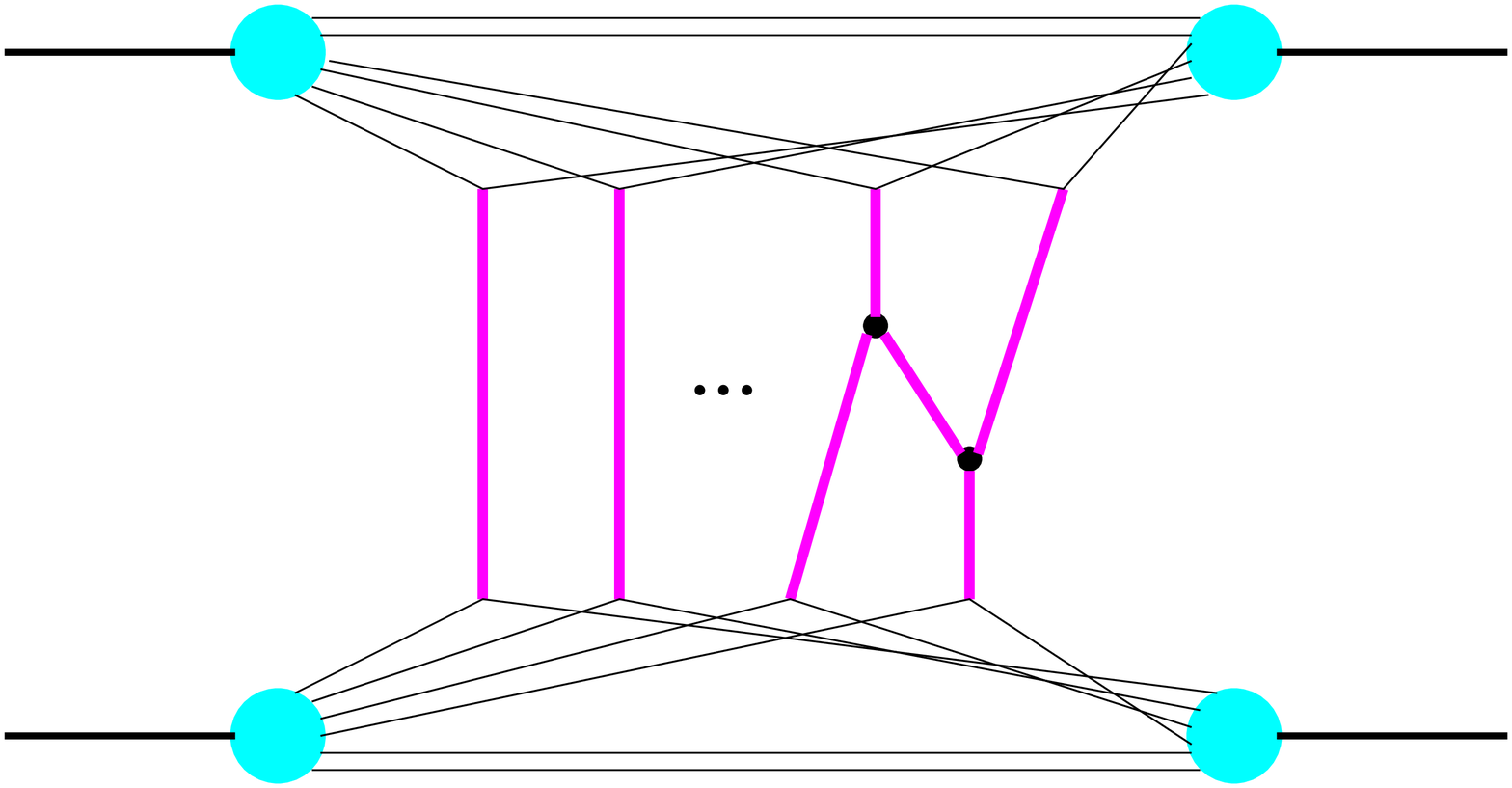}
\par\end{centering}

\caption{General multi-Pomeron contribution to hadron-hadron scattering amplitude;
elementary scattering processes correspond to Pomeron exchanges (vertical
thick lines) or to Pomeron-Pomeron interactions.\label{multiple}}

\end{figure}

In the RFT approach, high energy hadron-hadron scattering amplitude
is defined by multiple scattering graphs of the kind depicted in Fig.~\ref{multiple}.%
 The elementary rescattering contributions correspond to independent
parton cascades developing between the projectile and target hadrons,
which are described by Pomeron exchanges, and to parton cascades which
strongly overlap in the phase space and interact with each other,
which is described as Pomeron-Pomeron interactions. Thus, applying
the multichannel (Good-Walker-like) eikonal scheme \cite{goo60,kai79}
to account for contributions of small mass intermediate states between
Pomeron emissions, elastic hadron $a$ - hadron $d$ scattering amplitude
is defined as \cite{ost08}\begin{eqnarray}
f_{ad}(s,b)=i\,\sum_{j,k}C_{j/a}C_{k/d}\,\left[1-e^{-\frac{1}{2}\Omega_{ad(jk)}(s,b)}\right]\label{f_ad}\\
\Omega_{ad(jk)}(s,b)=2\chi_{ad(jk)}^{\mathbb{P}}(s,b)+2\chi_{ad(jk)}^{{\rm enh}}(s,b)\,.\label{eq:opac_ad}\end{eqnarray}
Here $s$ and $b$ are c.m.~energy squared and impact parameter for
the interaction, $C_{j/a}$ defines partial weight for hadron $a$
elastic scattering eigenstate $|j\rangle$ ($|a\rangle=\sum_{j}\sqrt{C_{j/a}}|j\rangle$,
$\sum_{j}C_{j/a}=1$), $\chi_{ad(jk)}^{\mathbb{P}}$ and $\chi_{ad(jk)}^{{\rm enh}}$
are eikonals corresponding to an exchange of a Pomeron or of an irreducible
enhanced (Pomeron-Pomeron interaction) graph between the projectile
and target hadrons, the latter being represented by eigenstates $|j\rangle$
and $|k\rangle$.

In this work, we use the ``semihard Pomeron'' approach \cite{qgs94,dre99}
to account for contributions of both nonperturbative soft processes
and of ``semihard'' ones, the latter corresponding to parton cascades
which develop at least partly in the perturbative region of relatively
high virtualities $|q^{2}|>Q_{0}^{2}$, $Q_{0}^{2}$ being some cutoff
for pQCD being applicable. Describing the former as phenomenological
soft Pomerons and the latter by ``semihard Pomeron'' exchanges, the
``general Pomeron'' eikonal is given by the sum of the two contributions:
\begin{equation}
\chi_{ad(jk)}^{\mathbb{P}}(s,b)
=\chi_{ad(jk)}^{\mathbb{P}_{{\rm soft}}}(s,b)
+\chi_{ad(jk)}^{\mathbb{P}_{{\rm sh}}}(s,b)\,.\label{eq:chi-pom-tot}
\end{equation}

The soft Pomeron eikonal $\chi_{ad(jk)}^{\mathbb{P}_{{\rm soft}}}$
is expressed via Pomeron emission vertices $N^{\mathbb{P}}$ and the
Pomeron propagator $D^{\mathbb{P}}$ as
\begin{equation}
\chi_{ad(jk)}^{\mathbb{P}_{{\rm soft}}}(s,b)
=\frac{1}{8\pi^{2}\, i\, s}\:\int\! d^{2}q\; e^{-i\,\vec{q}\,\vec{b}}\,
\int\! dx_{1}\, dx_{2}\; N_{j/a}^{\mathbb{P}}(x_{1},q^{2})\, 
N_{k/d}^{\mathbb{P}}(x_{2},q^{2})\, 
D^{\mathbb{P}}(x_{1}x_{2}s,q^{2})\,,\label{eq:chi-pom-soft}
\end{equation}
where 
\begin{equation}
D^{\mathbb{P}}(\hat{s},t)=8\pi\, i\,s_0\,(\hat{s}/s_{0})^{\alpha_{\mathbb{P}}}\: 
e^{\alpha'_{\mathbb{P}}\,\ln(s/s_{0})\: t},\label{eq:D-pom}
\end{equation}
with $\alpha_{\mathbb{P}}$ and $\alpha'_{\mathbb{P}}$ being the
intercept and the slope of the Pomeron Regge trajectory and $s_{0}\simeq1\;{\rm GeV}^{2}$
- the hadronic mass scale.

The Pomeron emission vertices are parametrized as\begin{equation}
N_{j/a}^{\mathbb{P}}(x,t)
=\gamma_{j/a}\, e^{\Lambda_{j/a}\, t}\, x^{-\alpha_{{\rm part}}}\,
(1-x)^{\alpha_{{\rm lead}}},\label{eq:N-pom}
\end{equation}
where the exponents $\alpha_{{\rm part}}\simeq0$ and $\alpha_{{\rm lead}}$
are related to intercepts of secondary Regge trajectories \cite{kai82,dre99}.

\begin{figure}[htb]
\begin{centering}
\includegraphics[width=6cm,height=3cm]{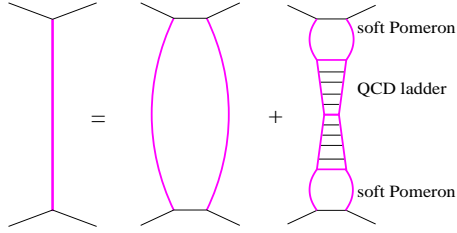}
\par\end{centering}

\caption{A {}``general Pomeron'' (l.h.s.) consists of the soft and semihard
ones - correspondingly the 1st and the 2nd contributions in the r.h.s.
\label{genpom} }

\end{figure}

The semihard contribution $\chi_{ad(jk)}^{\mathbb{P}_{{\rm sh}}}$
corresponds to a piece of QCD parton ladder sandwiched between two
soft Pomerons (see Fig.~\ref{genpom})%
 and is defined as \cite{dre99,dre01}
 \begin{eqnarray}
\chi_{ad(jk)}^{\mathbb{P}_{{\rm sh}}}(s,b)
=\frac{1}{32\pi^{2}s_0^2}\:\int\! d^{2}q\;
 e^{-i\,\vec{q}\,\vec{b}}\,\int\! dx_{1}\, dx_{2}\; 
 N_{j/a}^{\mathbb{P}}(x_{1},q^{2})\, N_{k/d}^{\mathbb{P}}(x_{2},q^{2})\,
 \int\! dx^{+}\, dx^{-}\nonumber \\
\times\:\mathfrak{Im}D^{\mathbb{P}}(s_{0}/x^{+},q^{2})\;
\mathfrak{Im}D^{\mathbb{P}}(s_{0}/x^{-},q^{2})
\sum_{I,J=g,q_{s}}g_{I}(x^{+})\; g_{J}(x^{-})\;
\sigma_{IJ}^{{\rm QCD}}(x^{+}x^{-}x_{1}x_{2}s,Q_{0}^{2})\,,\label{eq:chi-sh}
\end{eqnarray}
where the vertices $g_{I}$ for parton $I$ (gluon or sea quark)%
\footnote{For brevity, we do not discuss explicitly valence quark contributions
to the semihard eikonal; the corresponding description can be found
elsewhere \cite{dre99,dre01}.%
} coupling to the soft Pomeron are parametrized as
\begin{eqnarray}
g_{g}(z)=r_{g}\,(1-w_{qg})\,(1-z)^{\beta_{g}}\nonumber \\
g_{q_{s}}(z)=r_{g}\, w_{qg}\,\int_{z}^{1}\! dy\; 
y^{\alpha_{\mathbb{P}}-1}\: P_{qg}(y)\:(1-z/y)^{\beta_{g}},\label{eq:g-I}
\end{eqnarray}
with $P_{qg}(y)$ being the Altarelli-Parisi splitting function and
the constant $r_{g}$ being fixed by momentum conservation for parton
distribution functions (PDFs).

The contribution $\sigma_{IJ}^{{\rm QCD}}(x^{+}x^{-}s,Q_{0}^{2})$
of parton ladder with the virtuality cutoff $Q_{0}^{2}$ and with
the leg-parton types $I$, $J$ and light cone momentum fractions
$x^{+}$, $x^{-}$ is defined in a standard way (see, e.g.~\cite{dre99,dre01})
\begin{eqnarray}
\sigma_{IJ}^{{\rm QCD}}(\hat{s},Q_{0}^{2})=K\sum_{I',J'}\int\! dz^{+}\, dz^{-}\int dp_{t}^{2}\; E_{I\rightarrow I'}^{{\rm QCD}}(z^{+},Q_{0}^{2},M_{{\rm F}}^{2})\: E_{J\rightarrow J'}^{{\rm QCD}}(z^{-},Q_{0}^{2},M_{{\rm F}}^{2})\nonumber \\
\times\frac{d\sigma_{I'J'}^{2\rightarrow2}(z^{+}z^{-}\hat{s},p_{t}^{2})}{dp_{t}^{2}}\:\Theta(M_{{\rm F}}^{2}-Q_{0}^{2})\,,\label{eq:sigma-hard}\end{eqnarray}
with $d\sigma_{IJ}^{2\rightarrow2}/dp_{t}^{2}$ being the differential
parton-parton cross section, $p_{t}$ - parton transverse momentum
in the hard process, $M_{{\rm F}}^{2}$ - the factorization scale
(here $M_{{\rm F}}^{2}=p_{t}^{2}/4$), and with $E_{I\rightarrow I'}^{{\rm QCD}}(z,Q_{0}^{2},Q^{2})$
describing the evolution of the parton density from the virtuality
scale $Q_{0}^{2}$ to $Q^{2}$. The factor $K\simeq1.5$ is designed
to take effectively into account higher order QCD corrections.

The idea behind Eq.~(\ref{eq:chi-sh}) is to split parton evolution
in an elementary scattering process in two parts: i) nonperturbative
soft one described phenomenologically by the soft Pomeron asymptotics;
ii) parton cascading at $|q^{2}|>Q_{0}^{2}$, treated within the DGLAP
formalism. The former is characterized by a significant parton diffusion
in the transverse plane and, in the absence of nonlinear corrections,
forms parton (sea quark or gluon) momentum and impact parameter distributions
at the virtuality scale $Q_{0}^{2}$ \cite{dre99,dre01,ost06a}. During
the latter, parton transverse displacements can be neglected, leaving
only the momentum-dependent part, Eq.~(\ref{eq:sigma-hard}), which
is characterized by a stronger energy-rise compared to the soft
Pomeron amplitude and drives therefore the high energy behavior of
the semihard contribution (\ref{eq:chi-sh}).

To calculate enhanced diagram contributions, we adopt multi-Pomeron
vertices of the form \cite{kai86}\begin{equation}
G^{(m,n)}=G\,\gamma_{\mathbb{P}}^{m+n},\label{eq:g_mn}\end{equation}
where $m$ and $n$ are numbers of Pomerons connected to the vertex
from the projectile, respectively target, side ($m+n\geq3$) and the
constant $G$ is related to the triple-Pomeron coupling $r_{3\mathbb{P}}$
as $G=r_{3\mathbb{P}}/(4\pi\gamma_{\mathbb{P}}^{3})$. 

The eikonal $\chi_{a(j)}^{\mathbb{P}}(y,b)$ for a ``general Pomeron''
exchange between hadron $a$ (represented by eigenstate $|j\rangle)$
and a multi-Pomeron vertex, the two being separated from each other
by rapidity $y$ and transverse distance $b$, also receives contributions
from both soft and semihard processes
\begin{equation}
\chi_{a(j)}^{\mathbb{P}}(y,b)=
\chi_{a(j)}^{\mathbb{P}_{{\rm soft}}}(y,b)
+\chi_{a(j)}^{\mathbb{P}_{{\rm sh}}}(y,b)\,,\label{eq:chi-leg-pom}
\end{equation}
with the partial contributions $\chi_{a(j)}^{\mathbb{P}_{{\rm soft}}}$
and $\chi_{a(j)}^{\mathbb{P}_{{\rm sh}}}$ being defined similarly
to (\ref{eq:chi-pom-soft}), (\ref{eq:chi-sh}) \cite{ost06a}:
\begin{eqnarray}
\chi_{a(j)}^{\mathbb{P}_{{\rm soft}}}(y,b)=
\frac{\gamma_{\mathbb{P}}}{8\pi^{2}\, i\, s_{0}\,e^{y}}\:\int\! d^{2}q\;
 e^{-i\,\vec{q}\,\vec{b}}\,\int\! dx_{1}\;
  N_{j/a}^{\mathbb{P}}(x_{1},q^{2})\, 
  D^{\mathbb{P}}(x_{1}s_{0}\,e^{y},q^{2})\label{eq:chi-leg-soft}\\
\chi_{a(j)}^{\mathbb{P}_{{\rm sh}}}(y,b)=
\frac{\gamma_{\mathbb{P}}}{32\pi^{2}s_0^2}\:\int\! d^{2}q\; 
e^{-i\,\vec{q}\,\vec{b}}\,\int\! dx_{1}\; N_{j/a}^{\mathbb{P}}(x_{1},q^{2})\,
\int\! dx^{+}\, dx^{-}\:\mathfrak{Im}D^{\mathbb{P}}(s_{0}/x^{+},q^{2})
\nonumber \\
\times\;\mathfrak{Im}D^{\mathbb{P}}(s_{0}/x^{-},q^{2})
\sum_{I,J=g,q_{s}}g_{I}(x^{+})\; g_{J}(x^{-})\;
\sigma_{IJ}^{{\rm QCD}}(x^{+}x^{-}x_{1}s_{0}\,e^{y},Q_{0}^{2})\,,
\label{eq:chi-leg-sh}
\end{eqnarray}
where we included the vertex factors $\gamma_{\mathbb{P}}$ into the
definition of the eikonals.

Similarly, for a Pomeron exchange between two multi-Pomeron vertices
separated by rapidity and impact parameter distances $y$ and $b$
we use\begin{eqnarray}
\chi^{\mathbb{P}}(y,b)=\chi^{\mathbb{P}_{{\rm soft}}}(y,b)+
\chi^{\mathbb{P}_{{\rm sh}}}(y,b)\label{eq:chi-int-pom}\\
\chi^{\mathbb{P}_{{\rm soft}}}(y,b)=
\frac{\gamma_{\mathbb{P}}^{2}}{8\pi^{2}\, i\, s_{0}\,e^{y}}\:
\int\! d^{2}q\; e^{-i\,\vec{q}\,\vec{b}}\; D^{\mathbb{P}}(s_{0}\,e^{y},q^{2})
\label{eq:chi-int=soft}\\
\chi^{\mathbb{P}_{{\rm sh}}}(y,b)=\frac{\gamma_{\mathbb{P}}^{2}}
{32\pi^{2}s_0^2}\:\int\! d^{2}q\; e^{-i\,\vec{q}\,\vec{b}}\,
\int\! dx^{+}\, dx^{-}\:\mathfrak{Im}D^{\mathbb{P}}(s_{0}/x^{+},q^{2})\;
\mathfrak{Im}D^{\mathbb{P}}(s_{0}/x^{-},q^{2})\nonumber \\
\times\sum_{I,J=g,q_{s}}g_{I}(x^{+})\; g_{J}(x^{-})\;
\sigma_{IJ}^{{\rm QCD}}(x^{+}x^{-}s_{0}\,e^{y},Q_{0}^{2})\,.
\label{eq:chi-int-sh}\end{eqnarray}

\begin{figure}[t]
\begin{centering}
\includegraphics[width=7cm,height=2.5cm]{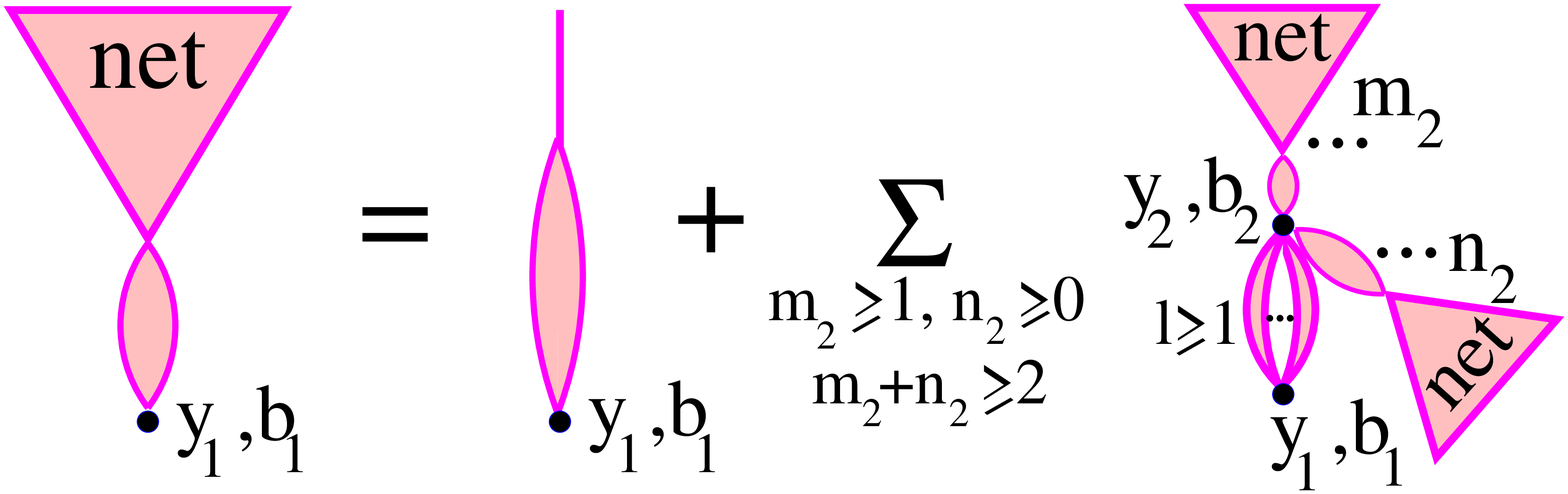}
\par\end{centering}

\caption{Recursive equation for projectile net-fan contribution $\chi_{a(j)|d(k)}^{{\rm net}}(y_{1},\vec{b}_{1}|Y,\vec{b})$;
$y_{1}$ and $b_{1}$ are rapidity and impact parameter distances
between the projectile proton and the vertex in the handle of the
fan. The vertex $(y_{2},\vec{b}_{2})$ couples together $m_{2}$ projectile
net-fans and $n_{2}$ target net-fans. In addition, there are $l\geq1$
irreducible 2-point sequences of Pomerons and Pomeron loops, exchanged
between the vertices $(y_{1},\vec{b}_{1})$ and $(y_{2},\vec{b}_{2})$.
\label{freve}}

\end{figure}

The above-defined eikonals can be used to calculate the total contribution
of irreducible enhanced Pomeron graphs $\chi_{ad(jk)}^{{\rm enh}}$.
As demonstrated in \cite{ost06,ost08}, the latter can be expressed
via contributions of subgraphs of certain structure,  so-called
``net-fans''. Those are defined by the Schwinger-Dyson equation of
Fig.~\ref{freve} %
 and correspond to arbitrary irreducible ``nets'' of Pomerons, with
neighboring net ``cells'' being connected to each other by 2-point
sequences of Pomerons and Pomeron loops and with one vertex in the
net having a fixed position in the rapidity and impact parameter space.%
\footnote{The 2-point sequence of Pomerons and Pomeron loops, coupled to this
vertex, will be referred to as the ``handle of the fan''.%
} The corresponding equation
\begin{eqnarray}
\chi_{a(j)|d(k)}^{{\rm net}}(y_{1},\vec{b}_{1}|Y,\vec{b})=
\chi_{a(j)}^{{\rm loop}}(y_{1},b_{1})+G\int_{\xi}^{y_{1}-
\xi}\! dy_{2}\int\! d^{2}b_{2}\;
\left(1-e^{-\chi^{{\rm loop}}(y_{1}-y_{2},|\vec{b}_{1}-\vec{b}_{2}|)}\right)
\nonumber \\
\times\left[\left(1-
e^{-\chi_{a(j)|d(k)}^{{\rm net}}(y_{2},\vec{b}_{2}|Y,\vec{b})}\right)\;
 e^{-\chi_{d(k)|a(j)}^{{\rm net}}(Y-y_{2},\vec{b}-\vec{b}_{2}|Y,\vec{b})}
 -\chi_{a(j)|d(k)}^{{\rm net}}(y_{2},\vec{b}_{2}|Y,\vec{b})\right]
 \label{net-fan}\end{eqnarray}
involves  the contribution 
$\chi^{{\rm loop}}(y_{1}-y_{2},|\vec{b}_{1}-\vec{b}_{2}|)$
of irreducible 2-point sequences of Pomerons and Pomeron loops, 
exchanged between the vertices $(y_{1},\vec{b}_{1})$ and $(y_{2},\vec{b}_{2})$
(2nd graph in the r.h.s.~of Fig.~\ref{freve}), and the 
 contribution $\chi_{a(j)}^{{\rm loop}}(y_{1},b_{1})$ of Pomeron loop
sequences exchanged between the vertex $(y_{1},\vec{b}_{1})$ and
hadron $a$,  with a single Pomeron coupled to hadron $a$ (1st graph in
the r.h.s.~of the Figure).
 The $y_{2}$ integration in the 2nd term in the r.h.s.~of (\ref{net-fan})
is performed between $\xi$ and $y_{1}-\xi$, with $\xi$ being the
minimal rapidity interval for the Pomeron asymptotics to be applicable.
As discussed in \cite{ost06a}, net-fan eikonals
 $\chi_{a(j)|d(k)}^{{\rm net}}$
are related to parton (sea quark and gluon) distributions which are
probed during hadron-hadron interaction and are thus influenced by
rescattering processes on the partner hadron.

\begin{figure}[t]
\begin{centering}
\includegraphics[width=6cm,height=6cm]{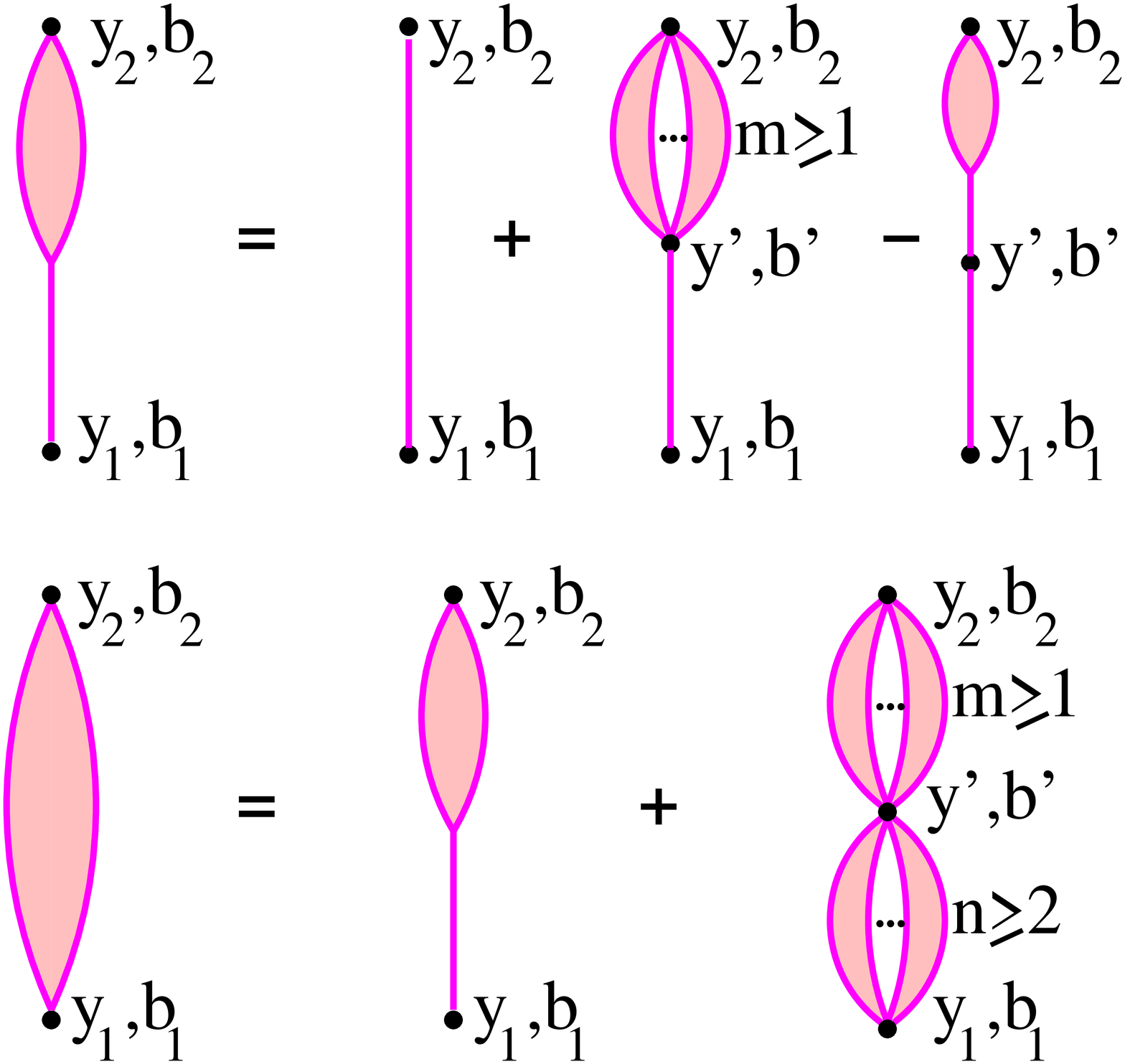}
\par\end{centering}

\caption{Recursive representations for the contributions of irreducible 2-point
sequences of Pomerons and Pomeron loops $\chi^{{\rm loop(1)}}$ (top)
and $\chi^{{\rm loop}}$ (bottom), exchanged between the vertices
$(y_{1},\vec{b}_{1})$ and $(y_{2},\vec{b}_{2})$.\label{fig: loops}}

\end{figure}

As in \cite{ost10}, we define the contribution
 $\chi^{{\rm loop}}(y_{1}-y_{2},|\vec{b}_{1}-\vec{b}_{2}|)$
and a part of it $\chi^{{\rm loop(1)}}(y_{1}-y_{2},|\vec{b}_{1}-\vec{b}_{2}|)$,
corresponding to Pomeron loop sequences which start from a single
Pomeron coupled to the vertex $(y_{1},\vec{b}_{1})$, via Schwinger-Dyson
equations of Fig.~\ref{fig: loops}, %
 which gives\begin{eqnarray}
\chi^{{\rm loop}(1)}(y_{2}-y_{1},|\vec{b}_{2}-\vec{b}_{1}|)=\chi^{\mathbb{P}}(y_{2}-y_{1},|\vec{b}_{2}-\vec{b}_{1}|)+G\int_{y_{1}+\xi}^{y_{2}-\xi}\! dy'\int\! d^{2}b'\;\chi^{\mathbb{P}}(y'-y_{1},|\vec{b}'-\vec{b}_{1}|)\nonumber \\
\times\left[1-e^{-\chi^{{\rm loop}}(y_{2}-y',|\vec{b}_{2}-\vec{b}'|)}-\chi^{{\rm loop(1)}}(y_{2}-y',|\vec{b}_{2}-\vec{b}'|)\right]\label{eq:loop(1)}\\
\chi^{{\rm loop}}(y_{2}-y_{1},|\vec{b}_{2}-\vec{b}_{1}|)=\chi^{{\rm loop}(1)}(y_{2}-y_{1},|\vec{b}_{2}-\vec{b}_{1}|)+G\int_{y_{1+}\xi}^{y_{2}-\xi}\! dy'\int\! d^{2}b'\;\nonumber \\
\times\left[1-e^{-\chi^{{\rm loop}}(y'-y_{1},|\vec{b}'-\vec{b}_{1}|)}-\chi^{{\rm loop}}(y'-y_{1},|\vec{b}'-\vec{b}_{1}|)\right]\left[1-e^{-\chi^{{\rm loop}}(y_{2}-y',|\vec{b_{2}}-\vec{b}'|)}\right].\label{eq:loop}\end{eqnarray}

In turn, for the contribution $\chi_{a(j)}^{{\rm loop}}(y_{1},b_{1})$
and a part of it $\chi_{a(j)}^{{\rm loop(1)}}(y_{1},b_{1})$, corresponding
to irreducible Pomeron loop sequences with a single Pomeron coupled
to the vertex $(y_{1},\vec{b}_{1})$, this leads
to recursive equations \cite{ost10}
\begin{eqnarray}
\chi_{a(j)}^{{\rm loop}}(y_{1},b_{1})=\chi_{a(j)}^{\mathbb{P}}(y_{1},b_{1})+G\int_{\xi}^{y_{1}-\xi}\! dy'\int\! d^{2}b'\;\chi_{a(j)}^{\mathbb{P}}(y',b')\nonumber \\
\times\left[1-e^{-\chi^{{\rm loop}}(y_{1}-y',|\vec{b}_{1}-\vec{b}'|)}-\chi^{{\rm loop(1)}}(y_{1}-y',|\vec{b}_{1}-\vec{b}'|)\right]\label{leg}\\
\chi_{a(j)}^{{\rm loop}(1)}(y_{1},b_{1})=\chi_{a(j)}^{\mathbb{P}}(y_{1},b_{1})+G\int_{\xi}^{y_{1}-\xi}\! dy'\int\! d^{2}b'\;\left[\chi_{a(j)}^{{\rm loop}}(y',b')-\chi_{a(j)}^{{\rm loop(1)}}(y',b')\right]\nonumber \\
\times\,\chi^{\mathbb{P}}(y_{1}-y',|\vec{b}_{1}-\vec{b}'|)\,.\label{leg(1)}\end{eqnarray}

Finally, for the total contribution of irreducible enhanced graphs
$\chi_{ad(jk)}^{{\rm enh}}$ (exchanged between eigenstates $|j\rangle$
and $|k\rangle$ of the projectile and of the target respectively)
to elastic scattering amplitude one obtains \cite{ost08,ost10}
\begin{eqnarray}
\chi_{ad(jk)}^{{\rm enh}}(s,b)
=G\int_{\xi}^{Y-\xi}\!\! dy_{1}\!\int\!\! d^{2}b_{1}\;
\left\{ \left[\left(1-e^{-\chi_{a(j)|d(k)}^{{\rm net}}}\right)
\left(1-e^{-\chi_{d(k)|a(j)}^{{\rm net}}}\right)
-\chi_{a(j)|d(k)}^{{\rm net}}\,\chi_{d(k)|a(j)}^{{\rm net}}\right]\right.
\nonumber \\
-\left[\chi_{a(j)|d(k)}^{{\rm net}}
-\chi_{a(j)}^{{\rm loop}}(Y-y_{1},|\vec{b}-\vec{b}_{1}|)\right]
\left[\left(1-e^{-\chi_{d(k)|a(j)}^{{\rm net}}}\right)
e^{-\chi_{a(j)|d(k)}^{{\rm net}}}-\chi_{d(k)|a(j)}^{{\rm net}}\right]
\nonumber \\
+\left.\chi_{d(k)}^{\mathbb{P}}(y_{1},b_{1})
\left[\chi_{a(j)}^{{\rm loop}}(Y-y_{1},|\vec{b}-\vec{b}_{1}|)
-\chi_{a(j)}^{{\rm loop(1)}}(Y-y_{1},|\vec{b}-\vec{b}_{1}|)\right]\right\}\!,
\label{eq:chi-enh}\end{eqnarray}
where $Y=\ln (s/s_0)$ and 
the omitted arguments of the eikonals read $\chi_{a(j)|d(k)}^{{\rm net}}
=\chi_{a(j)|d(k)}^{{\rm net}}(Y-y_{1},\vec{b}-\vec{b}_{1}|Y,\vec{b})$,
$\chi_{d(k)|a(j)}^{{\rm net}}=\chi_{d(k)|a(j)}^{{\rm net}}(y_{1},\vec{b}_{1}|Y,\vec{b})$.
As demonstrated in \cite{ost10}, Eqs.~(\ref{net-fan}-\ref{eq:chi-enh})
account for all important enhanced diagram contributions to elastic
scattering amplitude. 

The generalization of the treatment for nucleus-nucleus
scattering amplitude is described in Appendix~A.

\section{Configurations of final states\label{sec:Configurations-of-final}}

The knowledge of the elastic scattering amplitude is far insufficient
for the construction of a MC procedure for hadronic and nuclear inelastic
collisions. What we need are partial cross sections for particular
configurations of final states, which are defined by contributions
of the corresponding unitarity cuts of elastic scattering diagrams.
Those can be easily derived in the nonenhanced eikonal scheme using
the Abramovskii-Gribov-Kancheli (AGK) cutting rules \cite{agk}. Considering
diagrams with precisely $m\geq1$ Pomerons being cut, each cut Pomeron
corresponding to an elementary production process, and summing
over any number of uncut Pomerons which describe additional elastic
rescatterings, one obtains the so-called topological cross sections
for hadron-hadron scattering \cite{kai79,kai82}:\begin{equation}
\sigma_{ad}^{(m)}(s)=\int\! d^{2}b\;\sum_{j,k}C_{j/a}C_{k/d}\,\frac{\left(2\chi_{ad(jk)}^{\mathbb{P}}(s,b)\right)^{m}}{m!}\, e^{-2\chi_{ad(jk)}^{\mathbb{P}}(s,b)}.\label{eq:topol xsec}\end{equation}
The integrand in (\ref{eq:topol xsec}) can be interpreted as a probability
to have precisely $m$ elementary production processes in the hadron-hadron
collision at impact parameter $b$. On the other hand, combining diagrams
where the cut plane passes between $n\geq2$ Pomerons, none being
cut, and choosing either elastic or inelastic states in the cut plane
for the projectile and the target, one obtains either elastic $\sigma_{ad}^{{\rm el}}$
or various (low mass) diffraction cross sections. For example,
 for $\sigma_{ad}^{{\rm el}}$
and for single projectile hadron diffraction cross section one obtains
\cite{kai79}:\begin{eqnarray}
\sigma_{ad}^{{\rm el}}(s)=\int\! d^{2}b\;\left[\sum_{j,k}C_{j/a}C_{k/d}\,\left(1-e^{-\chi_{ad(jk)}^{\mathbb{P}}(s,b)}\right)\right]^{2}\label{eq:sig-el-ad}\\
\sigma_{ad}^{{\rm SD(proj)}}(s)=\int\! d^{2}b\;\sum_{j,k,l,m}(C_{j/a}\,\delta_{jl}-C_{j/a}C_{l/a})\, C_{k/d}C_{m/d}\; e^{-\chi_{ad(jk)}^{\mathbb{P}}(s,b)-\chi_{ad(lm)}^{\mathbb{P}}(s,b)}.\label{eq:sig-sd-ad}\end{eqnarray}
It is worth stressing that a configuration of the final state is defined
by the structure of the unitarity cuts, here - by the number of \textsl{cut
Pomeron }exchanges, and implies a resummation of all absorptive corrections
due to virtual rescattering processes - uncut Pomeron exchanges.

Taking into account enhanced Pomeron diagrams significantly complicates
the analysis and produces a variety of final state configurations,
including e.g.~ones with single or multiple large rapidity gaps (LRG)
not covered by secondary particle production. The complete set of
AGK-based cut enhanced diagrams has been derived in \cite{ost08},
the corresponding contributions being composed of various unitarity
cuts of net-fan subgraphs, and contains cut diagrams of two types.
The first class consists of cut diagrams characterized by a ``tree''-like
structure of cuts; such graphs are constructed coupling arbitrary
numbers of cut and uncut net-fan contributions in one central
(not necessarily unique) vertex, such that each cut net-fan subgraph
is characterized by a ``fan''-like structure of cuts \cite{ost08}.
The diagrams of the second kind are characterized by a ``zigzag''-like
structure of cuts; they are constructed in a similar way, with the
important difference that at least one of the cut net-fan subgraphs
has a zigzaglike structure of cuts, with subsequent Pomeron end rapidities
satisfying $y_{1}>y_{2}<y_{3}>\cdots$. Treelike cut diagrams give
important contributions to the total cross section and to partial
cross sections of various final states; they provide main corrections
to inclusive spectra of secondary particles. On the other hand, various
contributions of zigzaglike cut graphs to the total cross section
precisely cancel each other, moreover, as demonstrated in \cite{ost10},
they do not influence noticeably the rapidity gap structure of final
states. Nevertheless, such diagrams provide   contributions to
inclusive particle spectra and to partial cross sections for particular
final states. Therefore, our strategy will be to develop first a MC
scheme taking into consideration treelike cut enhanced graphs only.
After that, the procedure will be complemented by taking into account
zigzaglike cut contributions.

\begin{figure}[htb]
\begin{centering}
\includegraphics[width=15cm,height=6cm]{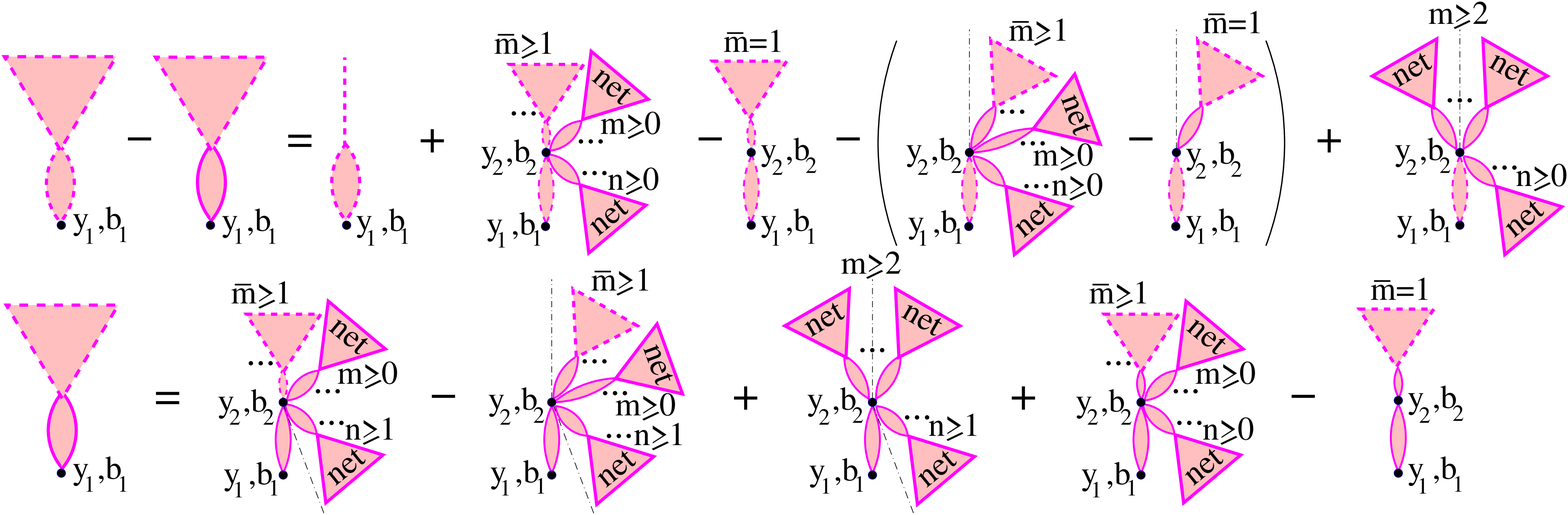}
\par\end{centering}

\caption{Recursive representations for cut net-fan diagrams characterized by
a fanlike structure of cuts. The top line of the figure defines the
contribution $2\hat{\chi}_{a(j)|d(k)}^{{\rm fan}}$ of the subset
of graphs in which the handle of the fan is cut; the bottom
line gives the one of the diagrams with uncut handle,
 $2\tilde{\chi}_{a(j)|d(k)}^{{\rm fan}}$.
\label{fan-cut-fig}}

\end{figure}

As discussed above, to obtain cross sections for various final state
configurations, we shall use as ``building blocks'' contributions
of various unitarity cuts of net-fan graphs. For the moment, we are
interested in the AGK-cuts of net-fans, characterized by a fanlike
structure of cuts, which are defined by the Schwinger-Dyson equations
of Fig.~\ref{fan-cut-fig}~%
\footnote{Here and in the following we use a slightly different graphic notation
compared to Figs.~\ref{freve},~\ref{fig: loops}: a shaded ellipse
with solid margins, positioned between the vertices $(y_{1},b_{1})$
and $(y_{2},b_{2})$, denotes a general (not necessarily irreducible)
2-point sequence of Pomerons and Pomeron loops exchanged between these
vertices, with the corresponding contribution $[1-\exp(-\chi^{{\rm loop}}(y_{1}-y_{2},|\vec{b}_{1}-\vec{b}_{2}|))]$.
Similarly, such an ellipse with dashed margins corresponds to the
AGK cuts of such a sequence, with the contribution $2[1-\exp(-\chi^{{\rm loop}}(y_{1}-y_{2},|\vec{b}_{1}-\vec{b}_{2}|))]$.%
} \cite{ost08,ost10}. %
 The top line of the Figure defines the contribution $2\hat{\chi}_{a(j)|d(k)}^{{\rm fan}}$
of the subset of graphs in which the Pomeron loop sequence coupled
to the  vertex $(y_{1},b_{1})$ (the handle of the fan) is crossed
by the cut plane. In turn, the equation in the bottom line gives the
one of the diagrams where the handle of the fan remains uncut,
 $2\tilde{\chi}_{a(j)|d(k)}^{{\rm fan}}$.
The total contribution of fanlike cuts of net-fans is thus
 $2\bar{\chi}_{a(j)|d(k)}^{{\rm fan}}=2\hat{\chi}_{a(j)|d(k)}^{{\rm fan}}
 +2\tilde{\chi}_{a(j)|d(k)}^{{\rm fan}}$.
The first graph in the r.h.s.~of the top line corresponds to all
possible AGK-based cuts of the single 2-point sequence of Pomerons
and Pomeron loops exchanged between the vertex $(y_{1},b_{1})$ and
the projectile hadron whereas the next two diagrams in the r.h.s.~of
the graphic equation describe the development of the cut Pomeron net:
The vertex $(y_{2},b_{2})$ couples together $\bar{m}\geq1$ cut projectile
net-fans, each one characterized by a fanlike structure of cuts,
and any numbers $m,n\geq0$ of uncut projectile and target net-fans,
such that $\bar{m}+m+n\geq2$. There one has to subtract the contributions
of the next two diagrams which correspond to configurations of non-AGK
type, where in all the $\bar{m}$ cut projectile net-fans connected
to the vertex $(y_{2},b_{2})$ the handles of the fans remain uncut
and are situated on the same side of the cut plane, together with
all the $m$ uncut projectile net-fans. Finally, in the last graph
in the top line of the Figure the cut plane passes between $m\geq2$
uncut projectile net-fans, with at least one remaining on either side
of the cut, such that a large rapidity gap is formed between the projectile
proton and the vertex $(y_{2},b_{2})$. The diagrams in the bottom
line of the Figure
 have a similar structure, as discussed in more detail in \cite{ost08,ost10}.

As demonstrated in \cite{ost08}, the total contribution of fanlike
cuts of net-fans coincides with twice the uncut one: \begin{eqnarray}
2\bar{\chi}_{a(j)|d(k)}^{{\rm fan}}(y_{1},\vec{b}_{1}|Y,\vec{b})=2\chi_{a(j)|d(k)}^{{\rm net}}(y_{1},\vec{b}_{1}|Y,\vec{b}),\label{equiv}\end{eqnarray}
whereas for $2\hat{\chi}_{a(j)|d(k)}^{{\rm fan}}$ one obtains the
recursive equation\begin{eqnarray}
2\hat{\chi}_{a(j)|d(k)}^{{\rm fan}}(y_{1},\vec{b}_{1}|Y,\vec{b})=2\chi_{a(j)}^{{\rm loop}}(y_{1},b_{1})+2G\int_{\xi}^{y_{1}-\xi}\! dy_{2}\int\! d^{2}b_{2}\;\left(1-e^{-\chi^{{\rm loop}}(y_{1}-y_{2},|\vec{b}_{1}-\vec{b}_{2}|)}\right)\nonumber \\
\times\left[\left(1-e^{-\hat{\chi}_{a(j)|d(k)}^{{\rm fan}}(y_{2},\vec{b}_{2}|Y,\vec{b})}\right)e^{-2\chi_{d(k)|a(j)}^{{\rm net}}(Y-y_{2},\vec{b}-\vec{b}_{2}|Y,\vec{b})}-\hat{\chi}_{a(j)|d(k)}^{{\rm fan}}(y_{2},\vec{b}_{2}|Y,\vec{b})\right]\!.\label{eq: fan-fan}\end{eqnarray}

In Appendix B we derive also alternative representations for $2\hat{\chi}_{a(j)|d(k)}^{{\rm fan}}$,
$2\tilde{\chi}_{a(j)|d(k)}^{{\rm fan}}$ which can be used in a MC
procedure to generate the cut Pomeron structure of an irreducible
cut diagram, to be discussed in Section \ref{sec:Monte-Carlo-sampling}.
In addition, we obtain there subcontributions $2\bar{\chi}_{a(j)|d(k)}^{{\rm loop}}$,
$2\tilde{\chi}_{a(j)|d(k)}^{{\rm loop}}$, which correspond to such
cuts of net-fan graphs in which just one \textsl{cut} Pomeron is coupled
to hadron $a$.

\begin{figure}[htb]
\begin{centering}
\includegraphics[width=15cm,height=10cm]{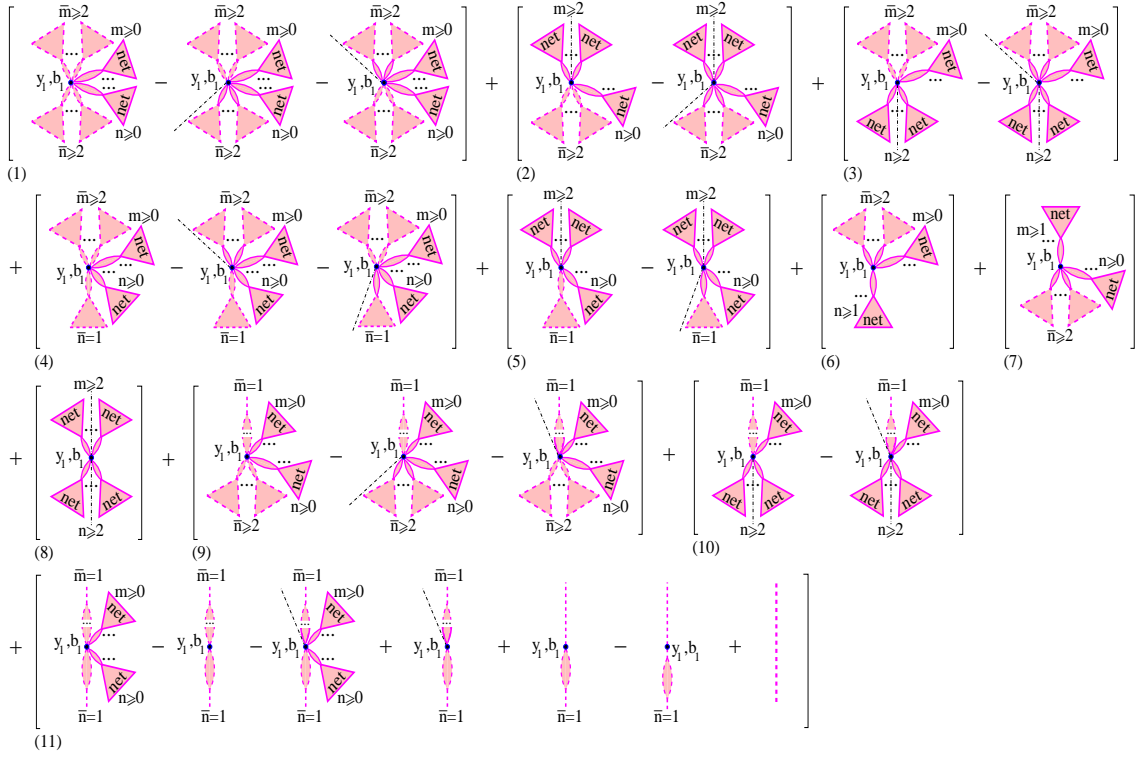}
\par\end{centering}

\caption{Complete set of irreducible cut diagrams characterized by a ``tree''-like
structure of cuts. \label{fig: tree-dif1}}

\end{figure}

Using the above-defined building blocks, the complete set of cut irreducible
graphs (with a treelike structure of cuts) for hadron-hadron
scattering is given in Fig.~\ref{fig: tree-dif1}~%
\footnote{As discussed in \cite{ost08}, the set of diagrams of Fig.~\ref{fig: tree-dif1}
can also be represented in a form explicitly symmetric with respect
to the projectile and the target.%
} \cite{ost08}.%
 Each square bracket in the Figure corresponds to a positively defined
contribution of a certain ``macro-configuration'' of the final state,
characterized by certain topology of \textsl{cut} Pomerons, hence,
by a definite pattern for secondary hadron production. For example,
the diagrams in the first square bracket in Fig.~\ref{fig: tree-dif1}
correspond to the configuration with at least two cut projectile and
target net-fans ($\bar{m}\geq2$, $\bar{n}\geq2$) coupled together
in the vertex $(y_{1},\vec{b}_{1})$, which results in a treelike
structure of the final state:~%
\footnote{The 2nd and 3rd graphs in the square bracket define the subtracted
contributions of the cuts of non-AGK types.%
} In each of the $\bar{m}$ cut projectile net-fans any cut Pomeron
may split into a few, forming a fanlike structure (composed of cut
Pomerons) developing towards the projectile hadron (in the 0s order
with respect to the triple-Pomeron coupling a cut net-fan is represented
by a single cut Pomeron exchanged between the hadron and the multi-Pomeron
vertex); in all the $\bar{n}$ cut target net-fans such fanlike structures
develop towards the target hadron. For the corresponding contribution
we obtain\begin{eqnarray}
\bar{\Omega}_{ad(jk)}^{(1)}(s,b)=\frac{G}{2}\int_{\xi}^{Y-\xi}\!\! dy_{1}\!\int\!\! d^{2}b_{1}\;\sum_{\bar{m}=2}^{\infty}\sum_{\bar{n}=2}^{\infty}\left\{ \frac{\left[2\hat{\chi}_{a(j)|d(k)}^{{\rm fan}}+2\tilde{\chi}_{a(j)|d(k)}^{{\rm fan}}\right]^{\bar{m}}}{\bar{m}!}\, e^{-2\chi_{a(j)|d(k)}^{{\rm net}}}\right.\nonumber \\
\times\,\frac{\left[2\hat{\chi}_{d(k)|a(j)}^{{\rm fan}}+2\tilde{\chi}_{d(k)|a(j)}^{{\rm fan}}\right]^{\bar{n}}}{\bar{n}!}\, e^{-2\chi_{d(k)|a(j)}^{{\rm net}}}-2\,\frac{\left[2\hat{\chi}_{a(j)|d(k)}^{{\rm fan}}+2\tilde{\chi}_{a(j)|d(k)}^{{\rm fan}}\right]^{\bar{m}}}{\bar{m}!}\, e^{-2\chi_{a(j)|d(k)}^{{\rm net}}}\nonumber \\
\times\left.\frac{\left[\tilde{\chi}_{d(k)|a(j)}^{{\rm fan}}\right]^{\bar{n}}}{\bar{n}!}\, e^{-\chi_{d(k)|a(j)}^{{\rm net}}}-2\,\frac{\left[\tilde{\chi}_{a(j)|d(k)}^{{\rm fan}}\right]^{\bar{m}}}{\bar{m}!}\, e^{-\chi_{a(j)|d(k)}^{{\rm net}}}\,\frac{\left[2\hat{\chi}_{d(k)|a(j)}^{{\rm fan}}+2\tilde{\chi}_{d(k)|a(j)}^{{\rm fan}}\right]^{\bar{n}}}{\bar{n}!}\, e^{-2\chi_{d(k)|a(j)}^{{\rm net}}}\right\} \nonumber \\
=\frac{G}{2}\int_{\xi}^{Y-\xi}\!\! dy_{1}\!\int\!\! d^{2}b_{1}\left\{ \left[1-e^{-2\chi_{a(j)|d(k)}^{{\rm net}}}\left(1+2\chi_{a(j)|d(k)}^{{\rm net}}\right)\right]\left[1-e^{-2\chi_{d(k)|a(j)}^{{\rm net}}}\left(1+2\chi_{d(k)|a(j)}^{{\rm net}}\right)\right]\right.\nonumber \\
-2\left[1-e^{-2\chi_{a(j)|d(k)}^{{\rm net}}}\left(1+2\chi_{a(j)|d(k)}^{{\rm net}}\right)\right]\left[e^{\tilde{\chi}_{d(k)|a(j)}^{{\rm fan}}}-1-\tilde{\chi}_{d(k)|a(j)}^{{\rm fan}}\right]e^{-\chi_{d(k)|a(j)}^{{\rm net}}}\nonumber \\
-\left.2\left[e^{\tilde{\chi}_{a(j)|d(k)}^{{\rm fan}}}-1-\tilde{\chi}_{a(j)|d(k)}^{{\rm fan}}\right]e^{-\chi_{a(j)|d(k)}^{{\rm net}}}\left[1-e^{-2\chi_{d(k)|a(j)}^{{\rm net}}}\left(1+2\chi_{d(k)|a(j)}^{{\rm net}}\right)\right]\right\} \!,\label{eq:chi-enh(1)}\end{eqnarray}
where the abbreviations are similar to the ones in (\ref{eq:chi-enh}).

The 2nd class of graphs corresponds to a LRG produced (in one particular
inelastic rescattering process) between the projectile hadron and
the vertex $(y_{1},\vec{b}_{1})$ and at least two cut target net-fans
($\bar{n}\geq2$) coupled to the vertex $(y_{1},\vec{b}_{1})$, while
in the third configuration the projectile and the target interchange
their places. The corresponding contributions read\begin{eqnarray}
\bar{\Omega}_{ad(jk)}^{(2)}(s,b)=\frac{G}{2}\int_{\xi}^{Y-\xi}\!\! dy_{1}\!\int\!\! d^{2}b_{1}\left[1-e^{-\chi_{a(j)|d(k)}^{{\rm net}}}\right]^{2}\nonumber \\
\times\left\{ \left[1-e^{-2\chi_{d(k)|a(j)}^{{\rm net}}}\left(1+2\chi_{d(k)|a(j)}^{{\rm net}}\right)\right]-2\left[e^{\tilde{\chi}_{d(k)|a(j)}^{{\rm fan}}}-1-\tilde{\chi}_{d(k)|a(j)}^{{\rm fan}}\right]e^{-\chi_{d(k)|a(j)}^{{\rm net}}}\right\} \label{eq:chi-enh(2_}\\
\bar{\Omega}_{ad(jk)}^{(3)}(s,b)=\frac{G}{2}\int_{\xi}^{Y-\xi}\!\! dy_{1}\!\int\!\! d^{2}b_{1}\left[1-e^{-\chi_{d(k)|a(j)}^{{\rm net}}}\right]^{2}\nonumber \\
\times\left\{ \left[1-e^{-2\chi_{a(j)|d(k)}^{{\rm net}}}\left(1+2\chi_{a(j)|d(k)}^{{\rm net}}\right)\right]-2\left[e^{\tilde{\chi}_{a(j)|d(k)}^{{\rm fan}}}-1-\tilde{\chi}_{a(j)|d(k)}^{{\rm fan}}\right]e^{-\chi_{a(j)|d(k)}^{{\rm net}}}\right\} \!.\label{eq:chi-enh(3)}\end{eqnarray}

The next two contributions are similar to the first pair, with the
difference that there is precisely one cut target net-fan ($\bar{n}=1$)
coupled to the vertex $(y_{1},\vec{b}_{1})$:\begin{eqnarray}
\bar{\Omega}_{ad(jk)}^{(4)}(s,b)=G\int_{\xi}^{Y-\xi}\!\! dy_{1}\!\int\!\! d^{2}b_{1}\left\{ \left[1-e^{-2\chi_{a(j)|d(k)}^{{\rm net}}}\left(1+2\chi_{a(j)|d(k)}^{{\rm net}}\right)\right]\left[\chi_{d(k)|a(j)}^{{\rm net}}\, e^{-2\chi_{d(k)|a(j)}^{{\rm net}}}\right.\right.\nonumber \\
-\left.\left.\tilde{\chi}_{d(k)|a(j)}^{{\rm fan}}\, e^{-\chi_{d(k)|a(j)}^{{\rm net}}}\right]-2\left[e^{\tilde{\chi}_{a(j)|d(k)}^{{\rm fan}}}-1-\tilde{\chi}_{a(j)|d(k)}^{{\rm fan}}\right]e^{-\chi_{a(j)|d(k)}^{{\rm net}}}\,\chi_{d(k)|a(j)}^{{\rm net}}\, e^{-2\chi_{d(k)|a(j)}^{{\rm net}}}\right\} \label{eq:chi-enh(4)}\\
\bar{\Omega}_{ad(jk)}^{(5)}(s,b)=G\int_{\xi}^{Y-\xi}\!\! dy_{1}\!\int\!\! d^{2}b_{1}\left[1-e^{-\chi_{a(j)|d(k)}^{{\rm net}}}\right]^{2}\left[\chi_{d(k)|a(j)}^{{\rm net}}\, e^{-2\chi_{d(k)|a(j)}^{{\rm net}}}-\tilde{\chi}_{d(k)|a(j)}^{{\rm fan}}\, e^{-\chi_{d(k)|a(j)}^{{\rm net}}}\right]\!.\label{eq:chi-enh(5)}\end{eqnarray}

In the 6th contribution, the secondary particles produced are separated
from the target hadron by a LRG which extends beyond the vertex $(y_{1},\vec{b}_{1})$:
In all the $\bar{m}\geq2$ cut projectile net-fans the handles of
the fans are uncut and positioned on the same side of the cut plane,
together with all the $m\geq0$ uncut projectile and $n\geq1$ target
net-fans. In the next graph the projectile and the target interchange
their roles, the two contributions being\begin{eqnarray}
\bar{\Omega}_{ad(jk)}^{(6)}(s,b)=2G\int_{\xi}^{Y-\xi}\!\! dy_{1}\!\int\!\! d^{2}b_{1}\left[e^{\tilde{\chi}_{a(j)|d(k)}^{{\rm fan}}}-1-\tilde{\chi}_{a(j)|d(k)}^{{\rm fan}}\right]e^{-\chi_{a(j)|d(k)}^{{\rm net}}}\left(1-e^{-\chi_{d(k)|a(j)}^{{\rm net}}}\right)\label{eq:chi-enh(6)}\\
\bar{\Omega}_{ad(jk)}^{(7)}(s,b)=2G\int_{\xi}^{Y-\xi}\!\! dy_{1}\!\int\!\! d^{2}b_{1}\left[e^{\tilde{\chi}_{d(k)|a(j)}^{{\rm fan}}}-1-\tilde{\chi}_{d(k)|a(j)}^{{\rm fan}}\right]e^{-\chi_{d(k)|a(j)}^{{\rm net}}}\left(1-e^{-\chi_{a(j)|d(k)}^{{\rm net}}}\right)\!.\label{eq:chi-enh(7)}\end{eqnarray}

In the graph in the 8th square bracket, there are only uncut net-fans
coupled to the vertex $(y_{1},\vec{b}_{1})$; particle production
emerges here from the cut multi-Pomeron vertex $(y_{1},\vec{b}_{1})$
only and is separated by large rapidity gaps from both the projectile
and the target.%
\footnote{In the following we shall neglect the production of such low mass
diffractive states at central rapidities, such that this particular
set of diagrams will contribute to (quasi-)elastic rescattering processes
only. As demonstrated in \cite{ost10}, such low mass diffractive
states produced at central rapidities do not provide significant contributions
to diffraction cross sections, with the sole exception of the central
diffraction (double Pomeron exchange).%
} For the corresponding contribution we easily obtain\begin{eqnarray}
\bar{\Omega}_{ad(jk)}^{(8)}(s,b)=\frac{G}{2}\int_{\xi}^{Y-\xi}\!\! dy_{1}\!\int\!\! d^{2}b_{1}\left[1-e^{-\chi_{a(j)|d(k)}^{{\rm net}}}\right]^{2}\left[1-e^{-\chi_{d(k)|a(j)}^{{\rm net}}}\right]^{2}\!.\label{eq:chi-enh(8)}\end{eqnarray}

The next set of cut enhanced diagrams reminds the one in the 4th square
bracket, being reversed upside-down, with the difference that the
vertex $(y_{1},\vec{b}_{1})$ is coupled to the projectile by a single
cut sequence of Pomerons and Pomeron loops, the corresponding contributions
$2\bar{\chi}_{a(j)|d(k)}^{{\rm loop}}$, $2\tilde{\chi}_{a(j)|d(k)}^{{\rm loop}}$
being defined in Appendix~B. Thus, the cut Pomeron ``tree'' develops
here towards the target while there is only one cut Pomeron coupled
to the projectile. The partial contribution of such a configuration
is\begin{eqnarray}
\bar{\Omega}_{ad(jk)}^{(9)}(s,b)=G\int_{\xi}^{Y-\xi}\!\! dy_{1}\!\int\!\! d^{2}b_{1}\left\{ \left[\bar{\chi}_{a(j)|d(k)}^{{\rm loop}}\, e^{-2\chi_{a(j)|d(k)}^{{\rm net}}}-\tilde{\chi}_{a(j)|d(k)}^{{\rm loop}}\, e^{-\chi_{a(j)|d(k)}^{{\rm net}}}\right]\right.\nonumber \\
\times\left.\left[1-e^{-2\chi_{d(k)|a(j)}^{{\rm net}}}\left(1+2\chi_{d(k)|a(j)}^{{\rm net}}\right)\right]-2\bar{\chi}_{a(j)|d(k)}^{{\rm loop}}\, e^{-2\chi_{a(j)|d(k)}^{{\rm net}}}\left[e^{\tilde{\chi}_{d(k)|a(j)}^{{\rm fan}}}-1-\tilde{\chi}_{d(k)|a(j)}^{{\rm fan}}\right]e^{-\chi_{d(k)|a(j)}^{{\rm net}}}\right\} \!.\label{eq:chi-enh(9)}\end{eqnarray}

In turn, the next contribution corresponds to a single cut sequence
of Pomerons and Pomeron loops, exchanged between the projectile and
the vertex $(y_{1},\vec{b}_{1})$ and being separated from the target
by a LRG:\begin{eqnarray}
\bar{\Omega}_{ad(jk)}^{(10)}(s,b)=G\int_{\xi}^{Y-\xi}\!\! dy_{1}\!\int\!\! d^{2}b_{1}\left[\bar{\chi}_{a(j)|d(k)}^{{\rm loop}}\, e^{-2\chi_{a(j)|d(k)}^{{\rm net}}}-\tilde{\chi}_{a(j)|d(k)}^{{\rm loop}}\, e^{-\chi_{a(j)|d(k)}^{{\rm net}}}\right]\left[1-e^{-\chi_{d(k)|a(j)}^{{\rm net}}}\right]^{2}.\label{eq:chi-enh(10)}\end{eqnarray}

Finally, the graphs in the last square bracket in Fig.~\ref{fig: tree-dif1}
describe an exchange of a single cut sequence of Pomerons and Pomeron
loops between the projectile and the target hadrons, which includes
also a single cut Pomeron exchange, with the contribution\begin{eqnarray}
\bar{\Omega}_{ad(jk)}^{(11)}(s,b)=2\chi_{ad(jk)}^{\mathbb{P}}(s,b)+2G\int_{\xi}^{Y-\xi}\!\! dy_{1}\!\int\!\! d^{2}b_{1}\left\{ \chi_{d(k)}^{{\rm loop}}(y_{1},b_{1})\left[\bar{\chi}_{a(j)|d(k)}^{{\rm loop}}\left(e^{-2\chi_{a(j)|d(k)}^{{\rm net}}-2\chi_{d(k)|a(j)}^{{\rm net}}}-1\right)\right.\right.\nonumber \\
-\left.\left.\tilde{\chi}_{a(j)|d(k)}^{{\rm loop}}\left(e^{-\chi_{a(j)|d(k)}^{{\rm net}}-2\chi_{d(k)|a(j)}^{{\rm net}}}-1\right)\right]+\chi_{a(j)}^{\mathbb{P}}(Y-y_{1},|\vec{b}-\vec{b}_{1}|)\left[\chi_{d(k)}^{{\rm loop}}(y_{1},b_{1})-\chi_{d(k)}^{{\rm loop(1)}}(y_{1},b_{1})\right]\right\} \!.\label{eq:chi-enh(11)}\end{eqnarray}

As shown in \cite{ost08}, one has the identity
\begin{equation}
\sum_{i=1}^{11}\bar{\Omega}_{ad(jk)}^{(i)}(s,b)=
\Omega_{ad(jk)}(s,b)\,,\label{eq:opac-iden}
\end{equation}
which relates the summary contribution of all the considered cut diagrams
to the total opacity $\Omega_{ad(jk)}$ for hadron-hadron scattering,
Eqs.~(\ref{eq:opac_ad}),~(\ref{eq:chi-enh}). Relation (\ref{eq:opac-iden})
is a direct consequence of the $s$-channel unitarity of the approach
and of the fact that contributions of zigzaglike cut graphs to the
elastic scattering amplitude precisely cancel each other \cite{ost08}.
Using (\ref{eq:opac-iden}), we can easily write down the absorptive
cross section which corresponds to multiple secondary hadron production,
including high mass diffraction processes:
\begin{eqnarray}
\sigma_{ad}^{{\rm abs}}(s)=
\int\! d^{2}b\;\sum_{N=1}^{\infty}
\frac{\left[\sum_{i=1}^{11}\bar{\Omega}_{ad(jk)}^{(i)}(s,b)\right]^{N}}{N!}\;
 e^{-\Omega_{ad(jk)}(s,b)}=\int\! d^{2}b\left[1-e^{-\Omega_{ad(jk)}(s,b)}\right]
 \!,\label{eq:sigma-ad-abs}
 \end{eqnarray}
where the factor $\left[\sum_{i=1}^{11}\bar{\Omega}_{ad(jk)}^{(i)}\right]^{N}/N!=\left[\Omega_{ad(jk)}\right]^{N}/N!$
comes from an exchange of precisely $N$ irreducible cut graphs whereas
the factor $\exp\!\left[-\Omega_{ad(jk)}\right]$ is obtained summing
over any number ($\geq0$) of elastic rescattering processes due to the
exchanges of uncut graphs. Proceeding as in Appendix~A,
the treatment can be generalized to the case of nucleus-nucleus (hadron-nucleus)
collisions, as outlined in Appendix C.

It is noteworthy that total inelastic cross section contains also
contributions from low mass diffraction of the projectile and/or target
hadrons:\begin{equation}
\sigma_{ad}^{{\rm inel}}(s)=\sigma_{ad}^{{\rm abs}}(s)+\sigma_{ad}^{{\rm SD(proj)}}(s)+\sigma_{ad}^{{\rm SD(targ)}}(s)+\sigma_{ad}^{{\rm DD}}(s)\,,\label{eq:sigma-inel-ad}\end{equation}
where the latter are defined as {[}cf.~(\ref{eq:sig-sd-ad}){]}\begin{eqnarray}
\sigma_{ad}^{{\rm SD(proj)}}(s)=\int\! d^{2}b\;\sum_{j,k,l,m}(C_{j/a}\,\delta_{jl}-C_{j/a}C_{l/a})\, C_{k/d}C_{m/d}\; e^{-\frac{1}{2}\Omega_{ad(jk)}(s,b)-\frac{1}{2}\Omega_{ad(lm)}(s,b)}\label{eq:sig-ad-proj-d}\\
\sigma_{ad}^{{\rm SD(targ)}}(s)=\int\! d^{2}b\;\sum_{j,k,l,m}C_{j/a}C_{l/a}(C_{k/d}\,\delta_{km}-C_{k/d}C_{m/d})\; e^{-\frac{1}{2}\Omega_{ad(jk)}(s,b)-\frac{1}{2}\Omega_{ad(lm)}(s,b)}\label{eq:sig-ad-targ-d}\\
\sigma_{ad}^{{\rm DD}}(s)=\int\! d^{2}b\;\sum_{j,k,l,m}(C_{j/a}\,\delta_{jl}-C_{j/a}C_{l/a})\,(C_{k/d}\,\delta_{km}-C_{k/d}C_{m/d})\; e^{-\frac{1}{2}\Omega_{ad(jk)}(s,b)-\frac{1}{2}\Omega_{ad(lm)}(s,b)}.\label{eq:sig-ad-dd}\end{eqnarray}

Eqs.~(\ref{eq:sigma-ad-abs}-\ref{eq:sig-ad-dd}) form the basis
for a MC treatment of inelastic hadron-hadron collisions. In particular,
using Eq.~(\ref{eq:sigma-ad-abs}), for a given geometrical configuration
of the collision (impact parameter $\vec{b}$ and elastic scattering
eigenstates $j$ and $k$ of the projectile and target hadrons) the
factor $\left[\Omega_{ad(jk)}\right]^{N}/N!\:\exp\!\left[-\Omega_{ad(jk)}\right]$
can be interpreted as the probability for precisely N elementary inelastic
interactions to take place in the collision. Each of the elementary
interactions may have different topologies (defined by the structure
of the unitarity cuts) of the kinds discussed above, characterized
by partial probabilities $\bar{\Omega}_{ad(jk)}^{(i)}/\Omega_{ad(jk)}$.
It is worth stressing that the probabilistic interpretation of Eq.~(\ref{eq:sigma-ad-abs})
and the positive-definiteness of the partial cut contributions $\bar{\Omega}_{ad(jk)}^{(i)}$
are due to the full resummation of absorptive corrections due to virtual
rescattering processes, in particular, due to the resummation of all
the irreducible cut diagrams characterized by a given topology of
the cuts, with any number of uncut Pomerons and any number of multi-Pomeron
vertices.

\section{Monte Carlo procedure\label{sec:Monte-Carlo-sampling}}

The obtained expressions allow a relatively straightforward MC implementation
of the approach, which is discussed below for the case of hadron-hadron
scattering. While   low mass diffraction processes, being sampled
according to the corresponding probabilities $\sigma_{ad}^{{\rm SD(proj)}}/\sigma_{ad}^{{\rm inel}}$,
$\sigma_{ad}^{{\rm SD(targ)}}/\sigma_{ad}^{{\rm inel}}$, and $\sigma_{ad}^{{\rm DD}}/\sigma_{ad}^{{\rm inel}}$
{[}Eqs.~(\ref{eq:sig-ad-proj-d})-(\ref{eq:sig-ad-dd}){]}, are treated
like in the original QGSJET model \cite{qgs97,qgs93} - assuming the
$\mathbb{PPR}$-asymptotics for the mass distribution of diffractive
states, the ``true inelastic'' interactions, which have the partial
probability $\sigma_{ad}^{{\rm abs}}/\sigma_{ad}^{{\rm inel}}$, are
simulated as follows. One starts from sampling the squared impact
parameter for the collision - uniformly in the area $b^{2}<b_{\max}^{2}$,
with $b_{max}$ chosen sufficiently large, corresponding to negligibly
small interaction probability at $b>b_{max}$. In addition, one generates
elastic scattering eigenstates $j$ and $k$ for the projectile and
target hadrons - according to their partial weights $C_{j/a}$, $C_{k/d}$.
In the specified geometry, one defines the number $N\geq0$ of elementary
inelastic processes according to the Poisson distribution with the
mean $\Omega_{ad(jk)}$ - Eq.~(\ref{eq:sigma-ad-abs}); in case $N=0$
the chosen geometry is rejected and the above-discussed steps are
repeated.

Next, for each of the $N$ elementary production processes one chooses
first the ``macro-structure'' of the contributing cut diagrams (as
defined in Fig.~\ref{fig: tree-dif1}) - according to the positively-defined
weights $\bar{\Omega}_{ad(jk)}^{(i)}/\Omega_{ad(jk)}$, and reconstructs
the configuration of cut Pomerons for the corresponding set of irreducible
cut graphs. For example, for the macro-configuration of the 1st square
bracket in Fig.~\ref{fig: tree-dif1} one chooses the rapidity $y_{1}$
and transverse vector $\vec{b}_{1}$ of the central multi-Pomeron
vertex $(y_{1},\vec{b}_{1})$ - according to the integrand of
 Eq.~(\ref{eq:chi-enh(1)}),
and samples the numbers of cut projectile and target net-fans $\bar{m}$,
$\bar{n}$ using the Poisson distribution with the corresponding mean values
$2\chi_{a(j)|d(k)}^{{\rm net}}$, $2\chi_{d(k)|a(j)}^{{\rm net}}$
(rejecting the cases $\bar{m},\bar{n}<2$) - see the 1st term in the
integrand of Eq.~(\ref{eq:chi-enh(1)}). For each of the $\bar{m}$
cut projectile net-fans one decides if the handle of the fan is cut
- with the probability $\hat{\chi}_{a(j)|d(k)}^{{\rm fan}}/\chi_{a(j)|d(k)}^{{\rm net}}$,
or   uncut (similarly for the $\bar{n}$ cut target net-fans);
the 2nd and 3rd terms in the integrand of Eq.~(\ref{eq:chi-enh(1)})
are accounted for via rejection in the case all the $\bar{m}$ cut
projectile net-fans and/or the $\bar{n}$ cut target net-fans have
their handles uncut. After that, the cut Pomeron structure for each
of the $\bar{m}+\bar{n}$ cut net-fans is reconstructed using an iterative
procedure, as discussed in Appendix~D. By the end of the procedure
one is left with cut Pomeron contributions of three types: i) stretched
between the projectile and target hadrons; ii) between a given (projectile
or target) hadron and a certain multi-Pomeron vertex; iii) between
a pair of multi-Pomeron vertices.

Each of those cut Pomeron contributions corresponds to an underlying
elementary parton cascade developing in the respective rapidity range;
hadronization of partons results in the production of secondary hadrons
which densely fill that rapidity interval. For example, a cut Pomeron
exchanged between the projectile and the target hadrons gives rise
to particle production in the whole range $[0,Y]$.%
\footnote{Constituent partons (``Pomeron ends'') are characterized by a relatively
hard light cone momentum distribution, hence, no LRGs arise from the
energy-momentum partition between those partons and the hadron ``remnant''
state. %
}A cut Pomeron exchanged between, say, projectile hadron and some multi-Pomeron
vertex $(y',\vec{b}')$ results in a chain of secondaries covering
the range $[y',Y]$, etc. It is noteworthy that we speak here about
cut Pomeron contributions in the sense of Fig.~\ref{fig:cutpom-line},
i.e.~accounting also for absorptive corrections for the corresponding
configuration of the final state: The cut Pomeron contribution includes
also the ones of diagrams with additional multi-Pomeron vertices placed
along the cut Pomeron line; those vertices are coupled to \textsl{uncut}
Pomerons which are in turn connected to the projectile and/or the
target and/or to other uncut Pomerons.

\begin{figure}[htb]

\begin{centering}
\includegraphics[width=9cm,height=3.5cm]{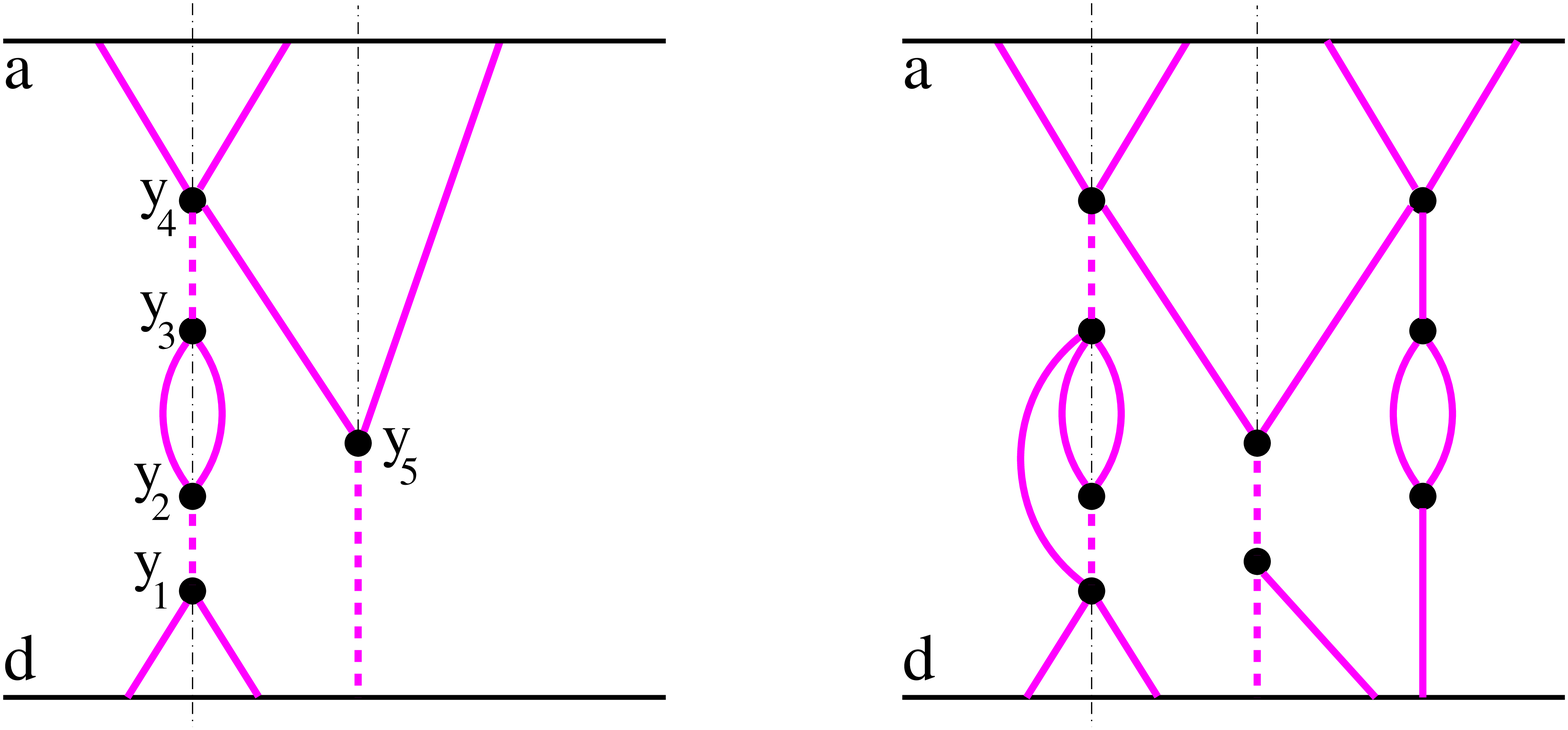}\caption{Example of a cut enhanced graph corresponding to a final state with
two LRGs (left) and a more complicated graph which describes absorptive
corrections to the same final state (right). \label{fig:lrg-ex}}

\par\end{centering}

\end{figure}

Certain configurations obtained may contain large rapidity gaps not
covered by secondary particles - when the corresponding rapidity intervals
are not spanned by any cut Pomeron. For example, in the configuration
of Fig.~\ref{fig:lrg-ex},%
 cut Pomerons cover rapidity intervals $[y_{1},y_{2}]$, $[y_{3},y_{4}]$,
and $[0,y_{5}]$, resulting in the production of chains of secondary
particles in those rapidity ranges. Hence, LRGs are produced in the
intervals $[y_{5},y_{3}]$ and $[y_{4},Y]$. On the other hand, there
is no rapidity gap in the interval $[0,y_{1}]$ which is covered by
secondaries emerging from the rightmost cut Pomeron in the graph.
While the diagram in Fig.~\ref{fig:lrg-ex}~(left) is the simplest
one corresponding to the discussed final state, the one in Fig.~\ref{fig:lrg-ex}~(right)
illustrates some absorptive corrections to the discussed configuration,
which are accounted for by the formalism.

In addition to the already generated configuration, which is based
on treelike cut enhanced graphs, an additional set of cut Pomerons
comes from zigzaglike cut diagrams. The latter are treated using
an effective procedure, as outlined in Appendix~E.

At the next step, both for the projectile and the target hadrons one
performs energy-momentum sharing between all the constituent partons
(Pomeron ends) connected to them and generates inelastic excitations
of the remaining remnant states. Finally, for each cut Pomeron contribution,
either exchanged between the projectile and the target, or between
a given (projectile or target) hadron and a multi-Pomeron vertex,
or between a pair of multi-Pomeron vertices, one chooses whether it
is represented by its soft or semihard Pomeron component. In the latter
case, like in the nonenhanced Pomeron scheme \cite{qgs97,dre01},
one samples the light cone momenta for the ``leg''-partons of the
QCD ladder and performs simulation of the development of the corresponding
perturbative parton cascade. One employs the standard treatment to
reconstruct the pattern of both initial and final state parton emission
using the forward evolution algorithms described in \cite{dre01}.
One ends up with the formation of strings stretched between the Pomeron
end-point partons in case of soft Pomerons; for semihard Pomerons
such strings are stretched also between the final $s$-channel partons
resulted from the perturbative cascades, following the direction of
the color flow.

The treatment is completed with the fragmentation of strings into
secondary hadrons, which is performed using the original procedure
of the QGSJET model \cite{qgs93}, using the algorithm described in \cite{wer89},
 with string fragmentation parameters
expressed via intercepts of secondary Regge trajectories \cite{kai84}.

\section{Some results and discussion\label{sec:Discussion}}

The basic model parameters have been calibrated from the combined
description of total and elastic hadron-proton cross sections, elastic
scattering slopes, and total and diffractive structure functions $F_{2}$,
$F_{2}^{{\rm D(3)}}$, the latter two being calculated as described
in \cite{ost06a}, generalizing the corresponding expressions to account
also for Pomeron loop contributions. 
In turn, the parameters for the
hadronization procedure have been tuned comparing with data on hadron
production in proton-proton interactions, using also new data sets
obtained at the Large Hadron Collider. 

\begin{figure}[t]
\begin{centering}
\includegraphics[width=15cm,height=7cm]{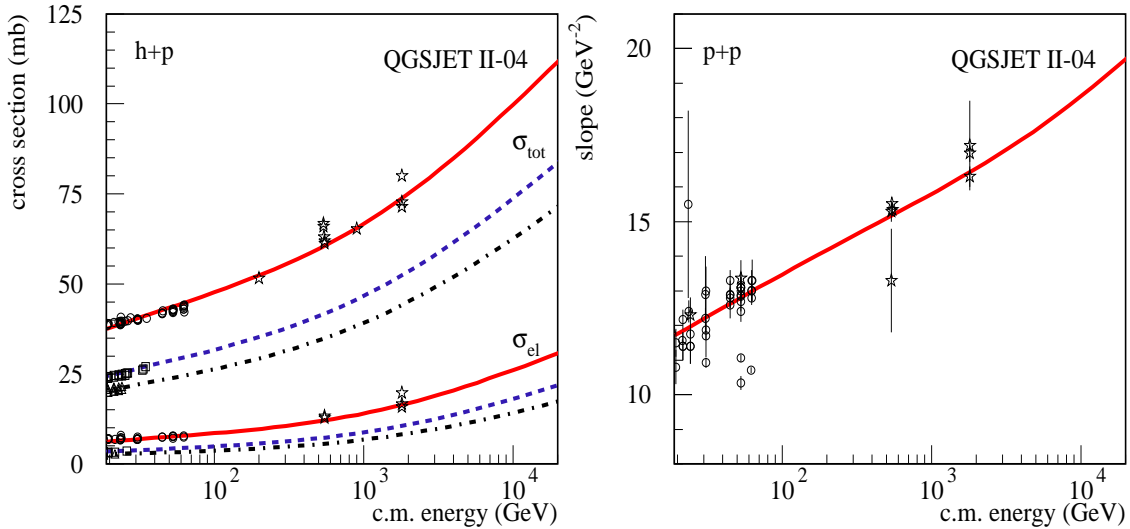}
\par\end{centering}

\caption{Left: Calculated total and elastic proton-proton, pion-proton, and
kaon-proton cross sections - respectively solid, dashed, and dotted-dashed
lines. Right: Calculated elastic scattering slope for proton-proton
scattering. The compilation of experimental data (points) is from
Ref.~\cite{pdg10}.\label{fig:sigtot}}

\end{figure}

\begin{figure}[htb]
\begin{centering}
\includegraphics[width=15cm,height=9.5cm]{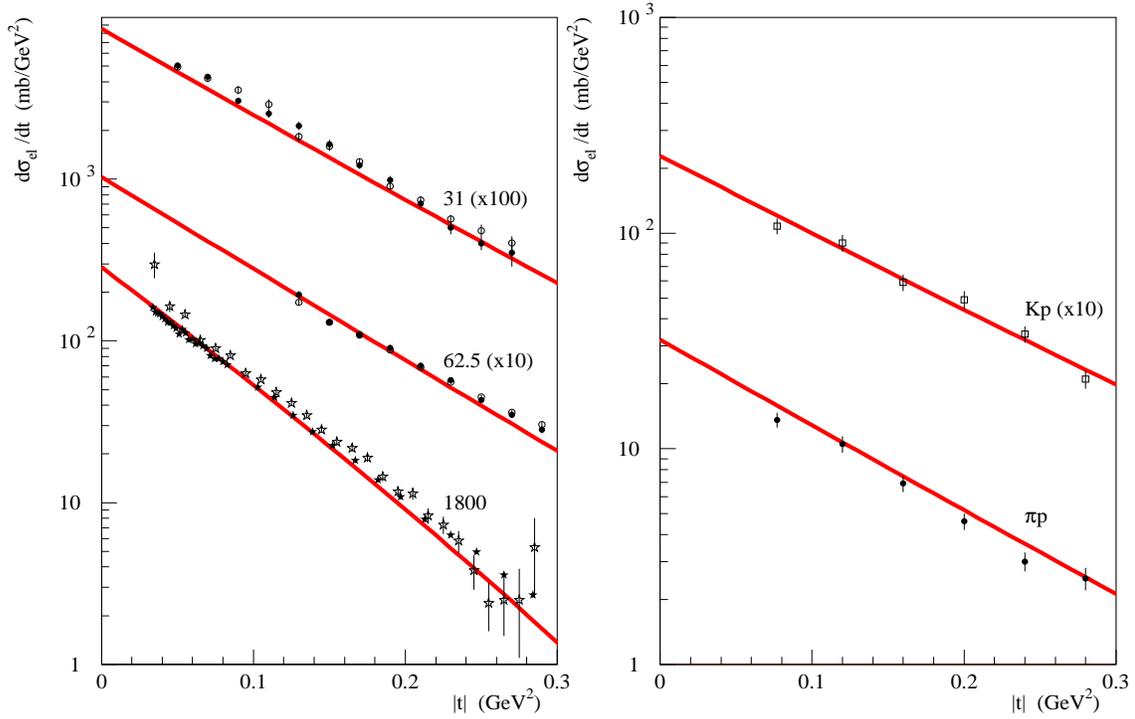}
\par\end{centering}

\caption{Left: Calculated differential elastic proton-proton cross section
for different $\sqrt{s}$ in GeV (as indicated in the plot). Right:
Calculated differential elastic pion-proton and kaon-proton cross
sections for the projectile lab.~momentum 250 GeV/c. Experimental
data are from Refs.~\cite{bre84,bat83,boz84,amo90,abe94b,ada87}.\label{fig:dsigel}}

\end{figure}

\begin{figure}[htb]
\begin{centering}
\includegraphics[width=15cm,height=5cm]{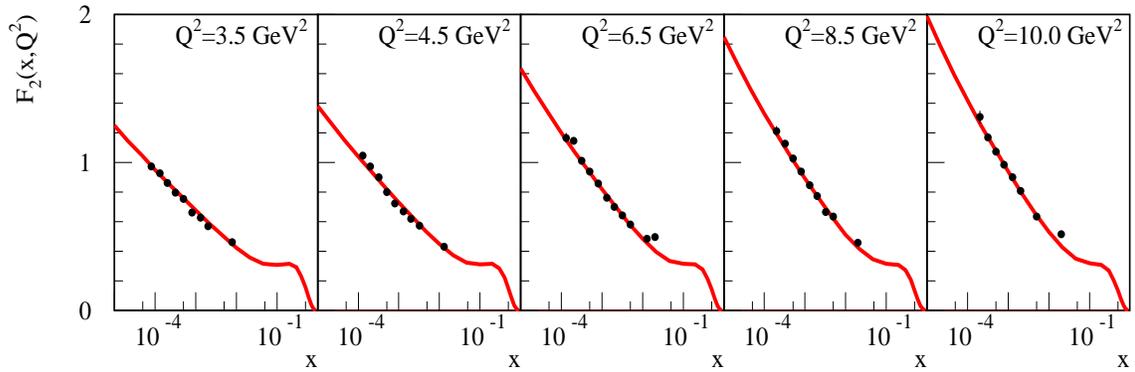}\caption{Calculated proton 
structure function $F_{2}(x,Q^{2})$ compared to HERA
data \cite{aar10}.\label{fig:f2}}

\par\end{centering}

\end{figure}

Using the virtuality cutoff $Q_{0}^{2}=3\;{\rm GeV}^{2}$ between
the soft and hard parton evolution, we obtained in particular for
the soft Pomeron intercept and slope $\alpha_{\mathbb{P}}=1.17$,
$\alpha_{\mathbb{P}}'=0.11$, while for the triple-Pomeron coupling
we got $r_{3\mathbb{P}}=0.1$ GeV, with $\gamma_{\mathbb{P}}=0.4$ GeV$^{-1}$.
 The corresponding results for
 $\sigma_{hp}^{{\rm tot}}(s)$,
$\sigma_{hp}^{{\rm el}}(s)$, $B_{pp}^{{\rm el}}(s)$,
 $d\sigma_{hp}^{{\rm el}}(s,t)/dt$,
and for proton structure function $F_{2}(x,Q^{2})$
are given in Figs.~\ref{fig:sigtot}, \ref{fig:dsigel}, \ref{fig:f2}%
 in comparison with experimental data. Using the simple exponential
form (\ref{eq:N-pom}) for the $t$-dependence of hadronic form factors,
the calculated differential elastic cross sections agree reasonably
well with measurements at small $|t|\lesssim0.3\;{\rm GeV}^{2}$ which
are responsible for the bulk of secondary hadron production. To have
a better agreement at larger values of $|t|$, a dipole parametrization
for the form factor would be more suitable. 

\begin{figure}[htb]
\centering{}\includegraphics[width=15cm,height=9cm]{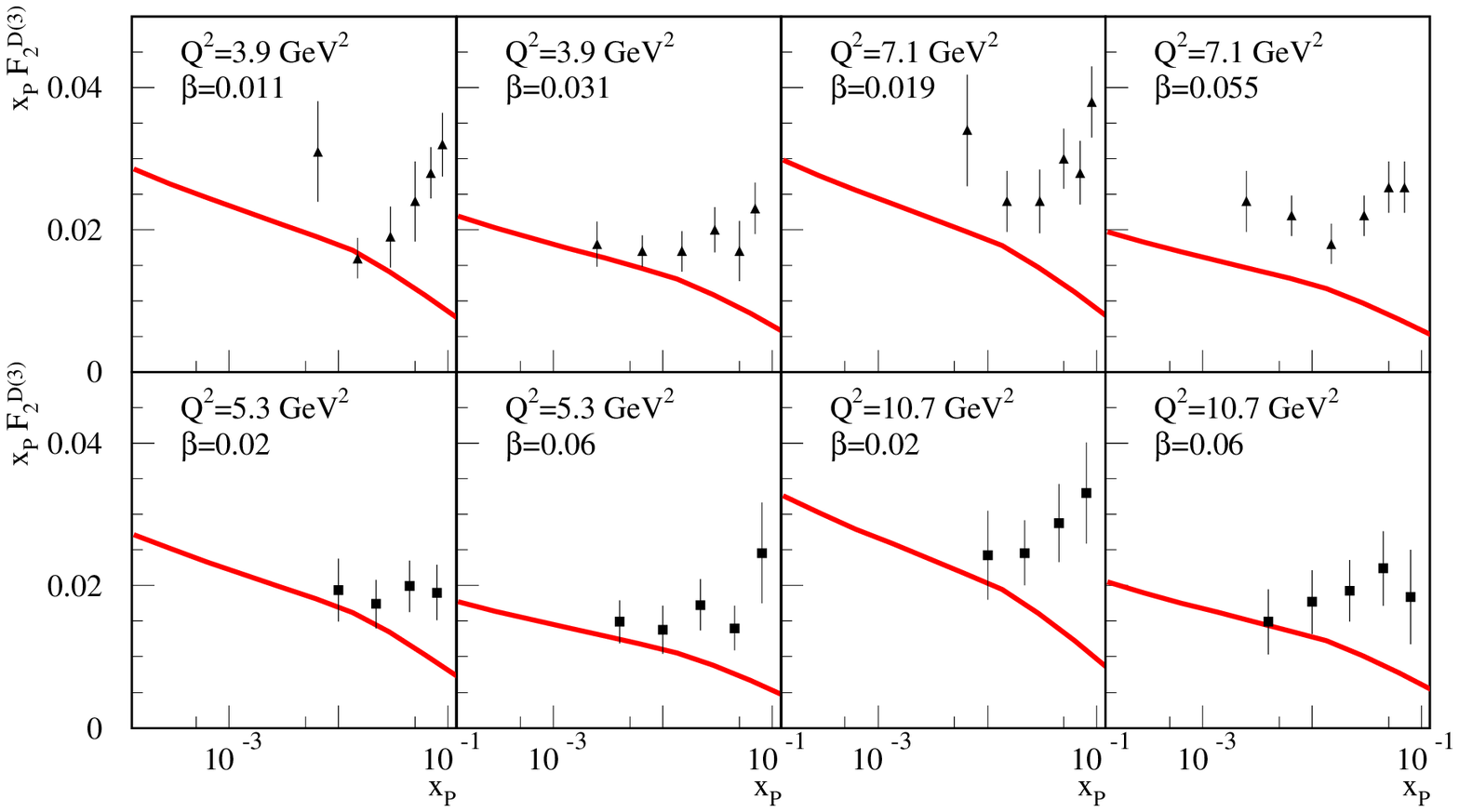}\caption{Calculated 
proton diffractive structure function 
$F_{2}^{{\rm D(3)}}(x,x_{\mathbb{P}},Q^{2})$
 compared to ZEUS  \cite{che09} (triangles) and H1 \cite{akt06} (squares) 
 data.\label{fig:f2d}}

\end{figure}

In Fig.~\ref{fig:f2d}, 
we compare the calculated proton diffractive structure
function  $F_{2}^{{\rm D(3)}}(x,x_{\mathbb{P}},Q^{2})$
(proton dissociation excluded) for small $Q^2$,
$x_{\mathbb{P}}$, $\beta=x/x_{\mathbb{P}}$ with HERA data. Our interest to this
observable is related to its strong sensitivity to the main parameter
for the enhanced Pomeron scheme -- the triple-Pomeron coupling. It is easy
to see that the model results for $F_{2}^{{\rm D(3)}}$ agree with
the measurements only in the limit of 
 small $Q^2$, $x_{\mathbb{P}}$,  and $\beta$.
With increasing $x_{\mathbb{P}}$, the $\mathbb{RRP}$ contribution to 
 $F_{2}^{{\rm D(3)}}$ becomes important while for larger $\beta$ and  $Q^2$
 so-called $q\bar{q}$ diffractive component (multiple Pomeron coupling to the
$q\bar{q}$-loop)  has to be accounted 
 for \cite{nik92}, both contributions neglected in the present 
 treatment.\footnote{Neither of the two neglected contributions involves the 
 triple-Pomeron coupling.} 
 Thus, diffractive HERA data set the {\em upper limit}
 on the value of the triple-Pomeron vertex.

\begin{figure}[t]
\begin{centering}
\includegraphics[width=7cm,height=6.5cm]{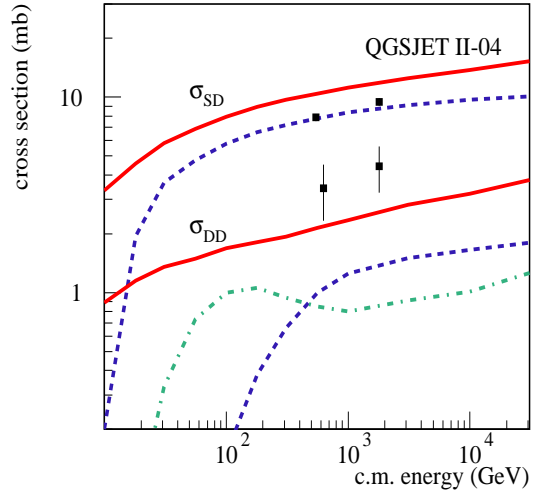}
\par\end{centering}

\caption{Calculated single $\sigma_{pp}^{{\rm SD}}(M_{X}^{2}/s<0.15)$ and
double $\sigma_{pp}^{{\rm DD}}(y_{{\rm gap}}^{(0)}\geq3)$ diffraction
proton-proton cross sections (solid lines), compared to CDF data (points)
\cite{abe94a,aff01}; high mass diffraction contributions to $\sigma_{pp}^{{\rm SD}}$,
$\sigma_{pp}^{{\rm DD}}$ (dashed lines); contribution to $\sigma_{pp}^{{\rm DD}}$
from a high mass diffraction of one proton and a low mass excitation
of the other one (dotted-dashed line).\label{fig:sigdifr}}

\end{figure}

In Fig.~\ref{fig:sigdifr} %
we plot the obtained energy dependence for single and double diffractive
proton-proton cross sections in comparison to CDF data \cite{abe94a,aff01},
showing also partial contributions of high mass diffractive states
to the discussed cross sections and the contribution of high mass
diffraction of one proton and a low mass excitation of the other one.
Here we adopt the experimental definitions for the diffractive cross
sections, applying the respective event selection triggers to hadronic
final states generated via a MC procedure: Single diffraction events
are obtained when either a projectile or target proton is separated
from the remaining final state of mass $M_{X}$ by a LRG and 
$M_{X}^{2}/s<0.15$ \cite{abe94a};
double diffraction events contain a central rapidity gap of size $y_{{\rm gap}}\geq3$,
which spans the central rapidity $y=\ln s/2$ point \cite{aff01}. 
Diffractive states
are classified as high mass ones when $M_{X}^{2}>25\;{\rm GeV}^{2}$
and as low mass excitations otherwise \cite{abe94a}. It is noteworthy
that at comparatively low energies ($\sqrt{s}\sim10$ GeV) certain
(theoretically) nondiffractive final states satisfy the imposed triggers,
constituting about half of the plotted $\sigma_{pp}^{{\rm SD}}$ and
most of the $\sigma_{pp}^{{\rm DD}}$. On the other hand, at sufficiently
high energies a part of the theoretical low mass diffraction, being
described by the $\mathbb{PPR}$-asymptotics, is classified as high
mass diffraction, which explains the energy dependence of the ``low-high''
double diffraction cross section (dotted-dashed line in Fig.~\ref{fig:sigdifr}).
The obtained values for $\sigma_{pp}^{{\rm SD}}$ agree reasonably well
with the measurements, taking the fact that most of the low mass diffraction
contribution could not be seen by the CDF detector \cite{abe94a}.
In turn, double diffraction is seriously   underestimated
by the model.

\begin{figure}[htb]
\begin{centering}
\includegraphics[width=15cm,height=6cm]{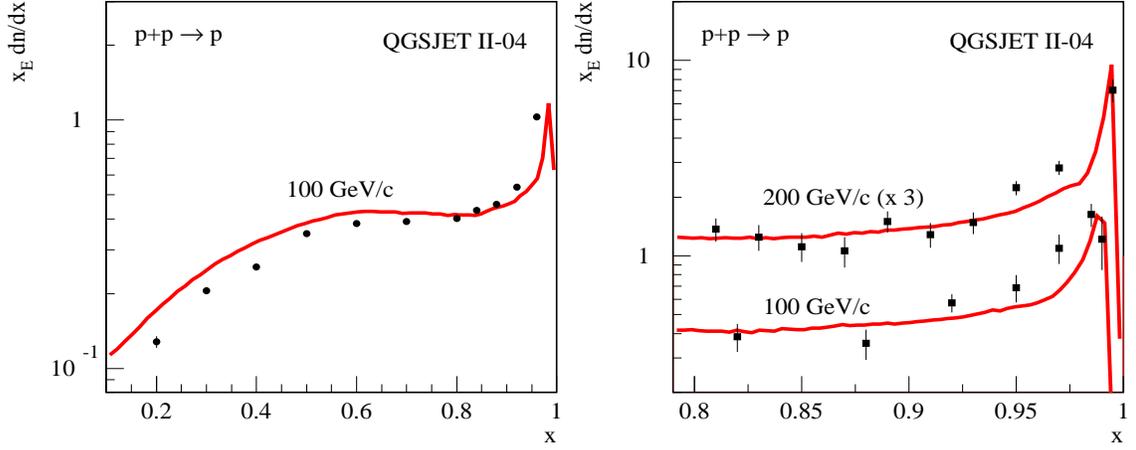}
\par\end{centering}
\caption{Calculated Feynman $x$ spectra  of secondary protons
in proton-proton collisions at 100 and 200  GeV/c lab.~momentum compared 
 to  data from Refs.~\cite{bre82} (circles) and \cite{whit74} 
 (squares).\label{fig:ppspec}}
\end{figure}

\begin{figure}[htb]
\begin{centering}
\includegraphics[width=15cm,height=6cm]{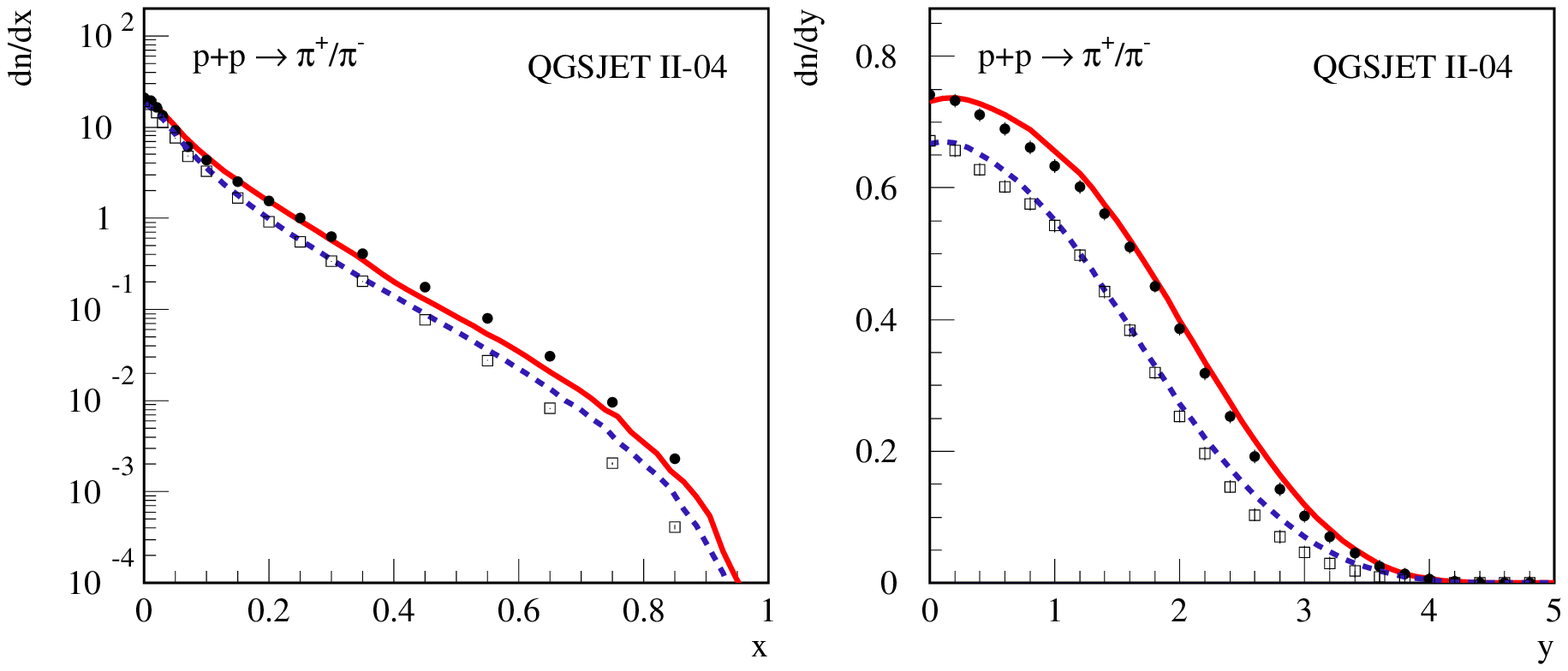}
\par\end{centering}
\caption{Calculated Feynman $x$ spectra (left) and rapidity distributions 
 (right) of positive (solid lines) and negative  (dashed lines)  pions
in proton-proton collisions at 158  GeV/c lab.~momentum compared 
 to NA49 data  \cite{na49-pp}.\label{fig:pp158}}
\end{figure}

\begin{figure}[t]
\begin{centering}
\includegraphics[width=15cm,height=12cm]{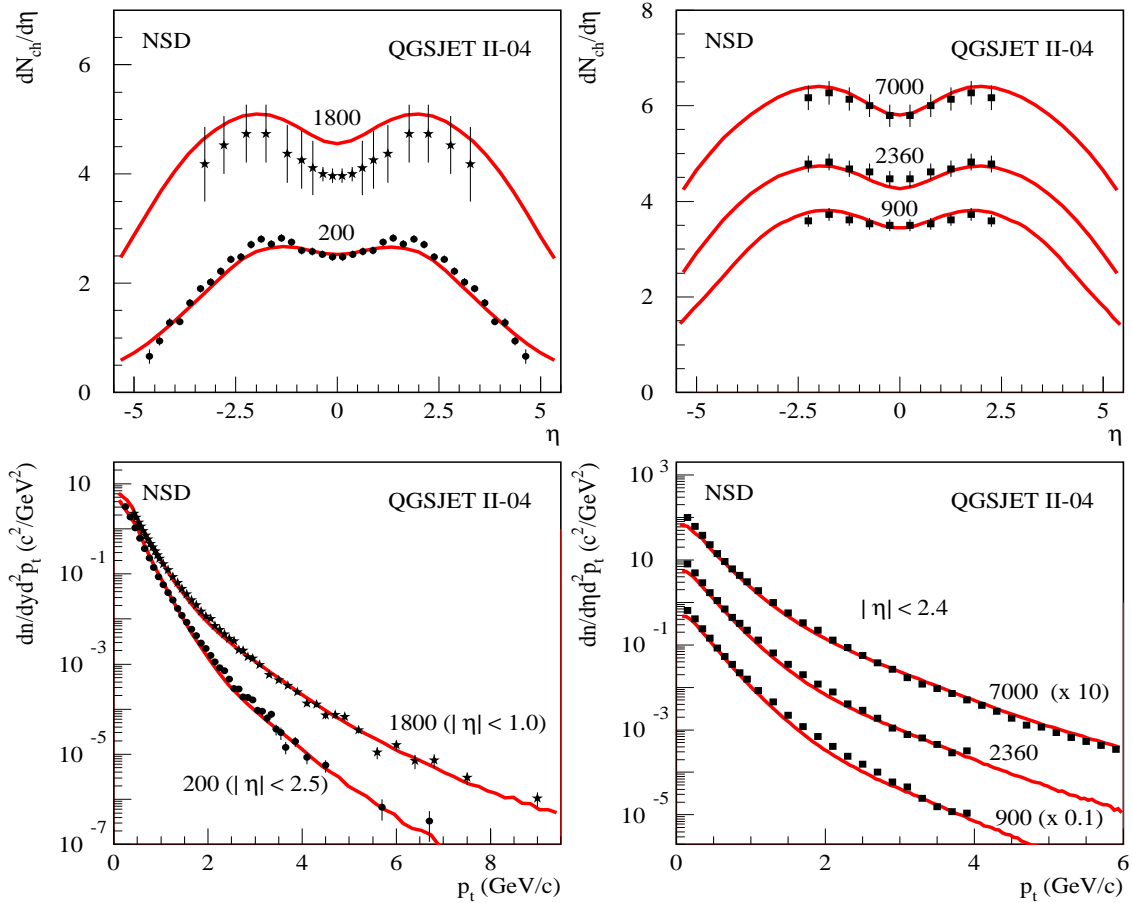}
\par\end{centering}
\caption{Calculated pseudorapidity density (top row)
and transverse momentum spectra (bottom row) of charged particles
produced in proton-antiproton (left panels) and proton-proton 
 (right panels) collisions
at different c.m.~energies in GeV (as indicated in the plots).
The data points are from 
Refs.~\cite{ua5-200,ua1-200,cdf-eta,cdf-pt,cms}.\label{fig:pp-coll}}
\end{figure}

A comparison with selected data on secondary particle production
in proton-proton and proton-antiproton collisions, 
which have been  used for the model calibration,
is presented in Figs.~\ref{fig:ppspec},~\ref{fig:pp158},~\ref{fig:pp-coll}. %
In Figs.~\ref{fig:ppspec} and \ref{fig:pp158} the calculated 
 Feynman $x$ spectra    of  protons and charged pions as well as
 pion  rapidity distributions are plotted together with  data
from fixed target experiments. In turn, Fig.~\ref{fig:pp-coll}  shows
 the results of calculations
of the pseudorapidity density and of transverse momentum spectra of charged
secondaries in  non-single-diffractive (NSD)
proton-antiproton and proton-proton collisions
over a broad  range of  energies $\sqrt s=0.2 \div 7$ TeV in
comparison with experimental data from the $Sp\bar pS$, Tevatron,
 and LHC colliders.\footnote{The calculations have been performed using the
 NSD triggers of the respective experiments.}
 More extensive compilation of the model results for secondary
particle production will be presented elsewhere \cite{ost10a}.

An interesting potential test for hadronic interaction models, which
could be performed using particle detectors at the LHC, has been proposed
in Ref.~\cite{kmr09c}. The idea was to compare event trigger rates
obtained by LHC experiments, using different combinations of the respective
charged particle scintillation counters. As such counters cover a
restricted range of pseudorapidities, $\eta_{1}<|\eta|<\eta_{2}$,
with $\eta_{1}=3.1$, $\eta_{2}=6.5$ for the TOTEM detector \cite{ber04}
and $\eta_{1}=2$, $\eta_{2}=4$ for ATLAS \cite{aad09}, the so-called
minimum-bias trigger (MBT) selections by the experiments will miss
a significant fraction of the inelastic proton-proton cross section,
which will include both the low mass diffraction and a part of the
high mass one. Using various combinations of such triggers, one gains
sensitivity both to the absolute value of $\sigma_{pp}^{{\rm inel}}$
and to the contributions of single and double high mass diffraction,
which have different selection efficiencies by such triggers. In Table~\ref{tab:mbt(14 TeV)}
we present our predictions for trigger rates by the TOTEM and ATLAS
experiments at $\sqrt{s}=14$ TeV, in comparison to the original calculations
of Ref.~\cite{kmr09c},%
\begin{table}[htb]
\centering{}\begin{tabular}{|l|cc|ccc|ccc|}
\hline 
 & \multicolumn{1}{c}{} &  &  & TOTEM &  &  & ATLAS & \tabularnewline
\cline{4-9} 
 & $\sigma_{{\rm tot}}$ & $\sigma_{{\rm inel}}$ & $\sigma_{{\rm MBT1}}$ & $\sigma_{{\rm MBT2}}$ & $\sigma_{{\rm MBT3}}$ & $\sigma_{{\rm MBT1}}$ & $\sigma_{{\rm MBT2}}$ & $\sigma_{{\rm MBT3}}$\tabularnewline
\hline
\hline 
this work & 105 & 76.8 & 66.6 & 69.7 & 63.4 & 63.9 & 66.4 & 61.3\tabularnewline
\cite{kmr09c} & 91.5 & 70.0 & 50.7 & 59.0 & 42.4 & 46.6 & 50.8 & 42.4\tabularnewline
\hline
\end{tabular}\caption{Calculated total, inelastic and minimum-bias (for different MBT selections)
proton-proton cross sections at $\sqrt{s}=14$ TeV.\label{tab:mbt(14 TeV)}}

\end{table}
for the selection of triggers proposed in that work: requiring a charge
particle hit at positive rapidities only (in the interval $[\eta_{1},\eta_{2}]$
covered by scintillators) - MBT1, or a signal at either positive ($[\eta_{1},\eta_{2}]$)
or negative ($[-\eta_{2},-\eta_{1}]$) rapidities - MBT2, or with
both detectors being fired - MBT3. The calculated trigger rates for
the presently attained LHC energy $\sqrt{s}=7$ TeV are collected
in Table~\ref{tab:mbt(7 TeV)}.%
\begin{table}[htb]
\centering{}\begin{tabular}{|cc|ccc|ccc|}
\hline 
\multicolumn{1}{|c}{} &  &  & TOTEM &  &  & ATLAS & \tabularnewline
\cline{3-8} 
$\sigma_{{\rm tot}}$ & $\sigma_{{\rm inel}}$ & $\sigma_{{\rm MBT1}}$ & $\sigma_{{\rm MBT2}}$ & $\sigma_{{\rm MBT3}}$ & $\sigma_{{\rm MBT1}}$ & $\sigma_{{\rm MBT2}}$ & $\sigma_{{\rm MBT3}}$\tabularnewline
\hline
93.3 & 69.7 & 60.8 & 64.3 & 57.4 & 58.1 & 60.8 & 55.3\tabularnewline
\hline
\end{tabular}\caption{Same as in Table~\ref{tab:mbt(14 TeV)} for $\sqrt{s}=7$ TeV.\label{tab:mbt(7 TeV)}}

\end{table}
Comparing our results with those of Ref.~\cite{kmr09c}, we observe
large differences concerning both the absolute magnitude of the calculated
minimum-bias cross sections and for their variations between different
trigger selections. Although such differences are partly due to hadronization
effects - as we apply the respective triggers to hadronic final states
generated via a MC procedure, the largest effect is related to a considerably
higher total (hence, also inelastic) cross section and to a much smaller
(by a factor of 2) single high mass diffraction cross section in our
approach compared to \cite{kmr09c}, as discussed in more detail in
\cite{ost10}. Nevertheless, the discussed triggers work in a similar
way for both model approaches, particularly for the MBT selections
of ATLAS: i) all the three triggers reject low mass diffraction; ii)
the MBT3 trigger misses also most of single high mass diffraction
events; iii) the MBT1 trigger rejects most of the target high mass diffraction.
As a consequence, the ratio $(\sigma_{{\rm MBT2}}-\sigma_{{\rm MBT3}})/(\sigma_{{\rm MBT2}}-\sigma_{{\rm MBT1}})$
is close to 2. Thus, the present study confirms that the method proposed
in \cite{kmr09c} will provide a powerful selection between different
model approaches to the treatment of minimum-bias hadronic collisions.

\begin{figure}[htb]
\begin{centering}
\includegraphics[width=6cm,height=5cm]{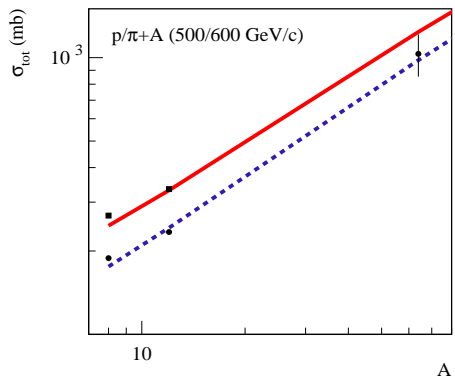}
\par\end{centering}
\caption{Calculated $A$-dependence of   total proton-nucleus (solid) and
pion-nucleus (dashed) cross sections at respectively 500 and 600 GeV/c 
lab.~momentum
 compared to experimental data \cite{der00}.\label{fig:sigha}}
\end{figure}

\begin{figure}[htb]
\begin{centering}
\includegraphics[width=15cm,height=6cm]{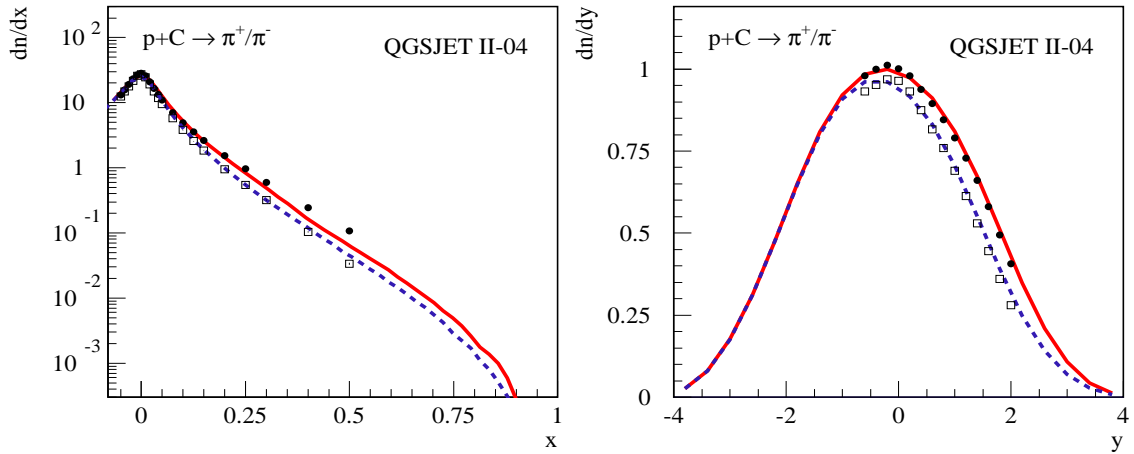}
\par\end{centering}
\caption{Calculated Feynman $x$ spectra (left) and rapidity distributions 
 (right) of positive (solid lines) and negative  (dashed lines)  pions
in proton-carbon collisions at 158  GeV/c lab.~momentum compared 
 to experimental data  \cite{na49-pc}.\label{fig:pcarbon}}
\end{figure}

As discussed above, the model generalization to hadron-nucleus and
nucleus-nucleus collisions does not involve any additional parameters.
As an illustration, we plot in Fig.~\ref{fig:sigha}%
 the calculated $A$-dependence of   total    hadron-nucleus
cross sections in comparison with experimental data.
In addition, in Fig.~\ref{fig:pcarbon}%
 the calculated Feynman $x$ spectra 
and rapidity distributions of secondary pions in proton-carbon collisions
 are compared to NA49 data \cite{na49-pc}.
 
\begin{figure}[htb]
\begin{centering}
\includegraphics[width=6cm,height=5cm]{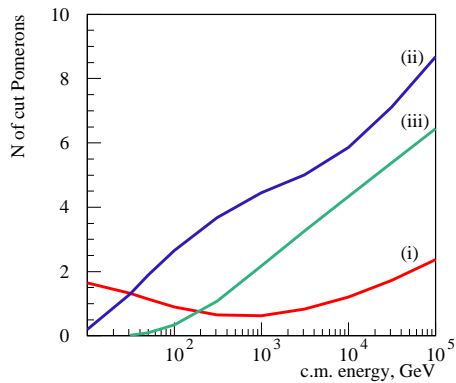}
\par\end{centering}

\caption{Energy dependence of the number of cut Pomerons of different types
for proton-proton interactions: exchanged between the projectile and
  target protons (i), exchanged between the projectile or target proton
and some multi-Pomeron vertex (ii), exchanged between a pair of multi-Pomeron
vertices (iii).\label{fig:npoms}}

\end{figure}

As demonstrated earlier in \cite{ost06,ost06a}, the high energy behavior
of hadronic cross sections is strongly affected by nonlinear interaction
effects. A similarly strong effect is observed when studying the generated
configurations for hadronic final states. For proton-proton collisions,
we plot in Fig.~\ref{fig:npoms}%
 the energy dependence of the number of ``elementary pieces'' of secondary
production - cut Pomerons, for the three possible contributions: i)
cut Pomerons exchanged between the projectile and   target protons
- which arise from eikonal Pomeron exchanges and obtain absorptive
corrections from various cut enhanced graphs characterized by the
same pattern of the final state; ii) ones exchanged between the projectile
or target proton and some multi-Pomeron vertex - which come from cut
enhanced graphs only (see the examples of the corresponding subgraphs
in Fig.~\ref{fig:cutpom-line}); iii) cut Pomerons exchanged between
a pair of multi-Pomeron vertices. While at moderately low energies
the configuration of the interaction is dominated by cut Pomerons
of the first type - as one would have in a (linear) eikonal Pomeron
scheme, at higher energies the corresponding contribution is 
damped by absorptive corrections and secondary hadron production comes
mainly from the other two contributions - which originate from cut
enhanced graphs, i.e.~from the treatment of nonlinear interactions.
The physical picture behind the observed trend is obvious: The contributions
of the 1st kind correspond to a number of elementary parton cascades
developing independently between the projectile and   target protons
and hadronizing into secondary hadrons. With the energy increasing,
a large number of such cascades is closely packed together in the
phase space, which forces them to overlap and to interact with each
other. Thus, production of single chains of secondaries covering the
available rapidity range $[0,Y${]} is strongly reduced by virtual
(elastic) rescattering of intermediate partons off the projectile
and target hadrons. More typical become configurations of fan (or
more complicated, see Fig.~\ref{fig: tree-dif1}) types, which correspond
to multiple {\em inelastic} rescattering of some intermediate partons off
the projectile and target hadrons, leading to a splitting or fusion
of the cut Pomeron lines, hence, to a branching of the chains of secondaries
and, generally, to a production of large rapidity gaps.

Let us finally stress that the complete all-order resummation
of the contributions of all significant enhanced diagrams
performed in this work was absolutely necessary for obtaining a
self-consistent description of hadronic cross sections and of
particle production in the very high energy limit. This is related
 to the fact that in the high energy asymptotics the diagrams
 of the highest considered order with respect to   Pomeron-Pomeron
 coupling (more precisely, the ones with
 the maximal number of Pomerons exchanged in parallel in some
 rapidity interval) dominate the elastic scattering amplitude
 \cite{baker}. Moreover, proceeding from one order to the next,
 one obtains sign-changing contributions. Thus, breaking the series
 at some given order, one obtains the total cross section which either
 falls down steeply above some energy or starts to rise in a powerlike
 way, violating the unitarity bound. Hence, a meaningful answer can 
 only be obtained after a resummation of the complete (infinite)
 series of diagrams. In fact, the situation is even more demanding
 when applied to calculations of partial cross sections for various
 (e.g.~diffractive)  hadronic final states -- as different unitarity
 cuts of the same enhanced graphs give positive contributions to
 some processes while providing (negative) screening corrections
 to others. The simplest example of the kind is the triple-Pomeron
 diagram which provides a steeply rising   contribution to high mass
 single diffraction cross section  and gives rise to a strong screening
 correction to the single cut Pomeron cross section,
the latter corresponding to a  single elementary production process.
An extensive analysis of such effects has been reported in the previous
work \cite{ost10}, where also the relative importance of various classes
of enhanced diagrams has been investigated.

A comment is in order on the adopted ansatz (\ref{eq:g_mn}) for 
multi-Pomeron vertices
and on the respective parameter $\gamma_{\mathbb{P}}$. As discussed already
in \cite{car74,kai86} and more recently in \cite{ost10}, in a scheme
based on a single Pomeron type, one is forced to choose $\gamma_{\mathbb{P}}$
such that $r_{3\mathbb{P}}/\gamma_{\mathbb{P}}<\Delta_{\mathbb{P}}$ --
in order to preserve the energy rise of the scattering amplitude.
In the present scheme, apart from  the usual soft Pomeron we have also
the ``semihard Pomeron'' contribution [Eqs.~(\ref{eq:chi-sh}),
 (\ref{eq:chi-leg-sh}), (\ref{eq:chi-int-sh})] which contains
a perturbative ``piece''. In addition, we assume that multi-Pomeron
coupling is dominated by low-$q^2$ processes ($|q^2|<Q_0^2$),
i.e.~such vertices are coupled only to soft Pomerons or to soft
``ends'' of semihard Pomerons,\footnote{The corresponding ``hard piece'' 
is always ``sandwiched'' between a pair of soft Pomerons -- see 
Eqs.~(\ref{eq:chi-sh}), (\ref{eq:chi-leg-sh}), (\ref{eq:chi-int-sh}) 
and the 2nd graph in the r.h.s.~of Fig.~\ref{genpom}.}
 as discussed in more detail in \cite{dre01,ost06a}.
 This allows us to choose $\gamma_{\mathbb{P}}$
such that $r_{3\mathbb{P}}/\gamma_{\mathbb{P}}>\Delta_{\mathbb{P}}$,
which leads to a saturation of soft processes in the dense limit (small
$b$ and large $s$)   and to a flattening of  PDFs {\em at the 
input scale} $Q_0^2$ \cite{ost05}. The energy rise of the scattering amplitude
is supported at very high energies by the increase of the semihard 
contribution.
However, neglecting hard
 ($|q^2|>Q_0^2$) Pomeron-Pomeron coupling, we are forced to choose
 the $Q_0$ cutoff high enough -- in order to safely neglect parton
 saturation effects at $|q^2|>Q_0^2$.\footnote{Note that for our choice
 of the factorization scale $M_{{\rm F}}^{2}=p_{t}^{2}/4$ in
  Eq.~(\ref{eq:sigma-hard}) the chosen cutoff corresponds to the minimal
  transverse momentum in parton hard process
   $p_{\perp}^{\min}=2Q_0\simeq 3.4$ GeV.}
   This in turn restricts our choice for the parameter $\gamma_{\mathbb{P}}$:
   For very small  $\gamma_{\mathbb{P}}$, in particular, in the limit of
   triple-Pomeron vertices only ($\gamma_{\mathbb{P}}\rightarrow 0$),
   having the triple-Pomeron coupling $r_{3\mathbb{P}}$ fixed by diffraction
   data,   the saturation of the soft particle production would be achieved at
   relatively low energies over a large range of impact parameters.
   As the semihard contribution is still inefficient there, this leads
   to an underestimation of secondary particle production and to a
    contradiction with observations. Whether or not one can restrict himself
    with just the triple-Pomeron vertices  can be investigated in the
    complete scheme only: Taking perturbative Pomeron-Pomeron coupling
    into consideration.

\section{Conclusions}

In this paper, we discussed in detail the MC procedure for modeling
hadronic collisions in the RFT framework, including the contributions
of enhanced Pomeron diagrams. The principal difference of the presented
Monte Carlo model compared to other generators of hadronic interactions
is the direct correspondence between the RFT treatment and the MC
implementation: Various hadronic final states are generated according
to their partial cross sections. The latter are defined by the contributions
of cut Pomeron diagrams characterized by the relevant structure of
the cuts. Defining the contributions of certain cut subgraphs by
means of recursive equations, we were able to generate the (generally
complicated) structure of hadronic final states in an iterative fashion.

The described model represents a self-consistent implementation of
the corresponding RFT treatment, providing, in particular, a close
link between the description of total and elastic hadron-proton cross
sections and the generation of hadronic final states. Indeed, while
the elastic scattering amplitude is defined by the contributions of
uncut nonenhanced and enhanced diagrams, partial cross sections for
various final states are defined by unitarity cuts of the very same
diagrams, the summary contribution of all the cuts being related to
the uncut one by   $s$-channel unitarity [Eqs.~(\ref{eq:opac-iden}),
 (\ref{opac-AB-iden})], as demonstrated explicitly in \cite{ost08}.
 On the other hand, the
generalization of the treatment to hadron-nucleus and nucleus-nucleus
case, both concerning cross section calculations and for modeling
particle production, does not involve additional adjustable parameters.

Being based on the RFT formalism, the present treatment shares most
of its usual assumptions, like the validity of the AGK cutting rules
and eikonal vertices for Pomeron-hadron (Pomeron-Pomeron) coupling.
It also has the usual drawback of neglecting energy-momentum correlations
between multiple scattering processes at the amplitude level \cite{bra90}.
Thus, the discussed model remains a phenomenological one and it is
experimental data which have to decide if it is suitable enough for
the treatment of very high energy hadronic collisions. While in the
current work we mainly addressed the construction of the MC generator
and showed only some representative results for secondary particle spectra,
a thorough comparison of the model predictions on particle production
with available experimental data 
 and the model applications for air shower
simulation will be presented elsewhere \cite{ost10a}.

\subsection*{Acknowledgments}

The author acknowledges the support of the European Commission under
the Marie Curie IEF program (Grant No.~220251) and of Norsk Forsknigsradet
under the program Romforskning.

\section*{Appendix A}
The generalization of the approach described in 
Section~\ref{sec:Hadron-hadron-scattering-amplitude} to hadron-nucleus
and nucleus-nucleus collisions is parameter free and formally straightforward.
Indeed, the only essential difference is that now different Pomerons
in a given irreducible enhanced graph may couple to different nucleons
of the projectile and/or target nuclei, whose positions in the impact
parameter plane should be chosen according to the corresponding nuclear
density profiles. Thus, in case of nucleus $A$ - nucleus $B$ interaction,
the net-fan contribution (\ref{net-fan}) should be generalized to\begin{eqnarray}
\chi_{A|B}^{{\rm net}}(y_{1},\vec{b}_{1}|Y,\vec{b})=\sum_{m=1}^{A}\chi_{p(j_{m}^{A})}^{{\rm loop}}(y_{1},|\vec{b}_{1}-\vec{b}_{m}^{A}|)+G\int_{\xi}^{y_{1}-\xi}\! dy_{2}\int\! d^{2}b_{2}\;\left(1-e^{-\chi^{{\rm loop}}(y_{1}-y_{2},|\vec{b}_{1}-\vec{b}_{2}|)}\right)\nonumber \\
\times\left[\left(1-e^{-\chi_{A|B}^{{\rm net}}(y_{2},\vec{b}_{2}|Y,\vec{b})}\right)\; e^{-\chi_{B|A}^{{\rm net}}(Y-y_{2},\vec{b}-\vec{b}_{2}|Y,\vec{b})}-\chi_{A|B}^{{\rm net}}(y_{2},\vec{b}_{2}|Y,\vec{b})\right]\!,\label{eq:chi-net-A}\end{eqnarray}
where $y_{1}$ is the rapidity distance between the given multi-Pomeron
vertex and the nucleus $A$, $\vec{b}_{1}$ - the position of the
vertex with respect to the center of the nucleus in the transverse
plane, $j_{m}^{A}$ - elastic scattering eigenstate of $m$-th nucleon,
and $\vec{b}_{m}^{A}$ - its transverse vector with respect to the
center of the nucleus. Thus, $\chi_{A|B}^{{\rm net}}$ depends on
the positions $\{\vec{b}^{A},\vec{b}^{B}\}$ and the elastic scattering
eigenstates $\{j^{A},k^{B}\}$ of all the $A+B$ nucleons of the interacting
nuclei, the corresponding indexes not shown explicitly in (\ref{eq:chi-net-A}).
The general contribution of irreducible enhanced graphs is generalized
in a similar way:\begin{eqnarray}
\chi_{AB}^{{\rm enh}}(s,b,\{j^{A},k^{B},\vec{b}^{A},\vec{b}^{B}\})
=G\int_{\xi}^{Y-\xi}\!\! dy_{1}\!\int\!\! d^{2}b_{1}\;
\left\{ \left[\left(1-e^{-\chi_{A|B}^{{\rm net}}}\right)
\left(1-e^{-\chi_{B|A}^{{\rm net}}}\right)-\chi_{A|B}^{{\rm net}}\,
\chi_{B|A}^{{\rm net}}\right]\right.\nonumber \\
-\left[\chi_{A|B}^{{\rm net}}-\sum_{m=1}^{A}
\chi_{p(j_{m}^{A})}^{{\rm loop}}(Y-y_{1},|\vec{b}-\vec{b}_{1}+\vec{b}_{m}^{A}|)\right]
\left[\left(1-e^{-\chi_{B|A}^{{\rm net}}}\right)e^{-\chi_{A|B}^{{\rm net}}}
-\chi_{B|A}^{{\rm net}}\right]\nonumber \\
+\left.\sum_{m=1}^{A}\sum_{n=1}^{B}
\chi_{p(k_{n}^{B})}^{\mathbb{P}}(y_{1},|\vec{b}_{1}-\vec{b}_{n}^{B}|)
\left[\chi_{p(j_{m}^{A})}^{{\rm loop}}(Y-y_{1},|\vec{b}-\vec{b}_{1}+\vec{b}_{m}^{A}|)
-\chi_{p(j_{m}^{A})}^{{\rm loop(1)}}(Y-y_{1},|\vec{b}-\vec{b}_{1}+\vec{b}_{m}^{A}|)\right]
\right\} \!,\label{eq:chi-full-AB}
\end{eqnarray}
where the abbreviations are similar to the ones in (\ref{eq:chi-enh}).

It is obvious that Eqs.~(\ref{eq:chi-net-A}-\ref{eq:chi-full-AB})
are impractical: for each particular configuration of the two nuclei,
i.e.~for each choice of the coordinates and elastic scattering eigenstates
of the nucleons, one has to calculate $\chi_{A|B}^{{\rm net}}$ recursively,
which is very time-consuming. Unlike hadron-hadron case, one can not
make a pretabulation of the corresponding contributions which now
depend on $2(A+B)$ coordinates of the nucleons, not counting the
numbers of their possible eigenstates and the variables shown explicitly
in (\ref{eq:chi-net-A}). 

To propose a suitable approximation for (\ref{eq:chi-net-A}-\ref{eq:chi-full-AB})
let us decompose $\chi_{A|B}^{{\rm net}}$ as \begin{equation}
\chi_{A|B}^{{\rm net}}(y_{1},\vec{b}_{1}|Y,\vec{b})=\sum_{m=1}^{A}\chi_{p(j_{m}^{A})}^{{\rm net}}(y_{1},|\vec{b}_{1}-\vec{b}_{m}^{A}|,\cdots)\,,\label{eq:chi-A-m}\end{equation}
with the aim to describe the dependence of $\chi_{p(j_{m}^{A})}^{{\rm net}}$
on the coordinates and eigenstates of all the $A+B-1$ projectile
and target nucleons but the current one, indicated symbolically by
the multidot in the r.h.s.~of (\ref{eq:chi-A-m}), by means of a
single factor. Substituting (\ref{eq:chi-A-m}) to (\ref{eq:chi-net-A})
and using the identity
\begin{equation}
1-\exp\!\left(-\sum_{m=1}^{A}\chi_{p(j_{m}^{A})}^{{\rm net}}\right)
=\sum_{m=1}^{A}\left(1-e^{-\chi_{p(j_{m}^{A})}^{{\rm net}}}\right)
e^{-\sum_{l=1}^{m-1}\chi_{p(j_{l}^{A})}^{{\rm net}}},\label{eq:chi-A-decomp}
\end{equation}
we obtain
\begin{eqnarray}
\sum_{m=1}^{A}\chi_{p(j_{m}^{A})}^{{\rm net}}(y_{1},|\vec{b}_{1}-\vec{b}_{m}^{A}|,\cdots)
\nonumber \\
=\sum_{m=1}^{A}\left\{ \chi_{p(j_{m}^{A})}^{{\rm loop}}(y_{1},|\vec{b}_{1}-\vec{b}_{m}^{A}|)
+G\int_{\xi}^{y_{1}-\xi}\! dy_{2}\int\! d^{2}b_{2}\;
\left(1-e^{-\chi^{{\rm loop}}(y_{1}-y_{2},|\vec{b}_{1}-\vec{b}_{2}|)}\right)\right.
\nonumber \\
\times\left.\left[\left(1
-e^{-\chi_{p(j_{m}^{A})}^{{\rm net}}(y_{2},|\vec{b}_{2}-\vec{b}_{m}^{A}|,\cdots)}\right)
Z_{A|B}^{(m)}(y_{2},\vec{b}_{2},Y,\vec{b},\{j^{A},k^{B},\vec{b}^{A},\vec{b}^{B}\})
-\chi_{p(j_{m}^{A})}^{{\rm net}}(y_{2},|\vec{b}_{2}-\vec{b}_{m}^{A}|,\cdots)\right]\right\} 
\label{eq:net-A-decomp}\\
Z_{A|B}^{(m)}(y_{2},\vec{b}_{2},Y,\vec{b},\{j^{A},k^{B},\vec{b}^{A},\vec{b}^{B}\})
\nonumber \\
=\exp\left[-\sum_{l=1}^{m-1}
\chi_{p(j_{l}^{A})}^{{\rm net}}(y_{2},|\vec{b}_{2}-\vec{b}_{l}^{A}|,\cdots)
-\sum_{n=1}^{B}
\chi_{p(k_{n}^{B})}^{{\rm net}}(Y-y_{2},|\vec{b}-\vec{b}_{2}+\vec{b}_{n}^{B}|,\cdots)\right]\!.
\label{eq:Z-factor}
\end{eqnarray}

Now we approximate the nuclear screening factor $Z_{A|B}^{(m)}$ for
$m$-th projectile nucleon by its value in the vertex $(y_{1},\vec{b}_{1})$:
\begin{equation}
Z_{A|B}^{(m)}(y_{2},\vec{b}_{2},Y,\vec{b},\{j^{A},k^{B},\vec{b}^{A},\vec{b}^{B}\})
\simeq Z_{A|B}^{(m)}(y_{1},\vec{b}_{1},Y,\vec{b},\{j^{A},k^{B},\vec{b}^{A},\vec{b}^{B}\})\,.
\label{eq:Z-approx}\end{equation}
Using this approximation, the solution of the nuclear net-fan equation
(\ref{eq:chi-net-A}) is
\begin{equation}
\chi_{A|B}^{{\rm net}}(y_{1},\vec{b}_{1}|Y,\vec{b})=\sum_{m=1}^{A}\chi_{p(j_{m}^{A})}^{{\rm net}}(y_{1},|\vec{b}_{1}-\vec{b}_{m}^{A}|,Z_{A|B}^{(m)}(y_{1},\vec{b}_{1},Y,\vec{b},\{j^{A},k^{B},\vec{b}^{A},\vec{b}^{B}\}))\,,\label{eq:A-m-decomp}\end{equation}
where a partial contribution $\chi_{p(j)}^{{\rm net}}$ of any of
the $A$ projectile nucleons is the solution of the recursive equation
[c.f.~(\ref{net-fan})]:
\begin{eqnarray}
\chi_{p(j)}^{{\rm net}}(y_{1},b',Z)=\chi_{p(j)}^{{\rm loop}}(y_{1},b')+G\int_{\xi}^{y_{1}-\xi}\! dy_{2}\int\! d^{2}b_{2}\;\left(1-e^{-\chi^{{\rm loop}}(y_{1}-y_{2},|\vec{b}'-\vec{b}_{2}|)}\right)\nonumber \\
\times\left[\left(1-e^{-\chi_{p(j)}^{{\rm net}}(y_{2},\vec{b}_{2},Z)}\right)\; Z-\chi_{p(j)}^{{\rm net}}(y_{2},\vec{b}_{2},Z)\right],\label{eq:chi-net-m}\end{eqnarray}
which can be easily tabulated as a function of its three arguments
$(y_{1},b',Z)$.

Substituting now (\ref{eq:A-m-decomp}) to (\ref{eq:chi-full-AB})
and using (\ref{eq:chi-A-decomp}), we obtain
\begin{eqnarray}
\chi_{AB}^{{\rm enh}}(s,b,\{j^{A},k^{B},\vec{b}^{A},\vec{b}^{B}\})
=\sum_{m=1}^{A}\sum_{n=1}^{B}
\chi_{mn}^{{\rm enh}}(s,b,\{j^{A},k^{B},\vec{b}^{A},\vec{b}^{B}\})
\label{eq:chi-AB-enh}\\
\chi_{mn}^{{\rm enh}}(s,b,\{j^{A},k^{B},\vec{b}^{A},\vec{b}^{B}\})
=G\int_{\xi}^{Y-\xi}\!\! dy_{1}\!\int\!\! d^{2}b_{1}\;
\left\{ \left[\left(1-e^{-\chi_{p(j_{m}^{A})}^{{\rm net}}}\right)
\left(1-e^{-\chi_{p(k_{n}^{B})}^{{\rm net}}}\right)\right.\right.
\nonumber \\
\times\left.e^{-\sum_{l=1}^{m-1}\chi_{p(j_{l}^{A})}^{{\rm net}}
-\sum_{i=1}^{n-1}\chi_{p(k_{i}^{B})}^{{\rm net}}}
-\chi_{p(j_{m}^{A})}^{{\rm net}}\:\chi_{p(k_{n}^{B})}^{{\rm net}}\right]
-\left[\chi_{p(j_{m}^{A})}^{{\rm net}}
-\chi_{p(j_{m}^{A})}^{{\rm loop}}(Y-y_{1},|\vec{b}-\vec{b}_{1}+\vec{b}_{m}^{A}|)\right]
\nonumber \\
\times\left[\left(1-e^{-\chi_{p(k_{n}^{B})}^{{\rm net}}}\right)
e^{-\sum_{l=1}^{A}\chi_{p(j_{l}^{A})}^{{\rm net}}
-\sum_{i=1}^{n-1}\chi_{p(k_{i}^{B})}^{{\rm net}}}-\chi_{p(k_{n}^{B})}^{{\rm net}}\right]
\nonumber \\
+\left.\chi_{p(k_{n}^{B})}^{\mathbb{P}}(y_{1},|\vec{b}_{1}-\vec{b}_{n}^{B}|)\:
\left[\chi_{p(j_{m}^{A})}^{{\rm loop}}(Y-y_{1},|\vec{b}-\vec{b}_{1}+\vec{b}_{m}^{A}|)
-\chi_{p(j_{m}^{A})}^{{\rm loop(1)}}(Y-y_{1},|\vec{b}-\vec{b}_{1}+\vec{b}_{m}^{A}|)
\right]\right\} ,\label{eq:chi-AB-mn}\end{eqnarray}
where the omitted arguments read 
$\chi_{p(k_{n}^{B})}^{{\rm net}}
=\chi_{p(k_{n}^{B})}^{{\rm net}}(y_{1},|\vec{b}_{1}-\vec{b}_{n}^{B}|
,Z_{B|A}^{(n)}(y_{1},\vec{b}_{1},Y,\vec{b},\{k^{B},j^{A},\vec{b}^{B},\vec{b}^{A}\}))$,
$\chi_{p(j_{m}^{A})}^{{\rm net}}
=\chi_{p(j_{m}^{A})}^{{\rm net}}(Y-y_{1},|\vec{b}-\vec{b}_{1}+\vec{b}_{m}^{A}|,Z_{A|B}^{(m)}(Y-y_{1},\vec{b}-\vec{b}_{1}+\vec{b}_{m}^{A},Y,\vec{b},\{j^{A},k^{B},\vec{b}^{A},\vec{b}^{B}\}))$.

Nucleus-nucleus elastic scattering amplitude $f_{AB}(s,b)$ can now
be defined taking into account contributions from any number of Pomerons
exchanged between an arbitrary pair of the projectile and target nucleons
and from exchanges of arbitrary enhanced graphs between the two nuclei:
\begin{eqnarray}
f_{AB}(s,b)=i\left\langle \left\langle 1-e^{-\sum_{m=1}^{A}\sum_{n=1}^{B}
\chi_{pp(j_{m}^{A}k_{n}^{B})}^{\mathbb{P}}(s,|\vec{b}+\vec{b}_{m}^{A}-\vec{b}_{n}^{B}|)
-\chi_{AB}^{{\rm enh}}(s,b,\{j^{A},k^{B},\vec{b}^{A},\vec{b}^{B}\})}\right
\rangle _{A}\right\rangle _{B}\nonumber \\
=i\left\langle \left\langle 1-\exp\!\left[-\frac{1}{2}\,\sum_{m=1}^{A}\sum_{n=1}^{B}
\Omega_{AB}^{(mn)}(s,b,\{j^{A},k^{B},\vec{b}^{A},\vec{b}^{B}\})\right]\right
\rangle _{A}\right\rangle _{B}\label{eq:f-AB}\\
\Omega_{AB}^{(mn)}(s,b,\{j^{A},k^{B},\vec{b}^{A},\vec{b}^{B}\})
=2\chi_{pp(j_{m}^{A}k_{n}^{B})}^{\mathbb{P}}(s,|\vec{b}+\vec{b}_{m}^{A}-\vec{b}_{n}^{B}|)
+2\chi_{mn}^{{\rm enh}}(s,b,\{j^{A},k^{B},\vec{b}^{A},\vec{b}^{B}\})\,,
\label{eq:opac-mn}\end{eqnarray}
where for averaging over transverse coordinates and elastic scattering
eigenstates of the nucleons we used the notation
\[
\left\langle h(\{j^{A},\vec{b}^{A}\})\right\rangle _{A}
=\sum_{j_{1}^{A}\ldots j_{A}^{A}}C_{j_{1}^{A}/p}\times\ldots\times C_{j_{A}^{A}/p}
\int\! d^{2}b_{1}^{A}\ldots d^{2}b_{A}^{A}\; 
T_{A}(\vec{b}_{1}^{A},\ldots,\vec{b}_{A}^{A})\; 
h_{j_{1}^{A}\ldots j_{A}^{A}}(\vec{b}_{1}^{A},\ldots,\vec{b}_{A}^{A})\,,
\]
with the profile function $T_{A}$ being expressed via nuclear ground
state density $\rho_{A}$ as \begin{equation}
T_{A}(\vec{b}_{1}^{A},\ldots,\vec{b}_{A}^{A})=\int\! dz_{1}^{A}\ldots dz_{A}^{A}\;\rho_{A}(\vec{r}_{1}^{A},\ldots,\vec{r}_{A}^{A})\,.\label{eq:profile}\end{equation}
Expression (\ref{eq:opac-mn}) for nucleus-nucleus scattering
amplitude reminds the usual multichannel eikonal form \cite{qgs93},
looking as a combination of binarylike nucleon-nucleon rescatterings.
In reality, each of the partial opacities $\Omega_{AB}^{(mn)}$ generally
depends on the transverse coordinates and elastic scattering eigenstates
of all the $A+B$ projectile and target nucleons and contains absorptive
corrections due to rescattering processes on those nucleons.

Knowing the elastic amplitude, one can easily calculate total and
elastic cross sections as \begin{eqnarray}
\sigma_{AB}^{{\rm tot}}(s)=2\int\! d^{2}b\,\mathfrak{Im}f_{AB}(s,b)\label{sigma-tot}\\
\sigma_{AB}^{{\rm el}}(s)=\int\! d^{2}b\,\left|f_{AB}(s,b)\right|^{2}.\label{eq:sigma-el}\end{eqnarray}

Hadron-nucleus scattering amplitude is obtained using in (\ref{eq:chi-AB-mn}-\ref{eq:opac-mn})
$A=1$, $T_{A}(\vec{b}_{1}^{A})=\delta^{(2)}(\vec{b}_{1}^{A})$ and
replacing the eikonals $\chi_{pp(j_{1}^{A}k_{n}^{B})}^{\mathbb{P}}$,
$\chi_{p(j_{1}^{A})}^{{\rm net}}$ with $\chi_{ap(j_{1}^{A}k_{n}^{B})}^{\mathbb{P}}$,
$\chi_{a(j_{1}^{A})}^{{\rm net}}$ for a given projectile hadron $a$;
the hadron-hadron case is recovered similarly.

\section*{Appendix B}
\begin{figure}[htb]
\begin{centering}
\includegraphics[width=14cm,height=5.5cm]{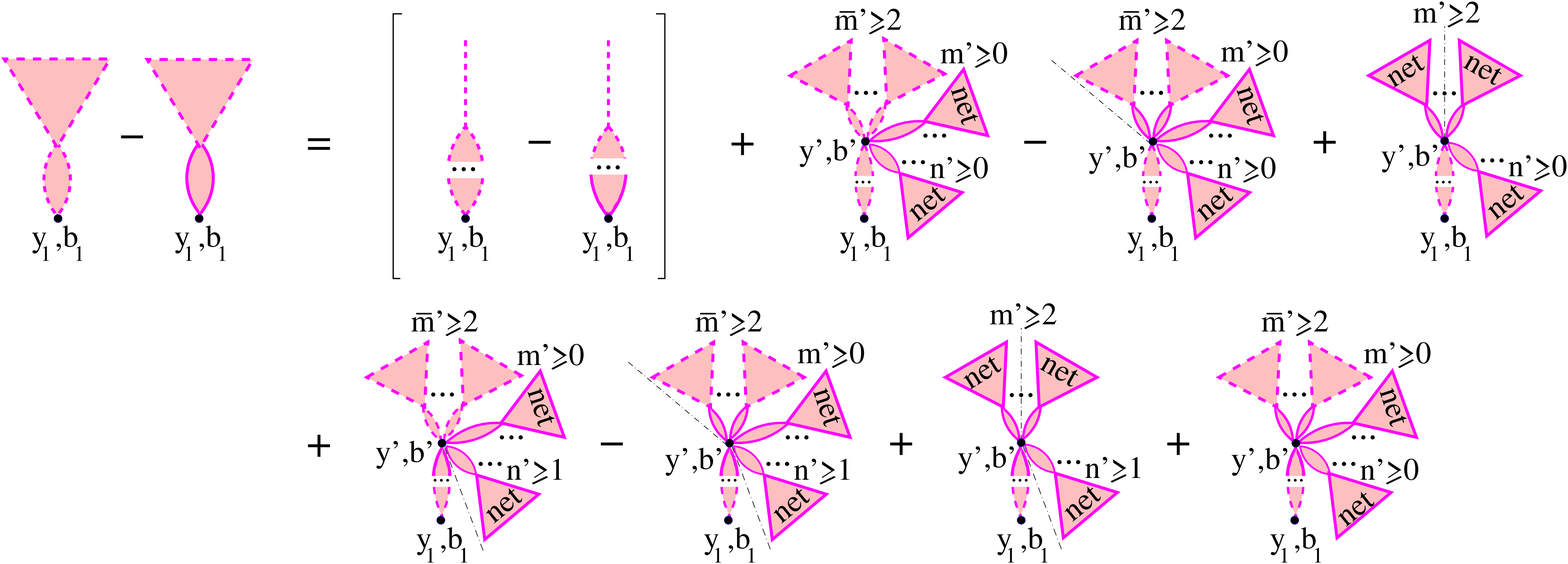}
\par\end{centering}

\caption{Alternative representation for the contribution $2\hat{\chi}_{a(j)|d(k)}^{{\rm fan}}$
of fanlike cuts of net-fans, the handle of the fan being cut. \label{fan-cut1-fig}}

\end{figure}

\begin{figure}[htb]
\begin{centering}
\includegraphics[width=11cm,height=5cm]{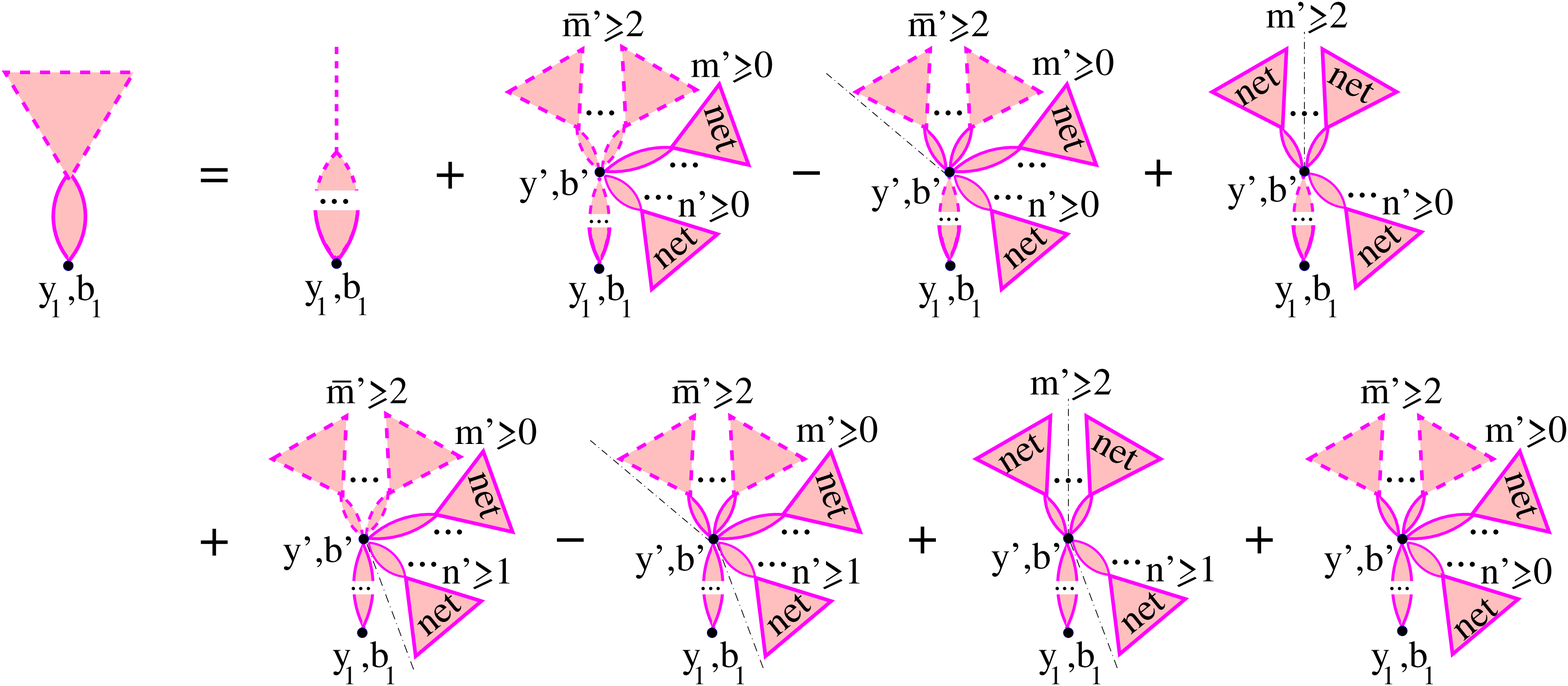}
\par\end{centering}

\caption{Alternative representation for the contribution $2\tilde{\chi}_{a(j)|d(k)}^{{\rm fan}}$
of fanlike cuts of net-fans, the handle of the fan being uncut.
\label{fan-cuth-fig}}

\end{figure}

To develop a MC procedure for sampling various configurations of hadronic
collisions we shall need alternative representations for the cut net-fan
contributions $2\hat{\chi}_{a(j)|d(k)}^{{\rm fan}}$, $2\tilde{\chi}_{a(j)|d(k)}^{{\rm fan}}$.
Those should allow us to generate recursively the cut Pomeron structure
for the corresponding subgraphs, with all the absorptive corrections
due to uncut Pomerons being summed up. Such representations are obtained
applying recursively the graphic equations of Fig.~\ref{fan-cut-fig}
to generate a $t$-channel sequence of multi-Pomeron vertices $(y_{2},\vec{b}_{2}),(y_{3},\vec{b}_{3}),\cdots$,
such that these vertices are coupled to uncut projectile and target
net-fans only ($m_{l}+n_{l}\geq1$, $\bar{m}_{l}=\bar{n}_{l}=0,$
$l=2,3,\cdots$) and they are connected to each other by either cut
or uncut 2-point sequences of Pomerons and Pomeron loops. The recursive
procedure stops in some vertex $(y',b')$ when: i) the vertex $(y',b')$
is connected to the given (here, projectile) hadron by a \textsl{cut}
2-point loop sequence, with a single \textsl{cut} Pomeron coupled
to the hadron, or ii) the vertex is coupled to $\bar{m}'\geq2$ \textsl{cut}
projectile net-fans, or iii) the vertex is coupled to $m'\geq2$ uncut
projectile net-fans, with the cut plane positioned between them, i.e.~there
is a diffractive cut between the projectile hadron and that vertex.%
\footnote{Note, however, that the corresponding rapidity gap may be filled by
particles resulting from other cut Pomerons produced in the corresponding
rapidity interval.%
} The resulting Schwinger-Dyson equations are depicted in
 Figs.~\ref{fan-cut1-fig}, \ref{fan-cuth-fig} 
 and read\begin{eqnarray}
2\hat{\chi}_{a(j)|d(k)}^{{\rm fan}}(y_{1},\vec{b}_{1}|Y,\vec{b})=\left[2\bar{\chi}_{a(j)|d(k)}^{{\rm loop}}(y_{1},\vec{b}_{1}|Y,\vec{b})-2\tilde{\chi}_{a(j)|d(k)}^{{\rm loop}}(y_{1},\vec{b}_{1}|Y,\vec{b})\right]+G\int_{\xi}^{y_{1}-\xi}\! dy'\int\! d^{2}b'\nonumber \\
\times\left\{ \left[\chi_{a(j)|d(k)}^{{\rm loop}_{{\rm cc}}}(y_{1},y',\vec{b}_{1},\vec{b}'|Y,\vec{b})\, e^{-2\chi_{d(k)|a(j)}^{{\rm net}}}-\chi_{a(j)|d(k)}^{{\rm loop}_{{\rm cu}}}(y_{1},y',\vec{b}_{1},\vec{b}'|Y,\vec{b})\left(1-e^{-\chi_{d(k)|a(j)}^{{\rm net}}}\right)e^{-\chi_{d(k)|a(j)}^{{\rm net}}}\right]\right.\nonumber \\
\times\left[\sum_{\bar{m}'=2}^{\infty}\frac{1}{\bar{m}'!}\left(\left(2\hat{\chi}_{a(j)|d(k)}^{{\rm fan}}+2\tilde{\chi}_{a(j)|d(k)}^{{\rm fan}}\right)^{\bar{m}'}e^{-2\chi_{a(j)|d(k)}^{{\rm net}}}-2\left(\tilde{\chi}_{a(j)|d(k)}^{{\rm fan}}\right)^{\bar{m}'}e^{-\chi_{a(j)|d(k)}^{{\rm net}}}\right)\right.\nonumber \\
+\left.\left.\left(1-e^{-\chi_{a(j)|d(k)}^{{\rm net}}}\right)^{2}\right]-2\chi_{a(j)|d(k)}^{{\rm loop}_{{\rm cu}}}(y_{1},y',\vec{b}_{1},\vec{b}'|Y,\vec{b})\, e^{-\chi_{a(j)|d(k)}^{{\rm net}}-\chi_{d(k)|a(j)}^{{\rm net}}}\sum_{\bar{m}'=2}^{\infty}\frac{\left(\tilde{\chi}_{a(j)|d(k)}^{{\rm fan}}\right)^{\bar{m}'}}{\bar{m}'!}\right\} \label{eq:chi-hat-alt}\\
2\tilde{\chi}_{a(j)|d(k)}^{{\rm fan}}(y_{1},\vec{b}_{1}|Y,\vec{b})=2\tilde{\chi}_{a(j)|d(k)}^{{\rm loop}}(y_{1},\vec{b}_{1}|Y,\vec{b})+G\int_{\xi}^{y_{1}-\xi}\! dy'\int\! d^{2}b'\nonumber \\
\times\left\{ \left[\chi_{a(j)|d(k)}^{{\rm loop}_{{\rm uc}}}(y_{1},y',\vec{b}_{1},\vec{b}'|Y,\vec{b})\, e^{-2\chi_{d(k)|a(j)}^{{\rm net}}}+\chi_{a(j)|d(k)}^{{\rm loop}_{{\rm uu}}}(y_{1},y',\vec{b}_{1},\vec{b}'|Y,\vec{b})\left(1-e^{-\chi_{d(k)|a(j)}^{{\rm net}}}\right)e^{-\chi_{d(k)|a(j)}^{{\rm net}}}\right]\right.\nonumber \\
\times\left[\sum_{\bar{m}'=2}^{\infty}\frac{1}{\bar{m}'!}\left(\left(2\hat{\chi}_{a(j)|d(k)}^{{\rm fan}}+2\tilde{\chi}_{a(j)|d(k)}^{{\rm fan}}\right)^{\bar{m}'}e^{-2\chi_{a(j)|d(k)}^{{\rm net}}}-2\left(\tilde{\chi}_{a(j)|d(k)}^{{\rm fan}}\right)^{\bar{m}'}e^{-\chi_{a(j)|d(k)}^{{\rm net}}}\right)\right.\nonumber \\
+\left.\left.\left(1-e^{-\chi_{a(j)|d(k)}^{{\rm net}}}\right)^{2}\right]+2\chi_{a(j)|d(k)}^{{\rm loop}_{{\rm uu}}}(y_{1},y',\vec{b}_{1},\vec{b}'|Y,\vec{b})\, e^{-\chi_{a(j)|d(k)}^{{\rm net}}-\chi_{d(k)|a(j)}^{{\rm net}}}\sum_{\bar{m}'=2}^{\infty}\frac{\left(\tilde{\chi}_{a(j)|d(k)}^{{\rm fan}}\right)^{\bar{m}'}}{\bar{m}'!}\right\} \!,\label{eq:chi-tild-alt}\end{eqnarray}
where the arguments of the eikonals in the integrand read $\chi_{a(j)|d(k)}^{{\rm net}}=\chi_{a(j)|d(k)}^{{\rm net}}(y',\vec{b}'|Y,\vec{b})$,
$\hat{\chi}_{a(j)|d(k)}^{{\rm fan}}=\hat{\chi}_{a(j)|d(k)}^{{\rm fan}}(y',\vec{b}'|Y,\vec{b})$,
$\tilde{\chi}_{a(j)|d(k)}^{{\rm fan}}=\tilde{\chi}_{a(j)|d(k)}^{{\rm fan}}(y',\vec{b}'|Y,\vec{b})$,
$\chi_{d(k)|a(j)}^{{\rm net}}=\chi_{d(k)|a(j)}^{{\rm net}}(Y-y',\vec{b}-\vec{b}'|Y,\vec{b})$.

\begin{figure}[htb]
\begin{centering}
\includegraphics[width=15cm,height=6cm]{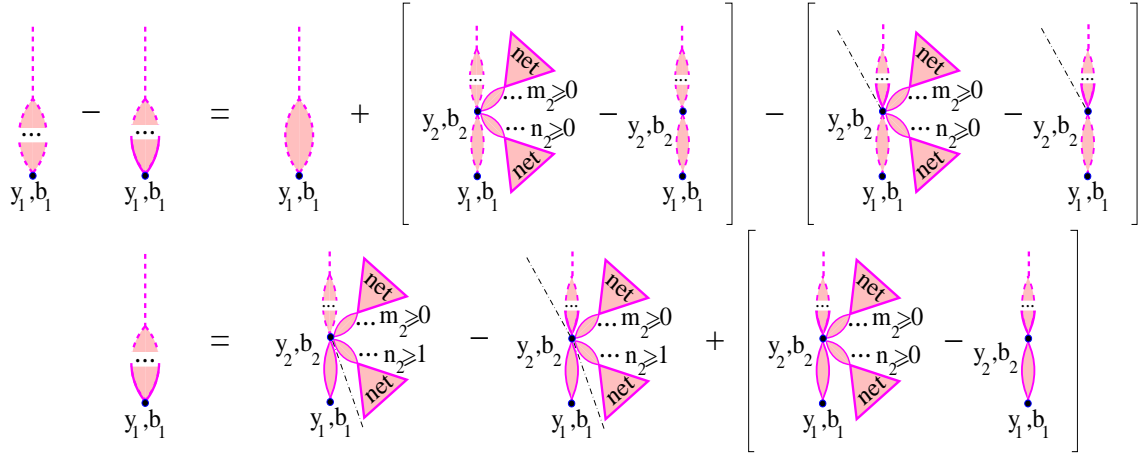}
\par\end{centering}

\caption{Recursive representations for the contributions
 $2\bar{\chi}_{a(j)|d(k)}^{{\rm loop}}-2\tilde{\chi}_{a(j)|d(k)}^{{\rm loop}}$
(top) and $2\tilde{\chi}_{a(j)|d(k)}^{{\rm loop}}$ (bottom) of the 
subsets of fanlike cuts of net-fans, which have 
  a single cut Pomeron coupled to the projectile hadron.\label{pom-legc-fig}}

\end{figure}

The contributions $2\bar{\chi}_{a(j)|d(k)}^{{\rm loop}}$ 
and $2\tilde{\chi}_{a(j)|d(k)}^{{\rm loop}}$
correspond to the subset of graphs obtained in   case (i) above,
being defined by the recursive equations (see Fig.~\ref{pom-legc-fig})%
 \begin{eqnarray}
\bar{\chi}_{a(j)|d(k)}^{{\rm loop}}(y_{1},\vec{b}_{1}|Y,\vec{b})=\chi_{a(j)}^{{\rm loop}}(y_{1},b_{1})+G\int_{\xi}^{y_{1}-\xi}\! dy_{2}\int\! d^{2}b_{2}\;\left(1-e^{-\chi^{{\rm loop}}(y_{1}-y_{2},|\vec{b}_{1}-\vec{b}_{2}|)}\right)\nonumber \\
\times\,\bar{\chi}_{a(j)|d(k)}^{{\rm loop}}(y_{2},\vec{b}_{2}|Y,\vec{b})\left(e^{-2\chi_{a(j)|d(k)}^{{\rm net}}(y_{2},\vec{b}_{2}|Y,\vec{b})-\chi_{d(k)|a(j)}^{{\rm net}}(Y-y_{2},\vec{b}-\vec{b}_{2}|Y,\vec{b})}-1\right)\label{eq:chi-loop-scr}\\
\tilde{\chi}_{a(j)|d(k)}^{{\rm loop}}(y_{1},\vec{b}_{1}|Y,\vec{b})=G\int_{\xi}^{y_{1}-\xi}\! dy_{2}\int\! d^{2}b_{2}\;\left(1-e^{-\chi^{{\rm loop}}(y_{1}-y_{2},|\vec{b}_{1}-\vec{b}_{2}|)}\right)\left[\bar{\chi}_{a(j)|d(k)}^{{\rm loop}}(y_{2},\vec{b}_{2}|Y,\vec{b})\right.\nonumber \\
\times\left(1-e^{-\chi_{d(k)|a(j)}^{{\rm net}}(Y-y_{2},\vec{b}-\vec{b}_{2}|Y,\vec{b})}\right)e^{-2\chi_{a(j)|d(k)}^{{\rm net}}(y_{2},\vec{b}_{2}|Y,\vec{b})-\chi_{d(k)|a(j)}^{{\rm net}}(Y-y_{2},\vec{b}-\vec{b}_{2}|Y,\vec{b})}\nonumber \\
+\left.\tilde{\chi}_{a(j)|d(k)}^{{\rm loop}}(y_{2},\vec{b}_{2}|Y,\vec{b})\left(e^{-\chi_{a(j)|d(k)}^{{\rm net}}(y_{2},\vec{b}_{2}|Y,\vec{b})-2\chi_{d(k)|a(j)}^{{\rm net}}(Y-y_{2},\vec{b}-\vec{b}_{2}|Y,\vec{b})}-1\right)\right]\!.\label{eq:chi-loopt-scr}\end{eqnarray}

In a similar way, for the contributions $\chi_{a(j)|d(k)}^{{\rm loop}_{xy}}(y_{1},y',\vec{b}_{1},\vec{b}'|Y,\vec{b})$
corresponding to $t$-channel sequences of multi-Pomeron vertices
positioned between $(y_{1},b_{1})$ and $(y',\vec{b}')$, coupled
to uncut projectile and target net-fans and connected to each other
and to the vertices $(y_{1},b_{1})$, $(y',\vec{b}')$ by cut or uncut
2-point loop sequences {[}the index $x$ ($y$) indicates whether
the down-most (uppermost) loop sequence is cut, $x={\rm c}$ ($y={\rm c}$),
or uncut, $x={\rm u}$ ($y={\rm u}$){]}, one obtains the equation
system\begin{eqnarray}
\chi_{a(j)|d(k)}^{{\rm loop}_{{\rm cc}}}(y_{1},y',\vec{b}_{1},\vec{b}'|Y,\vec{b})=\left[1-e^{-\chi^{{\rm loop}}(y_{1}-y',|\vec{b}_{1}-\vec{b}'|)}\right]+G\!\int_{y_{1}+\xi}^{y'-\xi}\!\! dy_{2}\int\!\! d^{2}b_{2}\left[1-e^{-\chi^{{\rm loop}}(y_{1}-y_{2},|\vec{b}_{1}-\vec{b}_{2}|)}\right]\nonumber \\
\times\left[\chi_{a(j)|d(k)}^{{\rm loop}_{{\rm cc}}}(y_{2},y',\vec{b}_{2},\vec{b}'|Y,\vec{b})\left(e^{-2\chi_{a(j)|d(k)}^{{\rm net}}-2\chi_{d(k)|a(j)}^{{\rm net}}}-1\right)-\chi_{a(j)|d(k)}^{{\rm loop}_{{\rm uc}}}(y_{2},y',\vec{b}_{2},\vec{b}'|Y,\vec{b})\right.\nonumber \\
\times\left.\left(1-e^{-\chi_{a(j)|d(k)}^{{\rm net}}}\right)e^{-\chi_{a(j)|d(k)}^{{\rm net}}-2\chi_{d(k)|a(j)}^{{\rm net}}}\right]\label{eq:loop-cc}\\
\chi_{a(j)|d(k)}^{{\rm loop}_{{\rm cu}}}(y_{1},y',\vec{b}_{1},\vec{b}'|Y,\vec{b})=G\!\int_{y_{1}+\xi}^{y'-\xi}\!\! dy_{2}\int\!\! d^{2}b_{2}\left[1-e^{-\chi^{{\rm loop}}(y_{1}-y_{2},|\vec{b}_{1}-\vec{b}_{2}|)}\right]\nonumber \\
\times\left[\chi_{a(j)|d(k)}^{{\rm loop}_{{\rm cu}}}(y_{2},y',\vec{b}_{2},\vec{b}'|Y,\vec{b})\left(e^{-2\chi_{a(j)|d(k)}^{{\rm net}}-2\chi_{d(k)|a(j)}^{{\rm net}}}-1\right)+\chi_{a(j)|d(k)}^{{\rm loop}_{{\rm uu}}}(y_{2},y',\vec{b}_{2},\vec{b}'|Y,\vec{b})\right.\nonumber \\
\times\left.\left(1-e^{-\chi_{a(j)|d(k)}^{{\rm net}}}\right)e^{-\chi_{a(j)|d(k)}^{{\rm net}}-2\chi_{d(k)|a(j)}^{{\rm net}}}\right]\label{eq:loop-cu}\\
\chi_{a(j)|d(k)}^{{\rm loop}_{{\rm uc}}}(y_{1},y',\vec{b}_{1},\vec{b}'|Y,\vec{b})=G\!\int_{y_{1}+\xi}^{y'-\xi}\!\! dy_{2}\int\!\! d^{2}b_{2}\left[1-e^{-\chi^{{\rm loop}}(y_{1}-y_{2},|\vec{b}_{1}-\vec{b}_{2}|)}\right]\nonumber \\
\times\left[\chi_{a(j)|d(k)}^{{\rm loop}_{{\rm cc}}}(y_{2},y',\vec{b}_{2},\vec{b}'|Y,\vec{b})\left(1-e^{-\chi_{d(k)|a(j)}^{{\rm net}}}\right)e^{-2\chi_{a(j)|d(k)}^{{\rm net}}-\chi_{d(k)|a(j)}^{{\rm net}}}+\chi_{a(j)|d(k)}^{{\rm loop}_{{\rm uc}}}(y_{2},y',\vec{b}_{2},\vec{b}'|Y,\vec{b})\right.\nonumber \\
\times\left.\left(e^{-\chi_{a(j)|d(k)}^{{\rm net}}-2\chi_{d(k)|a(j)}^{{\rm net}}}+e^{-2\chi_{a(j)|d(k)}^{{\rm net}}-\chi_{d(k)|a(j)}^{{\rm net}}}-e^{-2\chi_{a(j)|d(k)}^{{\rm net}}-2\chi_{d(k)|a(j)}^{{\rm net}}}-1\right)\right]\label{eq:loop-uc}\\
\chi_{a(j)|d(k)}^{{\rm loop}_{{\rm uu}}}(y_{1},y',\vec{b}_{1},\vec{b}'|Y,\vec{b})=\left[1-e^{-\chi^{{\rm loop}}(y_{1}-y',|\vec{b}_{1}-\vec{b}'|)}\right]+G\!\int_{y_{1}+\xi}^{y'-\xi}\!\! dy_{2}\int\!\! d^{2}b_{2}\left[1-e^{-\chi^{{\rm loop}}(y_{1}-y_{2},|\vec{b}_{1}-\vec{b}_{2}|)}\right]\nonumber \\
\times\left[-\chi_{a(j)|d(k)}^{{\rm loop}_{{\rm cu}}}(y_{2},y',\vec{b}_{2},\vec{b}'|Y,\vec{b})\left(1-e^{-\chi_{d(k)|a(j)}^{{\rm net}}}\right)e^{-2\chi_{a(j)|d(k)}^{{\rm net}}-\chi_{d(k)|a(j)}^{{\rm net}}}+\chi_{a(j)|d(k)}^{{\rm loop}_{{\rm uu}}}(y_{2},y',\vec{b}_{2},\vec{b}'|Y,\vec{b})\right.\nonumber \\
\times\left.\left(e^{-\chi_{a(j)|d(k)}^{{\rm net}}-2\chi_{d(k)|a(j)}^{{\rm net}}}+e^{-2\chi_{a(j)|d(k)}^{{\rm net}}-\chi_{d(k)|a(j)}^{{\rm net}}}-e^{-2\chi_{a(j)|d(k)}^{{\rm net}}-2\chi_{d(k)|a(j)}^{{\rm net}}}-1\right)\right]\!,\label{eq:loop-uu}\end{eqnarray}
where the omitted arguments of the eikonals $\chi_{a(j)|d(k)}^{{\rm net}}$,
$\chi_{d(k)|a(j)}^{{\rm net}}$ are the same as in Eqs.~(\ref{eq:chi-loop-scr}-\ref{eq:chi-loopt-scr}).

\section*{Appendix C}

In case of nucleus-nucleus scattering, different cut and uncut Pomerons
from the same irreducible graph may couple to different projectile
and target nucleons. We start again from the contributions of fanlike
cuts of net-fans, Fig.~\ref{fan-cut-fig}, for which we obtain \begin{eqnarray}
2\hat{\chi}_{A|B}^{{\rm fan}}(y_{1},\vec{b}_{1}|Y,\vec{b})=\sum_{m=1}^{A}2\chi_{p(j_{m}^{A})}^{{\rm loop}}(y_{1},|\vec{b}_{1}-\vec{b}_{m}^{A}|)+G\int_{\xi}^{y_{1}-\xi}\! dy_{2}\int\! d^{2}b_{2}\;\left(1-e^{-\chi^{{\rm loop}}(y_{1}-y_{2},|\vec{b}_{1}-\vec{b}_{2}|)}\right)\nonumber \\
\times\left\{ \left[\left(e^{2\bar{\chi}_{A|B}^{{\rm fan}}(y_{2},\vec{b}_{2}|Y,\vec{b})}-1\right)e^{-2\chi_{A|B}^{{\rm net}}(y_{2},\vec{b}_{2}|Y,\vec{b})-2\chi_{B|A}^{{\rm net}}(Y-y_{2},\vec{b}-\vec{b}_{2}|Y,\vec{b})}-2\bar{\chi}_{A|B}^{{\rm fan}}(y_{2},\vec{b}_{2}|Y,\vec{b})\right]\right.\nonumber \\
-2\left[\left(e^{\tilde{\chi}_{A|B}^{{\rm fan}}(y_{2},\vec{b}_{2}|Y,\vec{b})}-1\right)e^{-\chi_{A|B}^{{\rm net}}(y_{2},\vec{b}_{2}|Y,\vec{b})-2\chi_{B|A}^{{\rm net}}(Y-y_{2},\vec{b}-\vec{b}_{2}|Y,\vec{b})}-\tilde{\chi}_{A|B}^{{\rm fan}}(y_{2},\vec{b}_{2}|Y,\vec{b})\right]\nonumber \\
+\left.\left(1-e^{-\chi_{A|B}^{{\rm net}}(y_{2},\vec{b}_{2}|Y,\vec{b})}\right)^{2}e^{-2\chi_{B|A}^{{\rm net}}(Y-y_{2},\vec{b}-\vec{b}_{2}|Y,\vec{b})}\right\} \label{eq:chi-AB-fancut}\\
2\tilde{\chi}_{A|B}^{{\rm fan}}(y_{1},\vec{b}_{1}|Y,\vec{b})=G\int_{\xi}^{y_{1}-\xi}\! dy_{2}\int\! d^{2}b_{2}\;\left(1-e^{-\chi^{{\rm loop}}(y_{1}-y_{2},|\vec{b}_{1}-\vec{b}_{2}|)}\right)\nonumber \\
\times\left\{ \left(1-e^{-\chi_{B|A}^{{\rm net}}(Y-y_{2},\vec{b}-\vec{b}_{2}|Y,\vec{b})}\right)e^{-\chi_{B|A}^{{\rm net}}(Y-y_{2},\vec{b}-\vec{b}_{2}|Y,\vec{b})}\left[\left(e^{2\bar{\chi}_{A|B}^{{\rm fan}}(y_{2},\vec{b}_{2}|Y,\vec{b})}-1\right)e^{-2\chi_{A|B}^{{\rm net}}(y_{2},\vec{b}_{2}|Y,\vec{b})}\right.\right.\nonumber \\
-\left.2\left(e^{\tilde{\chi}_{A|B}^{{\rm fan}}(y_{2},\vec{b}_{2}|Y,\vec{b})}-1\right)e^{-\chi_{A|B}^{{\rm net}}(y_{2},\vec{b}_{2}|Y,\vec{b})}+\left(1-e^{-\chi_{A|B}^{{\rm net}}(y_{2},\vec{b}_{2}|Y,\vec{b})}\right)^{2}\right]\nonumber \\
+\left.2\left[\left(e^{\tilde{\chi}_{A|B}^{{\rm fan}}(y_{2},\vec{b}_{2}|Y,\vec{b})}-1\right)e^{-\chi_{A|B}^{{\rm net}}(y_{2},\vec{b}_{2}|Y,\vec{b})-\chi_{B|A}^{{\rm net}}(Y-y_{2},\vec{b}-\vec{b}_{2}|Y,\vec{b})}-\tilde{\chi}_{A|B}^{{\rm fan}}(y_{2},\vec{b}_{2}|Y,\vec{b})\right]\right\} \!,\label{eq:chi-hole-AB}\end{eqnarray}
where $2\hat{\chi}_{A|B}^{{\rm fan}}$ corresponds to cut graphs where
the handle of the fan is cut, $2\tilde{\chi}_{A|B}^{{\rm fan}}$ -
to the ones where it is uncut, and $2\bar{\chi}_{A|B}^{{\rm fan}}=2\hat{\chi}_{A|B}^{{\rm fan}}+2\tilde{\chi}_{A|B}^{{\rm fan}}$
is the total contribution of fanlike cuts of net-fans; the dependence
of the eikonals on the coordinates and eigenstates of all the nucleons
is not shown explicitly.

We are going to proceed like in Appendix~A,
expanding $\bar{\chi}_{A|B}^{{\rm fan}}$, $\hat{\chi}_{A|B}^{{\rm fan}}$,
and $\tilde{\chi}_{A|B}^{{\rm fan}}$ as

\begin{eqnarray}
\bar{\chi}_{A|B}^{{\rm fan}}(y_{1},\vec{b}_{1}|Y,\vec{b})=\sum_{m=1}^{A}\bar{\chi}_{p(j_{m}^{A})}^{{\rm fan}}(y_{1},|\vec{b}_{1}-\vec{b}_{m}^{A}|,\ldots)\label{eq:chi-fan-exp}\\
\hat{\chi}_{A|B}^{{\rm fan}}(y_{1},\vec{b}_{1}|Y,\vec{b})=\sum_{m=1}^{A}\hat{\chi}_{p(j_{m}^{A})}^{{\rm fan}}(y_{1},|\vec{b}_{1}-\vec{b}_{m}^{A}|,\ldots),\label{eq:chi-hat-exp}\\
\tilde{\chi}_{p(j_{m}^{A})}^{{\rm fan}}(y_{1},|\vec{b}_{1}-\vec{b}_{m}^{A}|,\ldots)=\bar{\chi}_{p(j_{m}^{A})}^{{\rm fan}}(y_{1},|\vec{b}_{1}-\vec{b}_{m}^{A}|,\ldots)-\hat{\chi}_{p(j_{m}^{A})}^{{\rm fan}}(y_{1},|\vec{b}_{1}-\vec{b}_{m}^{A}|,\ldots)\label{eq:chi-tilde-exp}\end{eqnarray}
and approximating the dependence on the coordinates and eigenstates
of all the $A+B-1$ projectile and target nucleons but the current
one by some factors.

Adding (\ref{eq:chi-AB-fancut}) to (\ref{eq:chi-hole-AB}), substituting
the decompositions (\ref{eq:chi-A-m}), (\ref{eq:chi-fan-exp}-\ref{eq:chi-tilde-exp}),
and using the identities
\begin{eqnarray}
\exp\!\left(2\sum_{m=1}^{A}\bar{\chi}_{p(j_{m}^{A})}^{{\rm fan}}\right)-1
=\sum_{m=1}^{A}\left(e^{2\bar{\chi}_{p(j_{m}^{A})}^{{\rm fan}}}-1\right)
e^{2\sum_{l=m+1}^{A}\bar{\chi}_{p(j_{l}^{A})}^{{\rm fan}}}
\label{eq:decomp-chi-bar}\\
\exp\!\left(\sum_{m=1}^{A}\tilde{\chi}_{p(j_{m}^{A})}^{{\rm fan}}\right)-1
=\sum_{m=1}^{A}\left(e^{\tilde{\chi}_{p(j_{m}^{A})}^{{\rm fan}}}-1\right)
e^{\sum_{l=m+1}^{A}\tilde{\chi}_{p(j_{l}^{A})}^{{\rm fan}}}
\label{eq:decomp-chi-tild}\\
\left[1-\exp\!\left(-\sum_{m=1}^{A}\chi_{p(j_{m}^{A})}^{{\rm net}}\right)\right]^{2}
=\sum_{m=1}^{A}\left[\left(1-e^{-\chi_{p(j_{m}^{A})}^{{\rm net}}}\right)^{2}
e^{-2\sum_{l=1}^{m-1}\chi_{p(j_{l}^{A})}^{{\rm net}}}\right.\nonumber \\
+\left.2\left(1-e^{-\chi_{p(j_{m}^{A})}^{{\rm net}}}\right)
\left(1-e^{-\sum_{l=1}^{m-1}\chi_{p(j_{l}^{A})}^{{\rm net}}}\right)
e^{-\sum_{l=1}^{m-1}\chi_{p(j_{l}^{A})}^{{\rm net}}}\right]\!,
\label{eq:decomp-diffr}\end{eqnarray}
 we obtain
 \begin{eqnarray}
\sum_{m=1}^{A}2\bar{\chi}_{p(j_{m}^{A})}^{{\rm fan}}(y_{1},|\vec{b}_{1}-\vec{b}_{m}^{A}|,\ldots)=\sum_{m=1}^{A}\left\{ 2\chi_{p(j_{m}^{A})}^{{\rm loop}}(y_{1},|\vec{b}_{1}-\vec{b}_{m}^{A}|)+G\int_{\xi}^{y_{1}-\xi}\! dy_{2}\int\! d^{2}b_{2}\right.\nonumber \\
\times\left(1-e^{-\chi^{{\rm loop}}}\right)\left[\left(e^{2(\bar{\chi}_{p(j_{m}^{A})}^{{\rm fan}}-\chi_{p(j_{m}^{A})}^{{\rm net}})}-e^{-2\chi_{p(j_{m}^{A})}^{{\rm net}}}\right)e^{2\sum_{l=m+1}^{A}(\bar{\chi}_{p(j_{l}^{A})}^{{\rm fan}}-\chi_{p(j_{l}^{A})}^{{\rm net}})}\left(Z_{A|B}^{(m)}\right)^{2}\right.\nonumber \\
-\left.\left.2\bar{\chi}_{p(j_{m}^{A})}^{{\rm fan}}+\left(1-e^{-\chi_{p(j_{m}^{A})}^{{\rm net}}}\right)^{2}\left(Z_{A|B}^{(m)}\right)^{2}+2\left(1-e^{-\chi_{p(j_{m}^{A})}^{{\rm net}}}\right)Z_{A|B}^{(m)}\left(1-Z_{A|B}^{(m)}\right)\right]\right\} \!,\label{eq:chi-fan-decomp}\end{eqnarray}
where the arguments of the eikonals in the integrand are the same
as in (\ref{eq:net-A-decomp}) and $Z_{A|B}^{(m)}$ is defined in
(\ref{eq:Z-factor}).

Comparing (\ref{eq:chi-fan-decomp}) to (\ref{eq:net-A-decomp}),
we obtain\begin{equation}
\bar{\chi}_{p(j_{m}^{A})}^{{\rm fan}}(y_{1},|\vec{b}_{1}-\vec{b}_{m}^{A}|,\ldots)=\chi_{p(j_{m}^{A})}^{{\rm net}}(y_{1},|\vec{b}_{1}-\vec{b}_{m}^{A}|,\ldots)\,.\label{eq:equiv}\end{equation}

Now, substituting (\ref{eq:chi-A-m}), (\ref{eq:chi-fan-exp}-\ref{eq:chi-tilde-exp})
into (\ref{eq:chi-AB-fancut}), using (\ref{eq:decomp-chi-bar}-\ref{eq:decomp-diffr}),
(\ref{eq:equiv}), and doing the same approximation as in (\ref{eq:Z-approx}),
we obtain
\begin{eqnarray}
\hat{\chi}_{A|B}^{{\rm fan}}(y_{1},\vec{b}_{1}|Y,\vec{b})
=\sum_{m=1}^{A}
\hat{\chi}_{p(j_{m}^{A})}^{{\rm fan}}\!\left(y_{1},|\vec{b}_{1}-\vec{b}_{m}^{A}|,
Z_{A|B}^{(m)}(y_{1},\vec{b}_{1},Y,\vec{b},\{j^{A},k^{B},\vec{b}^{A},\vec{b}^{B}\}),\right.
\nonumber \\
\left.\hat{Z}_{A}^{(m)}(y_{1},\vec{b}_{1},Y,\vec{b},\{j^{A},k^{B},\vec{b}^{A},\vec{b}^{B}\}),
Z_{B}(Y-y_{1},\vec{b}-\vec{b}_{1},Y,\vec{b},\{k^{B},j^{A},\vec{b}^{B},\vec{b}^{A}\})\right)\!,
\label{eq:chi-hat-AB}
\end{eqnarray}
where we introduced
\begin{eqnarray}
\hat{Z}_{A}^{(m)}(y_{1},\vec{b}_{1},Y,\vec{b},\{j^{A},k^{B},\vec{b}^{A},\vec{b}^{B}\})
=\exp\!\left[-\sum_{l=m+1}^{A}
\hat{\chi}_{p(j_{l}^{A})}^{{\rm fan}}(y_{1},|\vec{b}_{1}-\vec{b}_{l}^{A}|,
Z_{A|B}^{(l)},\hat{Z}_{A}^{(l)},Z_{B})\right]\label{eq:z-hat-A}\\
Z_{A}(y_{1},\vec{b}_{1},Y,\vec{b},\{j^{A},k^{B},\vec{b}^{A},\vec{b}^{B}\})
=\exp\!\left[-\sum_{l=1}^{A}
\chi_{p(j_{l}^{A})}^{{\rm net}}(y_{1},|\vec{b}_{1}-\vec{b}_{l}^{A}|,Z_{A|B}^{(m)})\right]
\label{eq:Z-A}\end{eqnarray}
and $\hat{\chi}_{p(j_{m}^{A})}^{{\rm fan}}$ is defined by the equation
\begin{eqnarray}
\hat{\chi}_{p(j)}^{{\rm fan}}(y_{1},b',Z_{1},Z_{2},Z_{3})=\chi_{p(j)}^{{\rm loop}}(y_{1},b')+G\int_{\xi}^{y_{1}-\xi}\! dy_{2}\int\! d^{2}b_{2}\;\left(1-e^{-\chi^{{\rm loop}}(y_{1}-y_{2},|\vec{b}'-\vec{b}_{2}|)}\right)\nonumber \\
\times\left[\left(1-e^{-\hat{\chi}_{p(j)}^{{\rm fan}}(y_{2},b_{2},Z_{1},Z_{2},Z_{3})}\right)\; Z_{1}\, Z_{2}\, Z_{3}-\hat{\chi}_{p(j)}^{{\rm fan}}(y_{2},b_{2},Z_{1},Z_{2},Z_{3})\right.\nonumber \\
+\left.\left(1-e^{-\chi_{p(j)}^{{\rm net}}(y_{2},b_{2},Z_1)}\right)\;
 Z_{1}\, Z_{3}\,(1-Z_{2})\right]\!.\label{eq:chi-hat-param}\end{eqnarray}

In a similar way we can obtain contributions of other subsets of cut
net-fan graphs, defined in Appendix~B, and apply them to decompose
partial opacities for various macro-configurations of nucleus-nucleus
collisions (as defined in Fig.~\ref{fig: tree-dif1}) in the
form resembling binarylike nucleon-nucleon collisions {[}cf.~(\ref{eq:chi-AB-enh}-\ref{eq:chi-AB-mn}){]}:\begin{equation}
\bar{\Omega}_{AB}^{(i)}(s,b,\{j^{A},k^{B},\vec{b}^{A},\vec{b}^{B}\})=\sum_{m=1}^{A}\sum_{n=1}^{B}\bar{\Omega}_{mn}^{(i)}(s,b,\{j^{A},k^{B},\vec{b}^{A},\vec{b}^{B}\})\,,\label{eq:chi-enh(i)-decomp}\end{equation}
which satisfy\begin{equation}
\sum_{i=1}^{11}\bar{\Omega}_{mn}^{(i)}(s,b,\{j^{A},k^{B},\vec{b}^{A},\vec{b}^{B}\})=\Omega_{AB}^{(mn)}(s,b,\{j^{A},k^{B},\vec{b}^{A},\vec{b}^{B}\})\,,\label{opac-AB-iden}\end{equation}
with $\Omega_{AB}^{(mn)}$ being defined in (\ref{eq:opac-mn}). Such
a decomposition is a useful technical trick for the MC implementation
of the approach: Each term in the decomposition corresponds to an
inelastic rescattering process between a given pair $(mn)$ of the
projectile and target nucleons but generally involves additional inelastic
rescatterings on other nucleons of the two nuclei.

For example, for the configuration defined by the graphs in the 1st
square bracket in Fig.~\ref{fig: tree-dif1} one obtains\begin{eqnarray}
\bar{\Omega}_{mn}^{(1)}=\frac{G}{2}\int_{\xi}^{Y-\xi}\!\! dy_{1}\!\int\!\! d^{2}b_{1}\left\{ \left[\left(1-e^{-2\chi_{p(j_{m}^{A})}^{{\rm net}}}\right)e^{-2\sum_{l=1}^{m-1}\chi_{p(j_{l}^{A})}^{{\rm net}}}-2\chi_{p(j_{m}^{A})}^{{\rm net}}\, e^{-2\sum_{l=1}^{A}\chi_{p(j_{l}^{A})}^{{\rm net}}}\right]\right.\nonumber \\
\times\left[\left(1-e^{-2\chi_{p(k_{n}^{B})}^{{\rm net}}}\right)e^{-2\sum_{i=1}^{n-1}\chi_{p(k_{i}^{B})}^{{\rm net}}}-2\chi_{p(k_{n}^{B})}^{{\rm net}}\, e^{-2\sum_{i=1}^{B}\chi_{p(k_{i}^{B})}^{{\rm net}}}\right]\nonumber \\
-2\left[\left(1-e^{-2\chi_{p(j_{m}^{A})}^{{\rm net}}}\right)e^{-2\sum_{l=1}^{m-1}\chi_{p(j_{l}^{A})}^{{\rm net}}}-2\chi_{p(j_{m}^{A})}^{{\rm net}}\, e^{-2\sum_{l=1}^{A}\chi_{p(j_{l}^{A})}^{{\rm net}}}\right]\nonumber \\
\times\left[\left(e^{\tilde{\chi}_{p(k_{n}^{B})}^{{\rm fan}}}-1\right)e^{\sum_{i=n+1}^{B}\tilde{\chi}_{p(k_{i}^{B})}^{{\rm fan}}}-\tilde{\chi}_{p(k_{n}^{B})}^{{\rm fan}}\right]e^{-\sum_{i=1}^{B}\chi_{p(k_{i}^{B})}^{{\rm net}}}\nonumber \\
-2\left[\left(e^{\tilde{\chi}_{p(j_{m}^{A})}^{{\rm fan}}}-1\right)e^{\sum_{l=m+1}^{A}\tilde{\chi}_{p(j_{l}^{A})}^{{\rm fan}}}-\tilde{\chi}_{p(j_{m}^{A})}^{{\rm fan}}\right]e^{-\sum_{l=1}^{A}\chi_{p(j_{l}^{A})}^{{\rm net}}}\nonumber \\
\times\left.\left[\left(1-e^{-2\chi_{p(k_{n}^{B})}^{{\rm net}}}\right)e^{-2\sum_{i=1}^{n-1}\chi_{p(k_{i}^{B})}^{{\rm net}}}-2\chi_{p(k_{n}^{B})}^{{\rm net}}\, e^{-2\sum_{i=1}^{B}\chi_{p(k_{i}^{B})}^{{\rm net}}}\right]\right\} \!.\label{eq:chi-AB-(1)}\end{eqnarray}

Using (\ref{opac-AB-iden}), we can easily write down the absorptive
nucleus-nucleus cross section:\begin{eqnarray}
\sigma_{AB}^{{\rm abs}}(s)=\int\! d^{2}b\left\langle \left\langle \prod_{m=1}^{A}\prod_{n=1}^{B}\left[\sum_{N_{mn}=0}^{\infty}\frac{\left[\sum_{i=1}^{11}\bar{\Omega}_{mn}^{(i)}\right]^{N_{mn}}}{N_{mn}!}\, e^{-\Omega_{AB}^{(mn)}(s,b,\{j^{A},k^{B},\vec{b}^{A},\vec{b}^{B}\})}\right]\right.\right.\nonumber \\
-\left.\left.\prod_{m=1}^{A}\prod_{n=1}^{B}e^{-\Omega_{AB}^{(mn)}(s,b,\{j^{A},k^{B},\vec{b}^{A},\vec{b}^{B}\})}\right\rangle _{A}\right\rangle _{B}=\int\! d^{2}b\,\left\langle \left\langle 1-e^{-\sum_{m=1}^{A}\sum_{n=1}^{B}\Omega_{AB}^{(mn)}(s,b,\{j^{A},k^{B},\vec{b}^{A},\vec{b}^{B}\})}\right\rangle _{A}\right\rangle _{B}\!,\label{eq:sigma_AB-abs}\end{eqnarray}
which can be applied for a MC treatment of inelastic nucleus-nucleus
(hadron-nucleus) collisions. It is worth stressing again that partial
opacities $\bar{\Omega}_{mn}^{(i)}$ generally correspond not only
to an inelastic rescattering between $m$-th projectile and $n$-th
target nucleons but involve also inelastic interactions with $(m+1)$-th,
$\cdots,$ $A$-th projectile and $(n+1)$-th, $\cdots,$ $B$-th
target nucleons.

\section*{Appendix D}

\begin{figure}[htb]
\begin{centering}
\includegraphics[width=10cm,height=3cm]{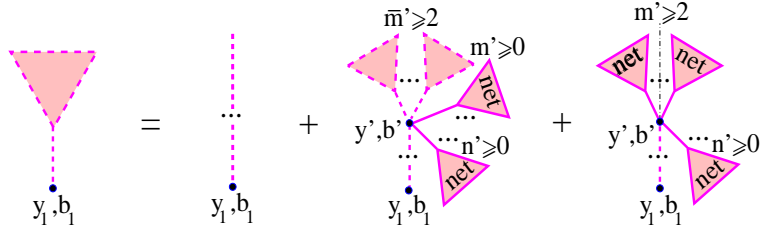}
\par\end{centering}

\caption{Recursive representation for the contribution of fanlike cuts of
net-fans, Pomeron loops and central rapidity gaps neglected.\label{netfan-noloop-fig}}

\end{figure}

\begin{figure}[htb]

\begin{centering}
\includegraphics[width=15cm,height=2.5cm]{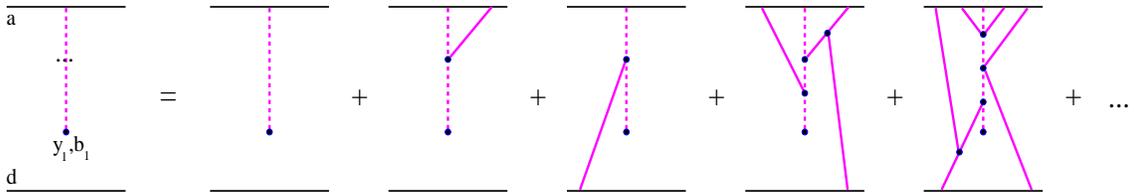}\caption{Examples of cut diagrams corresponding to a single $t$-channel sequence
of cut Pomerons exchanged between the vertex $(y_{1},\vec{b}_{1})$
and the projectile hadron.\label{fig:cutpom-line}}

\par\end{centering}

\end{figure}

The relations of Appendix~B allow one to reconstruct the cut Pomeron
structure for a cut net-fan contribution. For simplicity, we shall
illustrate the procedure neglecting the production of large rapidity
gaps at central rapidities, i.e.~neglecting the contributions of
Pomeron loops and the ones of fanlike cuts of net-fans, which leave
the handle of the fan uncut. Thus, we use 
$2\bar{\chi}_{a(j)|d(k)}^{{\rm fan}}=2\hat{\chi}_{a(j)|d(k)}^{{\rm fan}}$,
$2\tilde{\chi}_{a(j)|d(k)}^{{\rm fan}}=0$ and the representation of Fig.~\ref{fan-cut1-fig}
takes the form (see Fig.~\ref{netfan-noloop-fig})%
\begin{eqnarray}
2\bar{\chi}_{a(j)|d(k)}^{{\rm fan}}(y_{1},\vec{b}_{1}|Y,\vec{b})
=2\bar{\chi}_{a(j)|d(k)}^{\mathbb{P}}(y_{1},\vec{b}_{1}|Y,\vec{b})
+G\int_{\xi}^{y_{1}-\xi}\! dy'\int\! d^{2}b'\;
\chi_{a(j)|d(k)}^{\mathbb{P}_{{\rm cc}}}(y_{1},y',\vec{b}_{1},\vec{b}'|Y,\vec{b})\nonumber \\
\times
\;e^{-2\chi_{d(k)|a(j)}^{{\rm net}}}
 \left[\sum_{\bar{m}'=2}^{\infty}
\frac{\left(2\bar{\chi}_{a(j)|d(k)}^{{\rm fan}}\right)^{\bar{m}'}}{\bar{m}'!}\, 
e^{-2\chi_{a(j)|d(k)}^{{\rm net}}}
+\left(1-e^{-\chi_{a(j)|d(k)}^{{\rm net}}}\right)^{2}\right]\!,
\label{eq:fan-cut-nogap}\end{eqnarray}
where the omitted arguments of the eikonals in the integrand are the
same as in (\ref{eq:chi-hat-alt}) and the 1st term in the r.h.s.,
$2\bar{\chi}_{a(j)|d(k)}^{\mathbb{P}}(y_{1},\vec{b}_{1}|Y,\vec{b})$,
is the contribution of a $t$-channel sequence of cut Pomerons, exchanged
between the vertex $(y_{1},\vec{b}_{1})$ and the projectile hadron,
with the multi-Pomeron vertices which couple neighboring Pomerons
to each other being connected to at least one uncut projectile or
target net-fan. Similarly, $2\chi_{a(j)|d(k)}^{\mathbb{P}_{{\rm cc}}}(y_{1},y',\vec{b}_{1},\vec{b}'|Y,\vec{b})$
defines the contribution of such a cut Pomeron sequence exchanged
between the vertices $(y_{1},\vec{b}_{1})$ and $(y',\vec{b}')$.
The two contributions are defined by recursive equations 
{[}cf.~(\ref{eq:chi-loop-scr}),~(\ref{eq:loop-cc}){]}
\begin{eqnarray}
\bar{\chi}_{a(j)|d(k)}^{\mathbb{P}}(y_{1},\vec{b}_{1}|Y,\vec{b})
=\chi_{a(j)}^{\mathbb{P}}(y_{1},b_{1})
+G\int_{\xi}^{y_{1}-\xi}\! dy_{2}\int\! d^{2}b_{2}\;
\chi^{\mathbb{P}}(y_{1}-y_{2},|\vec{b}_{1}-\vec{b}_{2}|)\nonumber \\
\times\bar{\chi}_{a(j)|d(k)}^{\mathbb{P}}(y_{2},\vec{b}_{2}|Y,\vec{b})
\left[e^{-2\chi_{a(j)|d(k)}^{{\rm net}}(y_{2},\vec{b}_{2}|Y,\vec{b})
-2\chi_{d(k)|a(j)}^{{\rm net}}(Y-y_{2},\vec{b}-\vec{b}_{2}|Y,\vec{b})}-1\right]
\label{eq:pom-leg-cc}\\
\chi_{a(j)|d(k)}^{\mathbb{P}_{{\rm cc}}}(y_{1},y',\vec{b}_{1},\vec{b}'|Y,\vec{b})
=\chi^{\mathbb{P}}(y_{1}-y',|\vec{b}_{1}-\vec{b}'|)
+G\!\int_{y_{1}+\xi}^{y'-\xi}\!\! dy_{2}\int\!\! d^{2}b_{2}\;
\chi^{\mathbb{P}}(y_{1}-y_{2},|\vec{b}_{1}-\vec{b}_{2}|)\nonumber \\
\times\chi_{a(j)|d(k)}^{\mathbb{P}_{{\rm cc}}}(y_{2},y',\vec{b}_{2},\vec{b}'|Y,\vec{b})
\left[e^{-2\chi_{a(j)|d(k)}^{{\rm net}}(y_{2},\vec{b}_{2}|Y,\vec{b})
-2\chi_{d(k)|a(j)}^{{\rm net}}(Y-y_{2},\vec{b}-\vec{b}_{2}|Y,\vec{b})}-1\right]\!.
\label{eq:pom-int-cc}\end{eqnarray}
Examples of diagrams generated by Eq.~(\ref{eq:pom-leg-cc}) are
depicted in Fig.~\ref{fig:cutpom-line}. %
The corresponding piece of secondary hadron production is represented
by a single chain of particles produced between the projectile hadron
and the vertex $(y_{1},\vec{b}_{1})$, as defined by the 1st graph
in the r.h.s.~of the figure (single cut Pomeron exchange). All the
other graphs in the r.h.s.~have the same particle production pattern
and describe absorptive corrections to the process due to virtual
(elastic) rescatterings on the projectile and target hadrons of intermediate
partons of the underlying parton cascade.

Using Eq.~(\ref{eq:fan-cut-nogap}), one can easily generate the
cut structure for the contribution $2\bar{\chi}_{a(j)|d(k)}^{{\rm fan}}(y_{1},\vec{b}_{1}|Y,\vec{b})$.
With the probability $w_{1\mathbb{P}}=\bar{\chi}_{a(j)|d(k)}^{\mathbb{P}}(y_{1},\vec{b}_{1}|Y,\vec{b})/\bar{\chi}_{a(j)|d(k)}^{{\rm fan}}(y_{1},\vec{b}_{1}|Y,\vec{b})$,
the particle production pattern is the one of a single cut Pomeron
exchange between the projectile hadron and the vertex $(y_{1},\vec{b}_{1})$.
In the opposite case, sampled with the probability $1-w_{1\mathbb{P}}$,
one generates the rapidity $y'$ and transverse vector $\vec{b}'$
of the new multi-Pomeron vertex - according to the integrand of the
2nd term in the r.h.s.~of Eq.~(\ref{eq:fan-cut-nogap}). Then, with
the partial probability
\[
w_{{\rm gap}}=\frac{\left(1-e^{-\chi_{a(j)|d(k)}^{{\rm net}}}\right)^{2}}
{\left(1-e^{-\chi_{a(j)|d(k)}^{{\rm net}}}\right)^{2}
+\left(e^{2\bar{\chi}_{a(j)|d(k)}^{{\rm fan}}}-1
-2\bar{\chi}_{a(j)|d(k)}^{{\rm fan}}\right)
e^{-2\chi_{a(j)|d(k)}^{{\rm net}}}}\]
the corresponding piece of the final state consists of a single chain
of secondaries produced in the rapidity interval $[y_{1},y']$ {[}single
cut Pomeron exchange between $(y_{1},\vec{b}_{1})$ and $(y',\vec{b}')${]},
with the projectile hadron being separated from the particles produced
by a LRG. Alternatively, with the probability $1-w_{{\rm gap}}$,
one obtains a fanlike structure for the particle production pattern:
In addition to the above-mentioned chain of secondaries, produced
in the interval $[y_{1},y']$, secondary particles emerge from $\bar{m}'\geq2$
cut net-fans exchanged between the vertex $(y',\vec{b}')$ and the
projectile hadron. In such a case, one generates the number of cut
net-fans $\bar{m}'$ according to the Poisson distribution with the
mean $2\bar{\chi}_{a(j)|d(k)}^{{\rm fan}}$ (rejecting the cases $\bar{m}'\leq1$)
and applies the above-discussed procedure recursively for each of
the $\bar{m}'$ cut net-fans.

Taking into consideration Pomeron loops and central rapidity gaps,
the procedure remains qualitatively similar, being then based on Eqs.~(\ref{eq:chi-hat-alt}-\ref{eq:loop-uu}).
The difference compared to the above-discussed treatment is that instead
of $t$-channel sequences of cut Pomerons (as exemplified in Fig.~\ref{fig:cutpom-line})
one generally obtains cut $t$-channel sequences of Pomerons and Pomeron
loops, which are connected to each other by multi-Pomeron vertices
coupled to uncut net-fans. Hence, a similar algorithm is applied to
reconstruct the cut Pomeron structure of those cut loop sequences.

\section*{Appendix E}

\begin{figure}[htb]
\begin{centering}
\includegraphics[width=7cm,height=4cm]{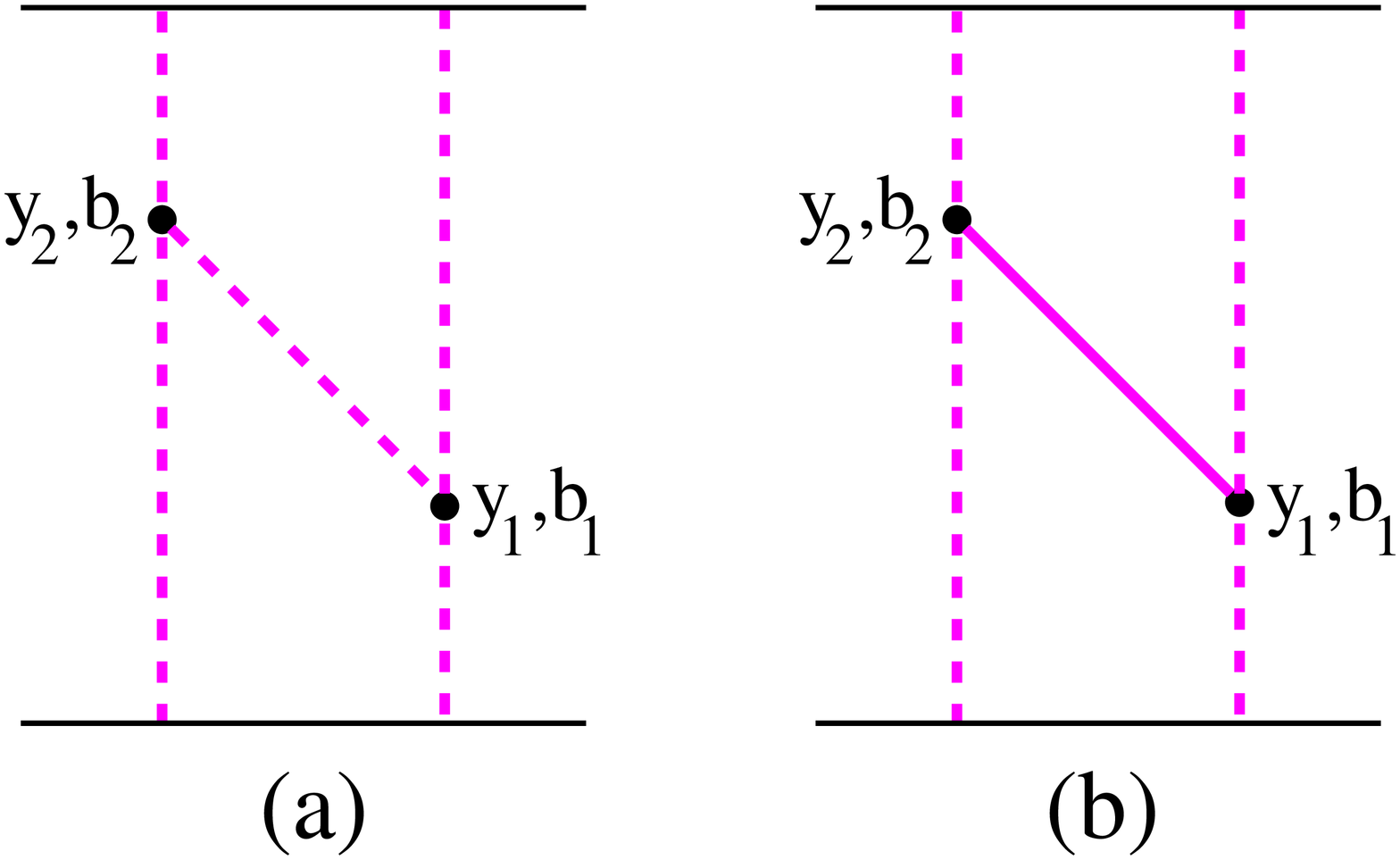}
\par\end{centering}

\caption{Lowest order zigzaglike cut diagrams. \label{fig:zigzag}}

\end{figure}

To illustrate the effect of zigzaglike cut contributions, let us
consider the simplest cut graphs of that kind shown in Fig.~\ref{fig:zigzag}~(a,b).%
 The contribution of the graph in Fig.~\ref{fig:zigzag}~(a) is
\begin{eqnarray}
\Delta\hat{\Omega}_{ad(jk)}^{{\rm zz}}(s,b)=8G^{2}\int\! d^{2}b_{1}d^{2}b_{2}\int_{\xi}^{Y-2\xi}\! dy_{1}\int_{y_{1}+\xi}^{Y-\xi}\! dy_{2}\;\chi_{a(j)}^{\mathbb{P}}(Y-y_{1},|\vec{b}-\vec{b}_{1}|)\,\nonumber \\
\times\chi_{a(j)}^{\mathbb{P}}(Y-y_{2},|\vec{b}-\vec{b}_{2}|)\,\chi_{d(k)}^{\mathbb{P}}(y_{1},b_{1})\,\chi_{d(k)}^{\mathbb{P}}(y_{2},b_{2})\,\chi^{\mathbb{P}}(y_{2}-y_{1},|\vec{b}_{2}-\vec{b}_{1}|)\,,\label{eq:zz(0)}\end{eqnarray}
and the one of the graph in Fig.~\ref{fig:zigzag}~(b) is defined
by the same expression up to a sign, $\Delta\tilde{\Omega}_{ad(jk)}^{{\rm zz}}(s,b)=-\Delta\hat{\Omega}_{ad(jk)}^{{\rm zz}}(s,b)$.
The diagram in Fig.~\ref{fig:zigzag}~(b) provides a (negative)
screening correction to the eikonal configuration with two cut Pomerons.
On the other hand, the one in Fig.~\ref{fig:zigzag}~(a) introduces
a new process, with the weight being equal to the one of the mentioned
screening contribution, and with the particle production pattern being
almost identical to the one of Fig.~\ref{fig:zigzag}~(b); the only
difference arises from the cut Pomeron exchanged between the vertices
($y_{1},b_{1}$) and ($y_{2},b_{2}$). Thus, the combined effect of
these two graphs is to provide additional particle production in the
rapidity interval $[y_{1},y_{2}]$. Hence, to account for the contributions
of the graphs of Fig.~\ref{fig:zigzag} to secondary particle production,
one has to select final state configurations with just two cut Pomerons
exchanged and, with the probability 
$w_{{\rm zz}}=\Delta\hat{\Omega}_{ad(jk)}^{{\rm zz}}(s,b)
/(\Omega_{ad(jk)}^{(2\mathbb{P})}(s,b)$
{[}$\Omega_{ad(jk)}^{(2\mathbb{P})}(s,b)$ being the partial weight
of the two cut Pomerons process{]} to add an additional cut Pomeron
exchange between the vertices ($y_{1},b_{1}$) and ($y_{2},b_{2}$),
with the rapidity and transverse coordinates of the vertices being
generated according to the integrand of Eq.~(\ref{eq:zz(0)}). At
sufficiently high energies
 $\Delta\hat{\Omega}_{ad(jk)}^{{\rm zz}}(s,b)>
 \Omega_{ad(jk)}^{(2\mathbb{P})}(s,b)$
due to the faster energy rise of the enhanced graph contributions.
A simple effective procedure would then be to consider $w_{zz}$ as
the mean number of additional Pomerons to be added to the initial
configuration.

The general treatment of zigzaglike cut graphs follows the above-discussed
logic. We restrict ourselves with the set of zigzaglike cut graphs
which provide nonzero contribution to \textsl{inclusive} particle
spectra and split it into two subsets whose contributions are equal
up to a sign: $\hat{\Omega}_{ad(jk)}^{{\rm zz}}(s,b)=-\tilde{\Omega}_{ad(jk)}^{{\rm zz}}(s,b)$,
where $\hat{\Omega}_{ad(jk)}^{{\rm zz}}$ can be written as\begin{equation}
\hat{\Omega}_{ad(jk)}^{{\rm zz}}(s,b)=\int\! d^{2}b_{1}d^{2}b_{2}\int_{\xi}^{Y-2\xi}\! dy_{1}\int_{y_{1}+\xi}^{Y-\xi}\! dy_{2}\;2\chi_{ad|jk}^{{\rm zz}}(y_{1},y_{2},\vec{b}_{1},\vec{b}_{2}|Y,\vec{b})\,.\label{eq:zz-tot}\end{equation}
The particle production pattern is almost identical for the contributions
$\hat{\Omega}_{ad(jk)}^{{\rm zz}}$ and $\tilde{\Omega}_{ad(jk)}^{{\rm zz}}$,
except that the former contains a \textsl{cut} sequence of Pomerons
and Pomeron loops exchanged between the vertices ($y_{1},b_{1}$)
and ($y_{2},b_{2}$) (with internal multi-Pomeron vertices in the
sequence being generally coupled to \textsl{uncut} projectile and/or
target net-fans) while the same sequence remains \textsl{uncut} in
the latter contribution. Thus, the combined effect of both subsets
of graphs is to add additional cut Pomerons resulting from this cut
loop sequence. For brevity, we shall not discuss the corresponding technical
implementation.

\end{document}